\documentclass[longauth]{aa}

\usepackage{graphicx}

\usepackage{txfonts}
\usepackage{natbib}
\usepackage[colorlinks=true, citecolor=blue, linkcolor=blue, urlcolor=blue]{hyperref}
\usepackage{siunitx}

\usepackage[utf8]{inputenc}
\usepackage[usenames, dvipsnames]{color}
\usepackage{amsmath}
\usepackage[english]{babel}
\usepackage{float}
\usepackage{url}
\usepackage{longtable}
\usepackage{fancyhdr}
\usepackage[version=4]{mhchem}
\usepackage{gensymb}

\usepackage{supertabular}

\newcommand{\tess}{\textit{TESS}\xspace}

\begin{document} 

   \title{TOI-1801\,b: A temperate mini-Neptune around a young M0.5 dwarf}

   \author{
 M.\,Mallorqu\'in\inst{\ref{i:iac},\ref{i:ull}}, 
 E.\,Goffo\inst{\ref{i:tls},\ref{i:unito}}, 
 E.\,Pall\'e\inst{\ref{i:iac},\ref{i:ull}}, 
 N.\,Lodieu\inst{\ref{i:iac},\ref{i:ull}},  
 V.\,J.\,S.\,B\'ejar\inst{\ref{i:iac},\ref{i:ull}}, 
 H.\,Isaacson\inst{\ref{i:ber}}, 
 M.\,R.\,Zapatero Osorio\inst{\ref{i:cab1}},    
 S.\,Dreizler\inst{\ref{i:got}},    
 S.\,Stock\inst{\ref{i:hei}},
 R.\,Luque\inst{\ref{i:chi}},   
 F.\,Murgas\inst{\ref{i:iac}, \ref{i:ull}}, 
 L.\,Peña\inst{\ref{i:iaa}}, 
 J.\,Sanz-Forcada\inst{\ref{i:esac}}, 
 G.\,Morello\inst{\ref{i:swe},\ref{i:iac}}, 
 D.\,R.\,Ciardi\inst{\ref{i:cal}}, 
 E.\,Furlan\inst{\ref{i:cal}}, 
 K.\,A.\,Collins\inst{\ref{i:cam}}, 
 E.\,Herrero\inst{\ref{i:ieec}}, 
 S.\,Vanaverbeke\inst{\ref{i:be1}, \ref{i:be2}, \ref{i:be3}}, 
 P.\,Plavchan\inst{\ref{i:george}}, 
 N.\,Narita\inst{\ref{i:meg}, \ref{i:mitaka}, \ref{i:iac}}, 
 A.\,Schweitzer\inst{\ref{i:ham}}, 
 M.\,P\'erez-Torres\inst{\ref{i:iaa}}, 
 A.\,Quirrenbach\inst{\ref{i:hei}}, 
 J.\,Kemmer\inst{\ref{i:hei}}, 
 A.\,P.\,Hatzes\inst{\ref{i:tls}},
 A.\,Howard\inst{\ref{i:ber}}, 
 M.\,Schlecker\inst{\ref{i:ari}}, 
 S.\,Reffert\inst{\ref{i:hei}}, 
 E.\,Nagel\inst{\ref{i:got}}, 
 J.\,C.\,Morales\inst{\ref{i:ieec}, \ref{i:ice}}, 
 J.\,Orell-Miquel\inst{\ref{i:iac}, \ref{i:ull}}, 
 C.\,Duque-Arribas\inst{\ref{i:ucm}}, 
 I.\,Carleo\inst{\ref{i:iac}, \ref{i:ull}}, 
 C.\,Cifuentes\inst{\ref{i:esac}}, 
 G.\,Nowak\inst{\ref{i:ncu}, \ref{i:iac},\ref{i:ull}}, 
 I.\,Ribas\inst{\ref{i:ieec},\ref{i:ice}}, 
 A.\,Reiners\inst{\ref{i:got}}, 
 P.\,J.\,Amado\inst{\ref{i:iaa}}, 
 J.\,A.\,Caballero\inst{\ref{i:esac}}, 
 Th.\,Henning\inst{\ref{i:max}}, 
 V.\,Pinter\inst{\ref{i:caha}}, 
 J.\,M.\,Akana\,Murphy\inst{\ref{tks12}, \ref{i:ca}}, 
 C.\,Beard\inst{\ref{tks8}, \ref{tks9}}, 
 S.\,Blunt\inst{\ref{tks4}}, 
 C.\,L.\,Brinkman\inst{\ref{tks10}, \ref{tks11}}, 
 B.\,Cale\inst{\ref{i:ipac}}, 
 A.\,Chontos\inst{\ref{tks14}, \ref{tks15}, \ref{tks10}}, 
 K.\,I.\,Collins\inst{\ref{i:george}}, 
 I.\,J.\,M.\,Crossfield\inst{\ref{tks7}}, 
 F.\,Dai\inst{\ref{tks2}, \ref{tks3}, \ref{tks4}}, 
 P.\,A.\,Dalba\inst{\ref{i:ca}, \ref{tks13}}, 
 S.\,Dufoer\inst{\ref{i:be2}}, 
 M.\,El Mufti\inst{\ref{i:george}}, 
 N.\,Espinoza\inst{\ref{i:bal}}, 
 T.\,Fetherolf\inst{\ref{tks5}, \ref{tks11}}, 
 A.\,Fukui\inst{\ref{i:meg}, \ref{i:iac}}, 
 S.\,Giacalone\inst{\ref{i:ber}}, 
 C.\,Gnilka\inst{\ref{i:mo}}, 
 E.\,Gonzales\inst{\ref{i:ca}}, 
 S.\,K.\,Grunblatt\inst{\ref{tks16}}, 
 S.\,Howell\inst{\ref{i:mo}}, 
 D.\,Huber\inst{\ref{tks10}, \ref{tks16}}, 
 S.\,R.\,Kane\inst{\ref{tks11}}, 
 J.\,P.\, de Le\'on\inst{\ref{i:meg}}, 
 J.\,Lubin\inst{\ref{tks9}}, 
 M.\,G.\,MacDougall\inst{\ref{tks1}}, 
 B.\,Massey\inst{\ref{i:la}}, 
 D.\,Montes\inst{\ref{i:ucm}}, 
 M.\,Mori\inst{\ref{i:meg}}, 
 H.\,Parviainen\inst{\ref{i:iac}, \ref{i:ull}}, 
 V.\,M.\,Passegger\inst{\ref{i:iac}, \ref{i:ull}, \ref{i:ham}, \ref{i:okl}}, 
 A.\,S.\,Polanski\inst{\ref{tks7}}, 
 P.\,Robertson\inst{\ref{tks9}}, 
 R.\,P.\,Schwarz\inst{\ref{i:cam}}, 
 G.\,Srdoc\inst{\ref{i:vi}}, 
 H.\,M.\,Tabernero\inst{\ref{i:cab1}}, 
 A.\,Tanner\inst{\ref{i:mis}}, 
 E.\,Turtelboom\inst{\ref{i:ber}}, 
 J.\,Van\,Zandt\inst{\ref{tks1}}, 
 L.\,Weiss\inst{\ref{tks6}}, 
 M.\,Zechmeister\inst{\ref{i:got}} 
     }
   \authorrunning{Mallorqu\'in, M., et al.}

  \institute{
        \label{i:iac} Instituto de Astrof\'isica de Canarias (IAC), Calle V\'ia L\'actea s/n, 38205 La Laguna, Tenerife, Spain
             \email{mmd@iac.es}
  	\and 
        \label{i:ull} Departamento de Astrof\'isica, Universidad de La Laguna (ULL), 38206 La Laguna, Tenerife, Spain
  	\and
        \label{i:tls} Th\"uringer Landessternwarte Tautenburg, D-07778 Tautenburg, Germany    
	\and 
        \label{i:unito} Dipartimento di Fisica, Universita degli Studi di Torino, via Pietro Giuria 1, I-10125, Torino, Italy
        \and
        \label{i:ber} Department of Astronomy, 501 Campbell Hall, University of California at Berkeley, Berkeley, CA 94720, USA
        \and
        \label{i:cab1} Centro de Astrobiolog\'ia (CSIC-INTA), Carretera de Ajalvir km 4, 28850 Torrej\'on de Ardoz, Madrid, Spain
        \and
        \label{i:got} Institut f\"ur Astrophysik und Geophysik, Georg-August-Universit\"at G\"ottingen, Friedrich-Hund-Platz 1, 37077 G\"ottingen, Germany
        \and
        \label{i:hei} Landessternwarte, Zentrum f\"ur Astronomie der Universit\"at Heidelberg, K\"onigstuhl 12, 69117, Heidelberg, Germany
        \and
        \label{i:chi} Department of Astronomy \& Astrophysics, University of Chicago, Chicago, IL 60637, USA
        \and
        \label{i:iaa} Instituto de Astrof\'isica de Andaluc\'ia (IAA-CSIC), Glorieta de la Astronom\'ia s/n, 18008, Granada, Spain
        \and
        \label{i:esac} Centro de Astrobiología (CAB), (CSIC-INTA), ESAC Campus, Camino bajo del castillo s/n, 28692, Villanueva de la Cañada, Madrid, Spain
        \and
        \label{i:swe} Department of Space, Earth and Environment, Chalmers University of Technology, SE-412 96 Gothenburg, Sweden
        \and
        \label{i:cal} NASA Exoplanet Science Institute, California Institute of Technology, Pasadena, CA 91106, USA
        \and
        \label{i:cam} Center for Astrophysics, Harvard \& Smithsonian, 60 Garden Street, Cambridge, MA 02138, USA
        \and
        \label{i:ieec} Institut d'Estudis Espacials de Catalunya (IEEC), Calle Gran Capita 2-4, 08034, Barcelona, Spain
        \and
        \label{i:be1} Public observatory ASTROLAB IRIS, Provinciaal Domein “De Palingbeek”, Verbrandemolenstraat 5, 8902 Zillebeke, Ieper, Belgium
        \and
        \label{i:be2} Vereniging Voor Sterrenkunde (VVS), Oostmeers 122 C, 8000 Brugge, Belgium
        \and
        \label{i:be3} Centre for Mathematical Plasma-Astrophysics, Department of Mathematics, KU Leuven, Celestijnenlaan 200B, 3001 Heverlee, Belgium
        \and
        \label{i:george} Department of Physics and Astronomy, George Mason University, 4400 University Drive, Fairfax, VA 22030, USA
        \and
        \label{i:meg} Komaba Institute for Science, The University of Tokyo, 3-8-1 Komaba, Meguro, Tokyo 153-8902, Japan
        \and
        \label{i:mitaka} Astrobiology Center, 2-21-1 Osawa, Mitaka, Tokyo 181-8588, Japan
        \and
        \label{i:ham} Hamburger Sternwarte, Gojenbergsweg 112, D-21029 Hamburg, Germany
        \and
        \label{i:ari} Steward Observatory and Department of Astronomy, The University of Arizona, Tucson, AZ 85721, USA
        \and
        \label{i:ice} Institut de Ci\'encies de l’Espai (CSIC-IEEC), Campus UAB, c/ de Can Magrans s/n, 08193 Bellaterra, Barcelona, Spain
        \and
        \label{i:ucm} Departamento de F\'sica de la Tierra y Astrof\'isica and IPARCOS-UCM (Instituto de F\'isica de Part\'iculas y del Cosmos de la UCM), Facultad de Ciencias F\'isicas, Universidad Complutense de Madrid, 28040, Madrid, Spain
        \and
        \label{i:ncu} Institute of Astronomy, Faculty of Physics, Astronomy and Informatics, Nicolaus Copernicus University, Grudzi\c{a}dzka 5, 87-100 Toru\'n, Poland
        \and
        \label{i:max} Max-Planck-Institut f\"ur Astronomie, K\"onigstuhl 17, 69117, Heidelberg, Germany
        \and
        \label{i:caha} Centro Astron\'omico Hispano en Andaluc\'ia (CAHA), Observatorio de Calar Alto, Sierra de los Filabres, 04550, G\'ergal, Spain
        \and
        \label{tks12} NSF Graduate Research Fellow
        \and
        \label{i:ca} Department of Astronomy and Astrophysics, University of California, Santa Cruz, Santa Cruz, CA, USA
        \and
        \label{tks8} NASA FINESST Fellow
        \and
        \label{tks9} Department of Physics \& Astronomy, The University of California, Irvine, Irvine, CA 92697, USA
        \and
        \label{tks4} Department of Astronomy, California Institute of Technology, Pasadena, CA 91125, USA
        \and
        \label{tks10} Institute for Astronomy, University of Hawai'i, 2680 Woodlawn Drive, Honolulu, HI 96822 USA
        \and
        \label{tks11} Department of Earth and Planetary Sciences, University of California, Riverside, CA 92521, USA
        \and
        \label{i:ipac} Caltech IPAC – NASA Exoplanet Science Institute 1200 E. California Ave, Pasadena, CA 91125, USA
        \and
        \label{tks14} Henry Norris Russell Fellow
        \and
        \label{tks15} Department of Astrophysical Sciences, Princeton University, 4 Ivy Lane, Princeton, NJ 08540, USA
        \and
        \label{tks7} Department of Physics and Astronomy, University of Kansas, Lawrence, KS 66045, USA
        \and
        \label{tks2} NASA Sagan Fellow
        \and
        \label{tks3} Division of Geological and Planetary Sciences, 1200 E California Blvd, Pasadena, CA, 91125, USA
        \and
        \label{tks13} Heising-Simons 51 Pegasi b Postdoctoral Fellow
        \and
        \label{i:bal} Space Telescope Science Institute, 3700 San Martin Drive, Baltimore, MD 21218, USA
        \and
        \label{tks5} UC Chancellor's Fellow
        \and
        \label{i:mo} NASA Ames Research Center, Moffett Field, CA 94035, USA
        \and
        \label{tks16} William H. Miller III Department of Physics and Astronomy, Johns Hopkins University, 3400 N Charles St, Baltimore, MD 21218, USA
        \and
        \label{tks1} Department of Physics \& Astronomy, University of California Los Angeles, Los Angeles, CA 90095, USA
        \and
        \label{i:la} Villa '39 Observatory, Landers, CA 92285, USA
        \and
        \label{i:okl} Homer L. Dodge Department of Physics and Astronomy, University of Oklahoma, 440 West Brooks Street, Norman, OK 73019, USA
        \and
        \label{i:vi} Kotizarovci Observatory, Sarsoni 90, 51216 Viskovo, Croatia
        \and
        \label{i:mis} Mississippi State University, 355 Lee Boulevard, Mississippi State, MS 39762, USA
        \and
        \label{tks6} Department of Physics and Astronomy, University of Notre Dame, Notre Dame, IN 46556, USA
        \and
        \label{tks17} Sydney Institute for Astronomy (SIfA), School of Physics, University of Sydney, NSW 2006, Australia
        }

  \date{Received 04 July 2023 / Accepted 04 October 2023}

  \abstract
   {
   We report the discovery, mass, and radius determination of TOI-1801\,b, a temperate mini-Neptune around a young M dwarf. TOI-1801\,b was observed in TESS sectors 22 and 49, and the alert that this was a TESS planet candidate with a period of 21.3 days went out in April 2020. However, ground-based follow-up observations, including seeing-limited photometry in and outside transit together with precise radial velocity (RV) measurements with CARMENES and HIRES revealed that the true period of the planet is 10.6 days. These observations also allowed us to retrieve a mass of 5.74\,$\pm$\,1.46 $M_\oplus$, which together with a  radius of 2.08\,$\pm$\,0.12 $R_\oplus$, means that TOI-1801\,b is most probably composed of water and rock, with an upper limit of 2\% by mass of H$_{2}$ in its atmosphere. The stellar rotation period of 16 days is readily detectable in our RV time series and in the ground-based photometry. We derived a likely age of 600--800 Myr for the parent star TOI-1801, which means that TOI-1801\,b is the least massive young mini-Neptune with precise mass and radius determinations. Our results suggest that if TOI-1801\,b had a larger atmosphere in the past, it must have been removed by some evolutionary mechanism on timescales shorter than 1 Gyr.
   }

  \maketitle

\section{Introduction}
\label{sec:introduction}

Although the first searches for exoplanets were mainly focused on solar-type stars \citep[][and references therein]{udry07}, it is necessary to extend these studies to stars of different masses to understand the processes of planet formation. In this sense, M dwarfs are very important because they are the most numerous stars in the solar neighbourhood \citep{henry18, reyle21}, and their lower masses and radii favour the detection of less massive planets. Thus, although the first planet searches made based on radial velocities (RVs) led to the discovery of gas giant planets at short distances, that is, the so-called hot Jupiters \citep{mayor95, butler96, marcy96}, the least massive planets were discovered mainly around low-mass stars \citep{rivera05}. Studies of planets around M-type stars were initially performed using ground-based facilities and the RV and transit techniques \citep{rivera05, char09, bonfils13, trifonov18, zech19}. However, with the arrival of satellites such as CoRoT \citep{Corot}, Kepler \citep{Kepler, K2}, and mainly TESS \citep{TESS}, numerous planets have recently been detected around M dwarfs with the transit technique. Their subsequent RV follow-up with spectrographs such as CARMENES \citep{CARMENES,CARMENES18}, which has provided prolific mass determinations of small TESS planet candidates \citep{luque19, luque22a, bluhm20, bluhm21, kemmer20, kemmer22, dreizler20, soto20, kossak21, gonz22, espinoza22, chatuverdi22}, has allowed the characterisation of the masses and radii of numerous planets, studies of comparative planetology, and investigation of the physical properties of their interiors \citep{Kane2021}.

The dominant mechanisms forming planets and the timescales of planetary migration and atmospheric evaporation remain to be probed observationally even though more than 5000 exoplanets have been discovered so far. Young exoplanets detected by both transit and RV techniques offer a unique opportunity to study the processes of formation and evolution. However, the primary stars of these young planets exhibit intense activity due to the presence of magnetic spots and flares, which dominate the shape and amplitude of the photometric and RV time series \citep{Trevor2019, plav20, cale21, Suarez2022}. Consequently, discovering new transits in the light curves and extracting the Keplerian RV signals associated with the planets, which typically are several times smaller than the stellar activity-induced photometric/spectroscopic imprint, are extremely challenging.

In the last years, several studies have found a paucity of low-mass planets at short orbital periods around 1.5--2\,R$_\oplus$ for solar-type stars \citep{fulton17} and a similar gap shifted to smaller sizes for less massive stars \citep{cloutier2020}. This is also known as the small-planet valley. The knowledge of the masses and radii of these close young planets would help us to understand whether this phenomenon is primordial \citep{lee16} or, as has been suggested, is due to atmospheric evaporation \citep{Sanz2011}. Currently, two main mechanisms have been suggested to explain the erosion of planet atmospheres: photo-evaporation by energetic radiation from the star \citep{owen17}, and atmospheric escape powered by the energy of the planetary core \citep{ginzburg18}. These two processes take place on different timescales of hundreds of millions or billions of years, respectively \citep{owen17, gupta20}, and hence, the study of planets at young ages can allow us to distinguish between the two mechanisms of atmospheric evolution. However, the recent study by \cite{luque22b} suggests that the distribution of planets transiting M dwarfs that are located in the valley depends on their composition. The distribution is divided into three classes: rocky (with a similar composition to that of Earth), water-rich (i.e. planets made of 50\% rocks and 50\% water ice by mass), and gas-rich planets (i.e. either rocky planets with massive H/He envelopes or water-rich planets with less massive envelopes). \cite{Venturini20} explained this trend by describing that planets accrete large amounts of water and gas beyond the ice line and subsequently migrate inward to shorter orbits. This sequence is a typical outcome of core-accretion models \citep{Mordasini18, Brugger20, Liu20, Burn21, Schlecker21b}. Then, some mechanisms such as those mentioned above remove part or all of the planet envelopes.

Recently, dozens of transiting planets with sizes smaller than Neptune and larger than the Earth have been discovered to orbit stellar members of the Hyades \citep{mann16, ciar18, mann18}, the young moving groups of Praesepe \citep{obermeier16, mann17, rizzuto18} and $\delta$\,Lyr \citep{bouma22a} open clusters, in the Ursa Major \citep{mann20}, $\beta$\,Pictoris \citep{plav20}, Pisces-Eridanus \citep{newton21}, Melange-1 \citep{toff21}, AB Doradus \citep{zhou22}, Melange-3 \citep{barber22}, and Melange-4 \citep{wood23}, in the Cepheus-Hercules complex \citep{bouma22b}, and around other young field stars \citep{david18a, david18b, zhou21, ment21, hedges21, kossak21, barr22, vach22, leon23, barros23, dai23, Desidera2023}.  However, only ten young planetary systems have measured densities, with masses derived from dedicated RV campaigns: AU\,Mic\,b and c \citep[$\sim$20\,Myr;][]{klein21, cale21, zicher22, klein22}, TOI-1807\,b \citep[$\sim$300\,Myr;][]{nardiello22}, TOI-179\,b \citep[$\sim$300\,Myr;][]{vines23, Desidera2023}, K2-233\,d \citep[$\sim$360\,Myr;][]{barr23}, HD\,63433\,c \citep[$\sim$400\,Myr;][]{Mallorquin23}, TOI-560\,b and c \citep[$\sim$490\,Myr;][]{barr22, elmufti2023}, TOI-1099\,b \citep[$\sim$520\,Myr;][]{barros23}, K2-25\,b \citep[$\sim$725\,Myr;][]{stef20}, K2-100\,b  \citep[$\sim$750\,Myr;][]{barr19}, and TOI-1201\,b \citep[600--800\,Myr;][]{kossak21}.

In this paper, we present the discovery and mass characterisation of a sub-Neptune planet orbiting the young M0.5\,V star TOI-1801 (LP\,375-23) with an orbital period of 10.6 days. This paper is organised as follows. In Sect.\,\ref{sec:tesslc} we describe the TESS photometry and planet discovery, while in Sect.\,\ref{sec:Observations} we present ground-based follow-up observations of the system. In Sect.\,\ref{sec:properties} we determine the physical properties of the star and its wide stellar companion (LP\,375-24). We perform a transit and RV analysis of the planet in Sect.\,\ref{sec:Analysis}. We discuss the composition of the planet and the main implications in Sect.\,\ref{sec:disc}. We summarize our main results in Sect.\,\ref{sec:Concl}.

\section{TESS photometry}
\label{sec:tesslc}

 TOI-1801 and its wide companion LP\,375-24 \citep[Sect.\,\ref{sec:widecomp},][]{Weis1991} were observed by TESS between 18 February and 18 March 2020 in sector 22, and between 26 February and 26 March 2022 in sector 49, during the TESS primary mission and its first extended mission. TOI-1801 was observed in 2-minute short-cadence integration. However, LP\,375-24 was observed in 30- and 10-minute integrations in sectors 22 and 49, respectively. At the time of writing, TESS is not scheduled to re-observe TOI-1801. 

\begin{figure}[ht!]
\includegraphics[width=0.95\linewidth]{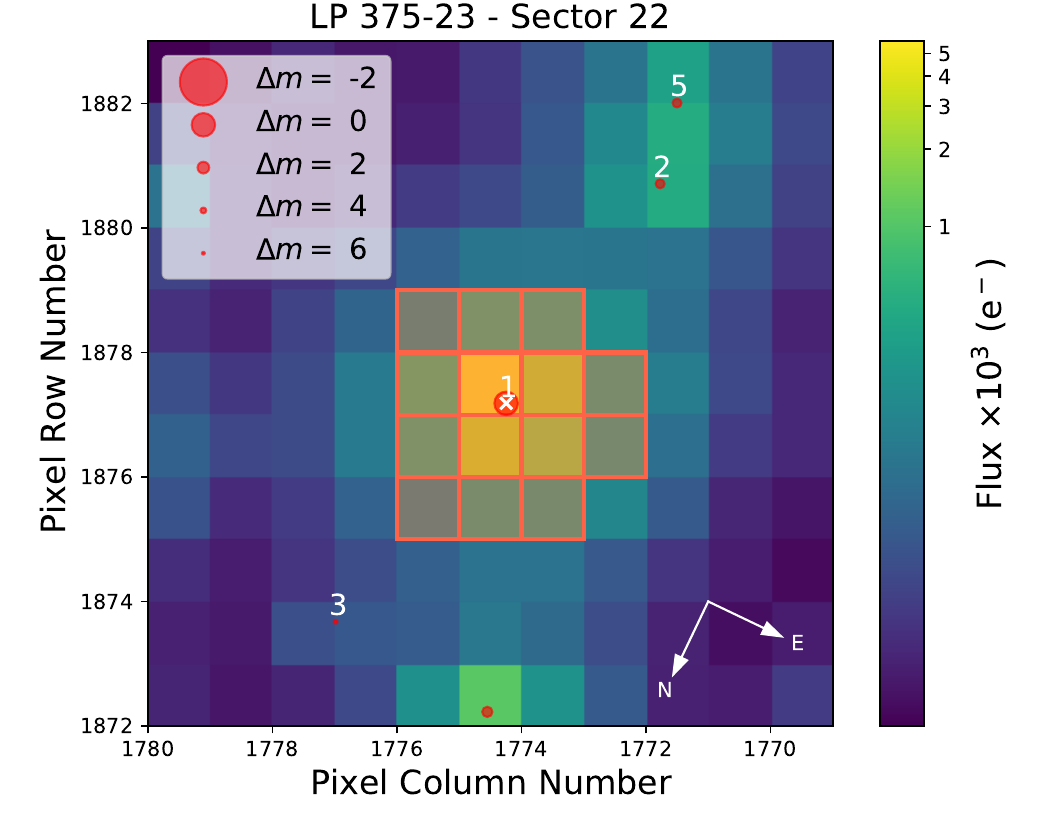}
\includegraphics[width=0.95\linewidth]{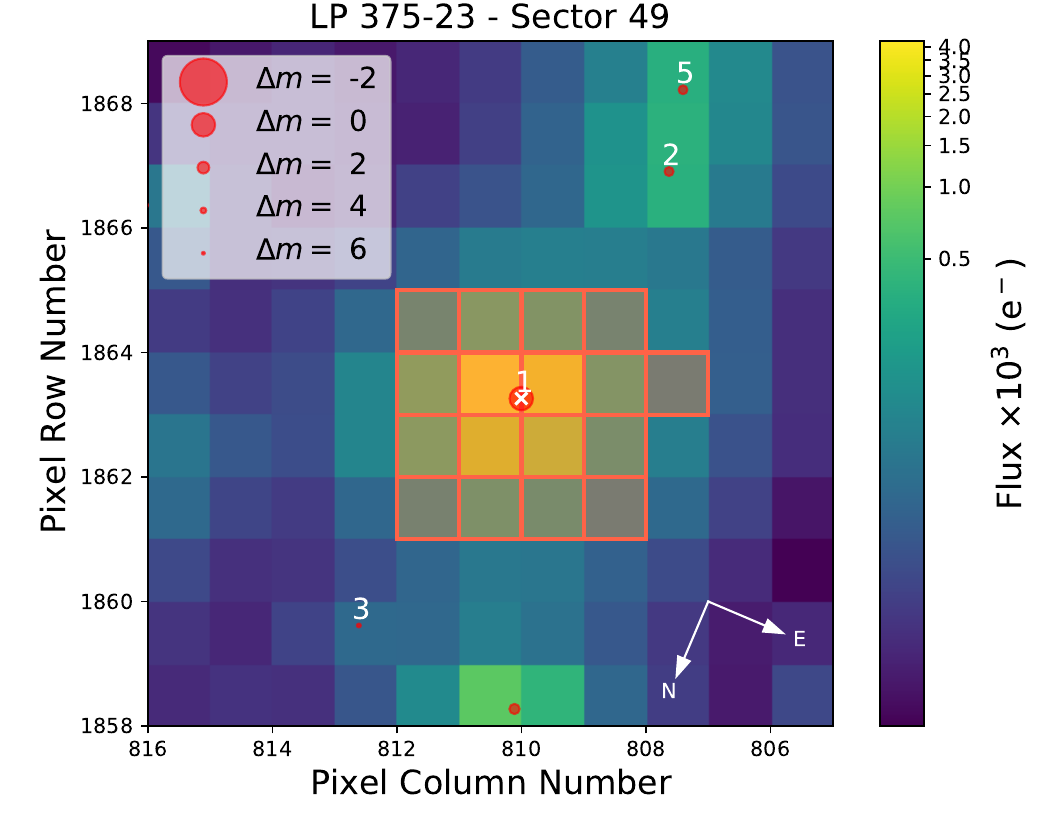}
\caption{TPF plot for TOI-1801 in sector 22 (top) and sector 49 (bottom). The red squares indicate the best-fit photometric aperture that was used to obtain the SAP flux. $G$-band magnitudes from $Gaia$ DR3 are shown with different sizes of red circles for all nearby stars up to 6 mag fainter.
\label{fig:TPF_Gaia}}
\end{figure}

\begin{figure*}[ht!]
\includegraphics[width=1\linewidth]{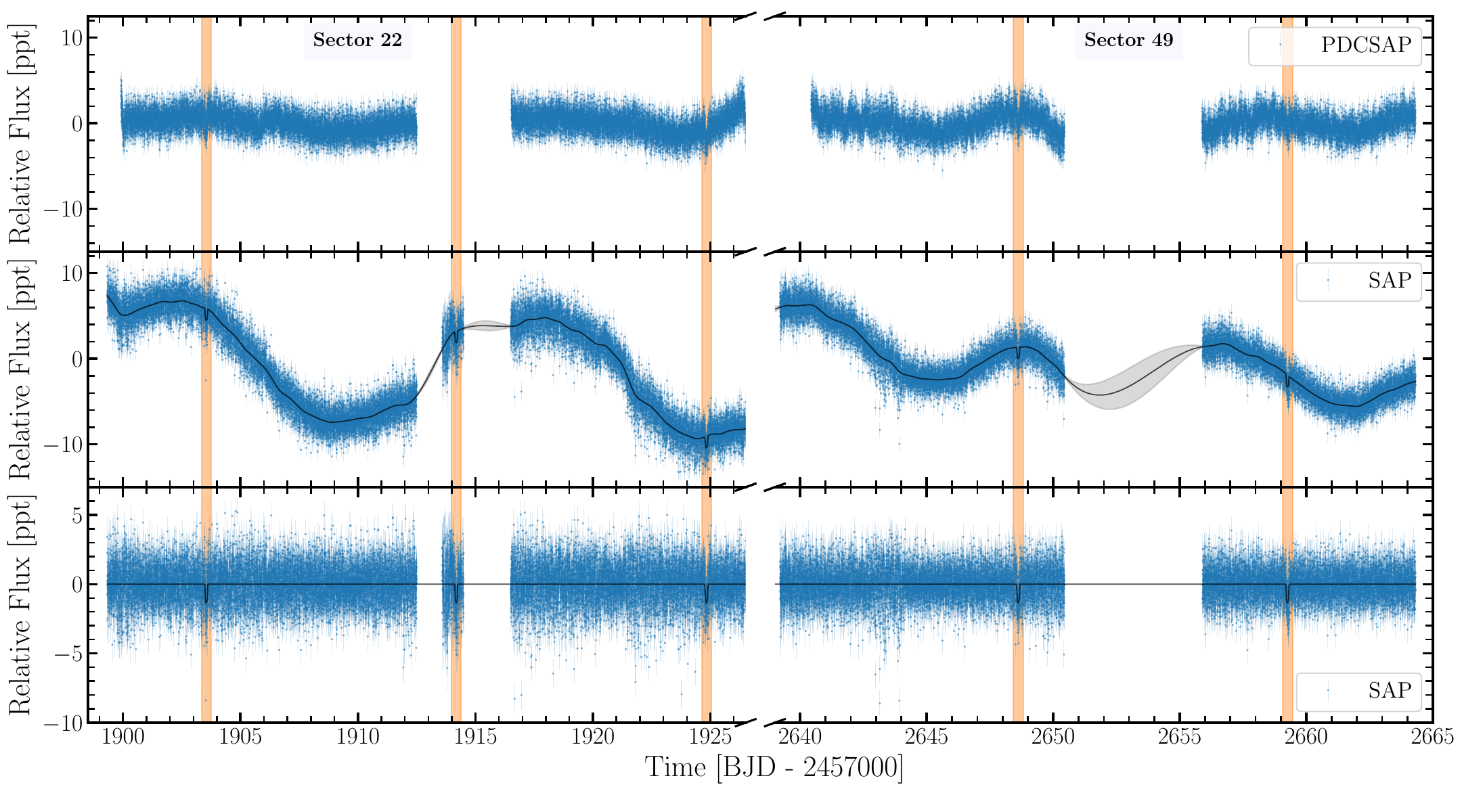}
\caption{TESS light curves of TOI-1801 (sectors 22 and 49, blue dots). The vertical orange lines indicate the location of the planetary transits of TOI-1801\,b. \textit{Top panel}: SPOC PDCSAP fluxes. \textit{Middle panel}: SPOC SAP fluxes; the black curve stands for the stellar activity plus planet transit model, the model 1$\sigma$ uncertainty is shown by the grey shaded region. \textit{Bottom panel}: Detrended SAP light curve; the black line shows the best fit from a transit-only model. All photometric data points are plotted together with their error bars.
\label{fig:PH_TESS}}
\end{figure*}

All sectors were processed by the Science Processing Operations Center (SPOC; \citealp{SPOC}) photometry and transit-search pipeline. Light curves and the TESS target pixel files (TPFs) were downloaded from the Mikulski Archive for Space Telescopes\footnote{\url{https://archive.stsci.edu/}} (MAST), which provides the simple aperture photometry (SAP) and the pre-search data conditioning SAP flux (PDCSAP), the latter being corrected for instrumental errors and crowding. Figure\,\ref{fig:TPF_Gaia}, produced with \texttt{tpfplotter}\footnote{\url{https://github.com/jlillo/tpfplotter}} \citep{aller2020}, illustrates the TPF and the optimal aperture delivering the best SAP fluxes for TOI-1801. The figure also includes the location of the target and other nearby stars according to  the $Gaia$ Data Release 3 (DR3) catalogue \citep{gaia2016, gaiadr3}. The PDCSAP algorithm may remove stellar activity signals (see the top and middle panels of Fig.\,\ref{fig:PH_TESS}). For this reason, we only used the SAP flux photometry in our study, although SAP fluxes might be contaminated by flux from nearby stars. We searched for possible contaminating sources using Fig.\,\ref{fig:TPF_Gaia} and verified the absence of stars within the selected photometric aperture down to 6 mag fainter. The SAP flux light curve of TOI-1801 is shown in the middle panel of Fig.\,\ref{fig:PH_TESS} along with the best model (Sect.\,\ref{sec:tronly}). The SAP data present some systematics at around BJD = 2458905.97, 2458922.5, and 2459644.0, which typically appear as jumps in the light curve each time the spacecraft resets the reaction wheel by firing a momentum dump. These defects are not entirely removed from the PDCSAP data (top panel of Fig.\,\ref{fig:PH_TESS}). However, we corrected for them by performing a linear fit before and after the jumps (with a baseline of one day) and  subtracting the relative flux offset from the SAP fluxes. The residuals of these defects are negligible in the SAP light curves shown in the middle and bottom panels of Fig.\,\ref{fig:PH_TESS}. The final TESS light curve shows a dispersion of $\sigma_{\mathrm{TESS}}$\,$\sim$\,4.9 parts per thousand (ppt), an average error bar of $\sim$1.0 ppt, and peak-to-peak variations up to 20 ppt. No obvious flares are present in the data.

On 15 April 2020, an alert was issued by the TESS Science Office (TSO) about a transit signal in TOI-1801\@. In the first report, two transit-like features were identified with an orbital period of 21.2844\,$\pm$\,0.0033 days and a depth of 1.546\,$\pm$\,0.161 ppt, but the period was also consistent with a transit signal at half the reported value, that is, 10.6422 days, when considering the TESS data gap in the middle of the sector. However, after carefully inspecting sector 22 SAP photometry without relying on any quality flags, we realised that the observations actually registered a third transit (chronologically, it is the second transit) at BJD = 2458914.2. The data around this date were assigned a quality flag of "scattered light" (flag 4096) and were not used in the first TESS report. In order to preserve the valuable information given by this transit, we ignored the quality flags around this specific date and incorporated the data in the final TESS light curve (bottom and middle panels of Fig.\,\ref{fig:PH_TESS}). This additional transit-like feature in sector 22, together with the transit-like features observed in sector 49, does confirm that the periodicity of the signal is 10.6422 days. In total, there are five planetary transit-like signatures in the TESS light curve (three in sector 22 and two in sector 49). The TESS original report named this planet candidate TOI-1801.01 and provided a radius of 2.08\,$\pm$\,0.12 R$_{\oplus}$ and an equilibrium temperature of 440\,K\@. So far, we are not aware of  further alerts on additional planet candidates transiting LP\,375-24\@.

\section{Ground-based follow-up observations}
\label{sec:Observations}

\subsection{High-resolution imaging}
\label{sec:ao}

\begin{figure}[ht!]
\includegraphics[width=1\linewidth]{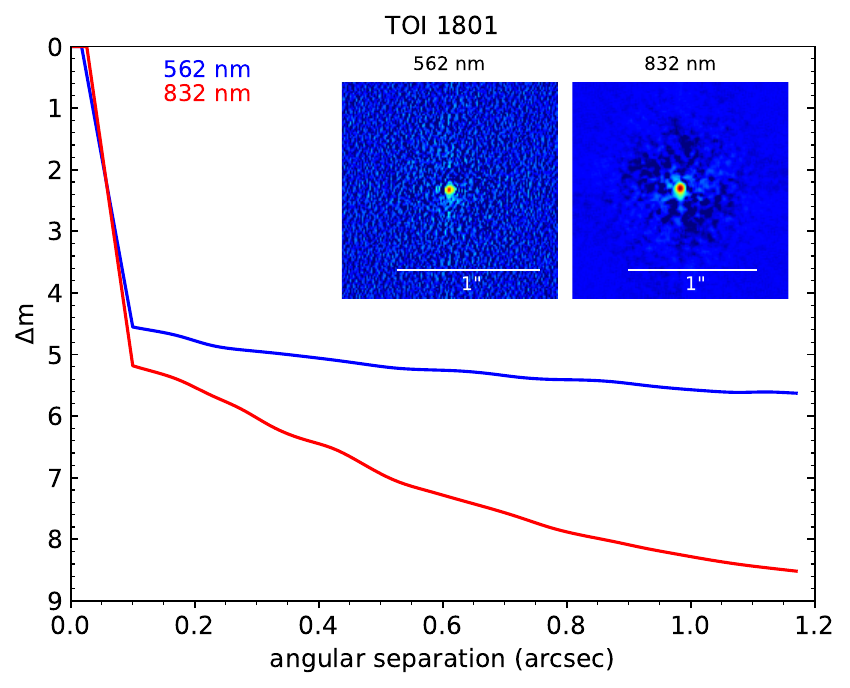}
\includegraphics[width=1\linewidth]{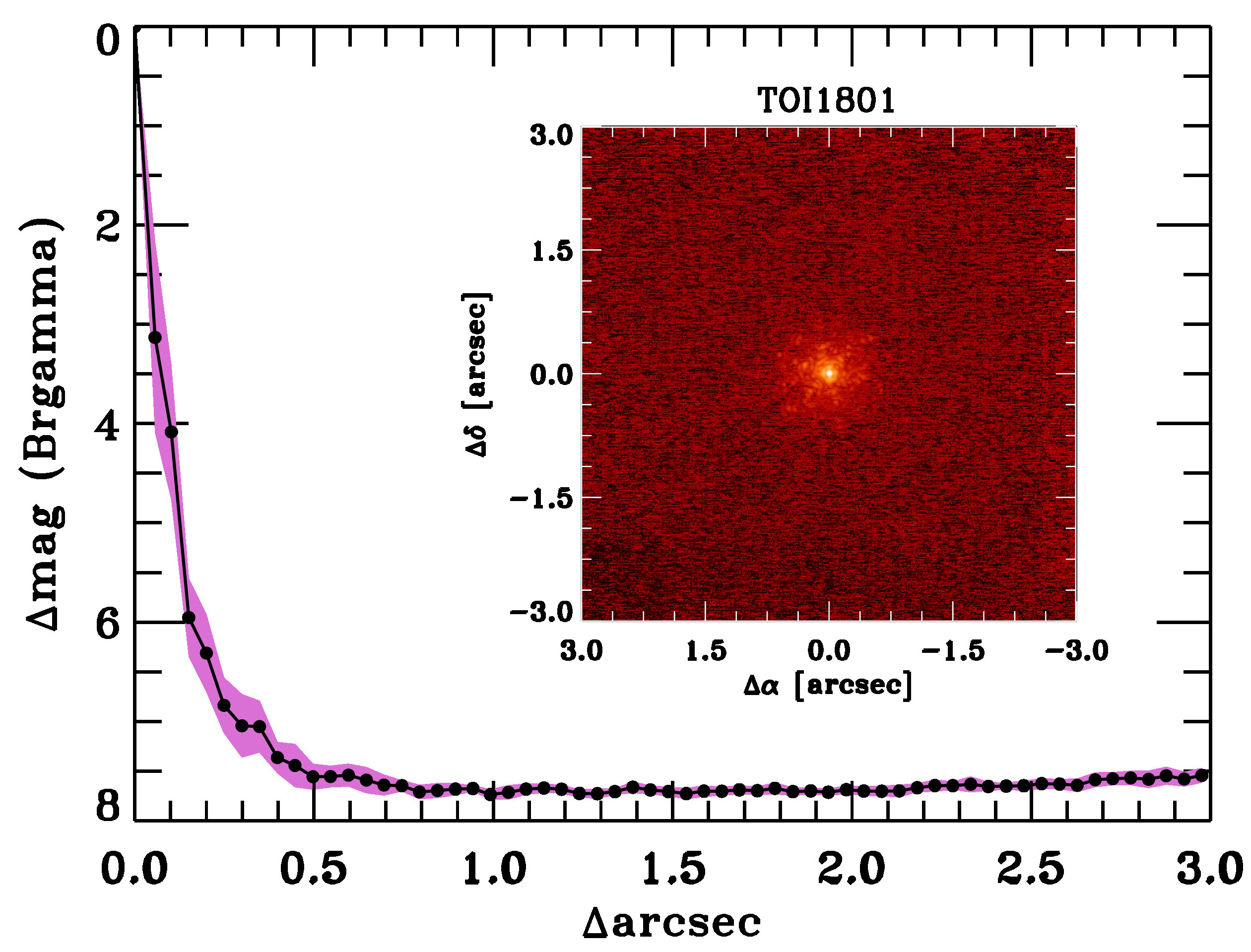}
\caption{Contrast curves for TOI-1801 from 'Alopeke speckle imaging (top) and from the NIRC2 instrument (bottom). The blue, red, and black lines correspond to the filters of 562 nm, 832 nm, and  2.1686 $\mu$m, respectively. The insets show the reconstructed 1.25$\times$1.25\arcsec{}$^2$ for 'Alopeke speckle imaging and 6.0$\times$6.0\arcsec{}$^2$ for the NIRC2 instrument for each of filters. None of them reveals a close companion within 0.1\arcsec{} down to $\Delta$\,$=$\,4--5 mag with respect to the target.  
\label{fig:AO}}
\end{figure}

TOI-1801 was observed on 28 May 2020 with high spatial resolution imaging at Keck Observatory on the top of Mauna Kea, Hawai'i (USA). The observations were carried out with the NIRC2 instrument on the 10\,m Keck II telescope. TOI-1801 was also observed on 19 June 2020 with the 'Alopeke speckle instrument mounted on the 8\,m Gemini North telescope, also on Mauna Kea. The IR (infrared) / AO (adaptive optics) imaging was performed with an integration time of 2\,s using the narrow-band Br-$\gamma$ filter ($\lambda_0$\,$=$\,2.1686 $\mu$m; $\Delta \lambda$\,$=$\,0.0326 $\mu$m), and the speckle imaging was carried out in two simultaneous narrow-band filters ($\lambda_1$\,$=$\,562 nm; $\Delta \lambda_1$\,$=$\,54 nm and $\lambda_2$\,$=$\,832 nm; $\Delta \lambda_2$\,$=$\,40 nm). We obtained 5000 images with an exposure time of 0.06\,s each in each channel. These high-resolution IR and visible-light observations allowed us to rule out unresolved companions, and we can therefore support the hypothesis that the candidate is indeed a planet. We searched for nearby sources (Fig.\,\ref{fig:AO}) and detected no companion to TOI-1801 within 0.02--0.1\arcsec{} down to 4--5 magnitudes fainter at 5$\sigma$ contrast.

\subsection{LCOGT transit photometry}
\label{sec:lco}

We observed two full transit windows of TOI-1801 in the Pan-STARRS $z$-short filter on 13 March 2021 and 13 April 2021 from the Las Cumbres Observatory Global Telescope (LCOGT; \citealp{Brown2013}) 1.0\,m network node at Cerro Tololo Inter-American Observatory (CTIO), Chile. We also observed in the 13 March 2021 window using an LCOGT 0.4\,m network telescope at CTIO in the Sloan $i'$ filter. The 1\,m telescopes are equipped with 4096$\times$4096 SINISTRO cameras with an image scale of 0.389\arcsec{} per pixel, resulting in a field of view of 26\arcmin$\times$26\arcmin{}. The 0.4\,m telescopes are equipped with 2048$\times$3072 SBIG STX6303 cameras with an image scale of 0.57\arcsec{} pixel$^{-1}$, resulting in a field of view of 19\arcmin$\times$29\arcmin{}. The images were processed by the standard LCOGT {\tt BANZAI} pipeline \citep{McCully2018}, and photometric data were extracted using {\tt AstroImageJ} \citep{Collins2017}. The images were focused and have typical stellar point spread functions with a full width at half maximum (FWHM) of roughly 2.0\arcsec{}, and circular apertures with radius 6.6\arcsec{} were used to extract the differential photometry. The observations taken on 13 March 2021 were affected by variable sky transparency and poor guiding, and they are not sensitive at the expected 1.5 ppt depth level. They did rule out nearby eclipsing binaries within 2.5\arcmin{} of TOI-1801, however, which could have caused the TESS detection due to blending in the TESS photometric aperture. The data from 13 April 2021 were observed under good conditions, and we detected a $\sim$\,1.3 hour late $\sim$\,1.5 ppt transit-like signal on target. The transit data are included in the transit-only fit and in the global model described in Sects.\,\ref{sec:tronly} and \ref{sec:jfit}, respectively.

\subsection{MuSCAT2 transit photometry}
\label{sec:muscat2}

TOI-1801 was observed on 5 February 2022 with the multi-band imager MuSCAT2 \citep{Narita2019} mounted on the 1.5\,m Telescopio Carlos S\'{a}nchez (TCS) at the Observatorio del Teide, Spain. MuSCAT2 is equipped with four CCDs and can obtain simultaneous images in $g'$, $r'$, $i'$, and $z_s$ bands with short read-out times. Each CCD has 1024$\times$1024 pixels with a field of view of 7.4\arcmin$\times$7.4\arcmin{}.

The observations were made while the telescope was slightly defocused. The $i'$-band camera had connection issues and could not be used for the observations. The exposure times were set to 50, 25, and 10 seconds in $g'$, $r'$, and $z_s$, respectively. The full transit observation could not be completed, and the observations had to be interrupted due to high levels of dust at the observatory. The raw data were reduced by the MuSCAT2 pipeline \citep{Parviainen2019}. The pipeline performs dark and flat-field calibrations, aperture photometry, and transit model fitting, including instrumental systematics. The data are included in the transit-only and global fits in Sects.\,\ref{sec:tronly} and \ref{sec:jfit}, respectively, and are shown in Fig.\,\ref{fig:PH_folded} with the best transit model.

\subsection{Long-term photometry}
\label{sec:longphot}

We gathered archival time-series photometry from the ASAS-SN, SuperWASP, and MEarth public ground-based surveys, and took additional data with TJO, e-EYE, and LCOGT of TOI-1801, as described in the following subsections. We used all of these data for an accurate determination of the true rotational period of the parent star.

\subsubsection{ASAS-SN}

TOI-1801 was observed by the All-Sky Automated Survey for Supernovae (ASAS-SN; \citealp{Shappee2014, Kochanek2017}) project. ASAS-SN consists of a network of 24 robotic telescopes with a diameter of 14 cm, which is distributed around the globe in both the northern and southern hemispheres. It can survey the entire sky down to $V$\,$\approx$\,18\,mag. We retrieved the ASAS-SN time-series photometry of TOI-1801 from the web page of the project\footnote{\url{https://asas-sn.osu.edu}.}. The data cover a baseline of $\sim$\,5 years and consist of a total of 221 $V$-band measurements acquired between 18 November 2013 and 28 November 2018. The root mean square (rms) of the data is $\sim$0.008 mag.

\subsubsection{SuperWASP}

TOI-1801 was photometrically monitored between 2 May 2004 and 16 May 2007 by the SuperWASP-North survey \citep{Pollacco2006}, a multi-camera system located at Roque de los Muchachos Observatory in La Palma, Spain. At the time of the observations, SuperWASP-North consisted of an array of eight 200 mm $f/1.8$ telescopes, each equipped with a 2048$\times$2048 pixel back-illuminated CCD camera. SuperWASP-North gathered a total of 5939 broad-band photometric measurements (400--700\,nm). We retrieved the SuperWASP photometry of TOI-1801 from the web page of the project\footnote{\url{https://wasp.cerit-sc.cz/form}.}. The rms of the data is $\sim$0.010\,mag.

\subsubsection{MEarth}

The MEarth project \citep{irwin2015mearthnorth} observes the entire sky with two robotic arrays of telescopes in the northern and southern hemispheres. The MEarth-North telescope array is located at the Fred Lawrence Whipple Observatory (FLWO) in Arizona. One of the eight 40 cm telescopes of the MEarth-North array photometrically monitored TOI-1801 from 3 November 2011 to 10 November 2015, acquiring a total of 425 data points. We downloaded the MEarth photometric data from the web page of the project\footnote{\url{https://lweb.cfa.harvard.edu/MEarth/DataDR10.html}.}. The rms of the data is $\sim$0.006\,mag.

\subsubsection{TJO}

We observed TOI-1801 from 20 February 2021 to 7 June 2022 with the 0.8\,m Joan Or\'o telescope \citep[TJO;][]{colome2010} at the Montsec Observatory in Lleida, Spain. We obtained a total of 330 images on 42 different nights with an exposure time of 60\,s each using the Johnson $R$ filter of the LAIA imager, a 4k$\times$4k CCD with a field of view of 30\arcmin{} and a scale of 0.4\arcsec{} pixel$^{-1}$. The images were calibrated with darks, bias, and flat fields with the {\tt ICAT} pipeline \citep{colome2006} of the TJO\@. The differential photometry was extracted with {\tt AstroImageJ} \citep{Collins2017} using the aperture size that minimised the rms of the resulting relative fluxes, and a selection of the 20 brightest comparison stars in the field that did not show variability. Then, we used our own pipelines to remove outliers and measurements affected by poor observing conditions or with a low signal-to-noise ratio (S/N). The resulting rms of the differential photometry from the TJO in the $R$ filter is $\sim$\,0.005 mag.

\subsubsection{e-EYE}
TOI-1801 was observed from e-EYE (shorthand for Entre Encinas y Estrellas)\footnote{\url{https://www.e-eye.es/}}, a telescope-hosting facility located at Fregenal de la Sierra in Badajoz, Spain. Observations in the $V$ filter were taken between May 2022 and July 2022 using a 16\arcsec{} ODK corrected-Dall-Kirkham reflector with a Kodak KAF-16803 CCD chip on an ASA DDM85 mount. The CCD camera is equipped with Astrodon filters. The effective pixel scale is 2.04\arcsec{}/pixel with 3$\times$3 binning. The images and differential aperture photometry of the target and several reference stars were reduced using the {\tt Lesve} photometry package\footnote{\url{http://www.dppobservatory.net}}. The rms of the data is $\sim$0.041 mag.

\subsubsection{LCOGT}
TOI-1801 was observed in the $V$ band with the 40\,cm telescopes of LCOGT (Sect.\,\ref{sec:lco}) at the McDonald and Haleakala\=a observatories between 4 May 2022 and 12 July 2022. We obtained ten individual exposures of 40\,s for a total of 46 observing epochs. Weather conditions at the observatories were mostly clear, and the average seeing varies from 1.0\arcsec{} to 3.0\arcsec{}. Raw data were processed using the {\tt BANZAI} pipeline \citep{McCully2018}, which includes bad pixel, bias, dark, and flat field corrections for each individual night. We performed differential aperture photometry of TOI-1801 with respect to three references stars of similar brightness in the same field of view. An optimal aperture of 10 pixels ($\sim$6\arcsec{}), which minimises the dispersion of the differential light curve, was adopted. The rms of the data is $\sim$0.014 mag.

\subsection{CARMENES spectroscopic observations}
\label{sec:carmrvs}

We collected 88 spectra between 30 January 2021 (UT) and 11 June 2022 (UT) with the CARMENES instrument installed at the 3.5\,m telescope of Calar Alto Observatory, Almer\'ia, Spain. The CARMENES spectrograph has two channels \citep{CARMENES,CARMENES18}, a visible (VIS) channel covering the spectral range 0.52--0.96\,$\mu$m and a near-infrared (NIR) channel covering the spectral range 0.96--1.71\,$\mu$m. Three spectra were discarded because the drift correction was missing and another 5 and 20 were removed due to their low S/N ($<$\,30) or outliers from the VIS and NIR channels, respectively. The final data sets contain 80 spectra in the VIS range and 68 spectra in the NIR range. These observations were taken with exposure times of 1800\,s obtaining an S/N per pixel at 745 nm in the range 32--122, and at 1221 nm in the range of 34--143. The CARMENES performance, data reduction and wavelength calibration were described by \citet{2016SPIE.9910E..0EC}, \citet{trifonov18}, and \citet{2018A&A...618A.115K}. Relative RVs and activity indicators such as the chromatic index (CRX), differential line width (dLW), H$\alpha$ index, the \ion{Ca}{ii} IR triple (IRT), and the \ion{Na}{i} D values were obtained using {\tt serval}\footnote{\url{https://github.com/mzechmeister/serval}} \citep{2018A&A...609A..12Z}. 
The RV measurements were corrected for barycentric motion, secular acceleration, nightly zero-points, and for telluric lines as described by \cite{Nagel23}. The typical dispersion of the RV measurements are $\sigma_{\mathrm{CARMENES\ VIS}}$\,$\sim$\,6.3 m\,s$^{-1}$ and $\sigma_{\mathrm{CARMENES\ NIR}}$\,$\sim$\,9.3\,m\,s$^{-1}$. The uncertainties of the measured RVs are in the range 1.6--5.2\,m\,s$^{-1}$ with a median value of 2.4\,m\,s$^{-1}$ for the VIS and between 4.6--22\,m\,s$^{-1}$ with a median value of 8.9\,m\,s$^{-1}$ for the NIR measurements.

\begin{figure*}[ht!]
\includegraphics[width=1\linewidth]{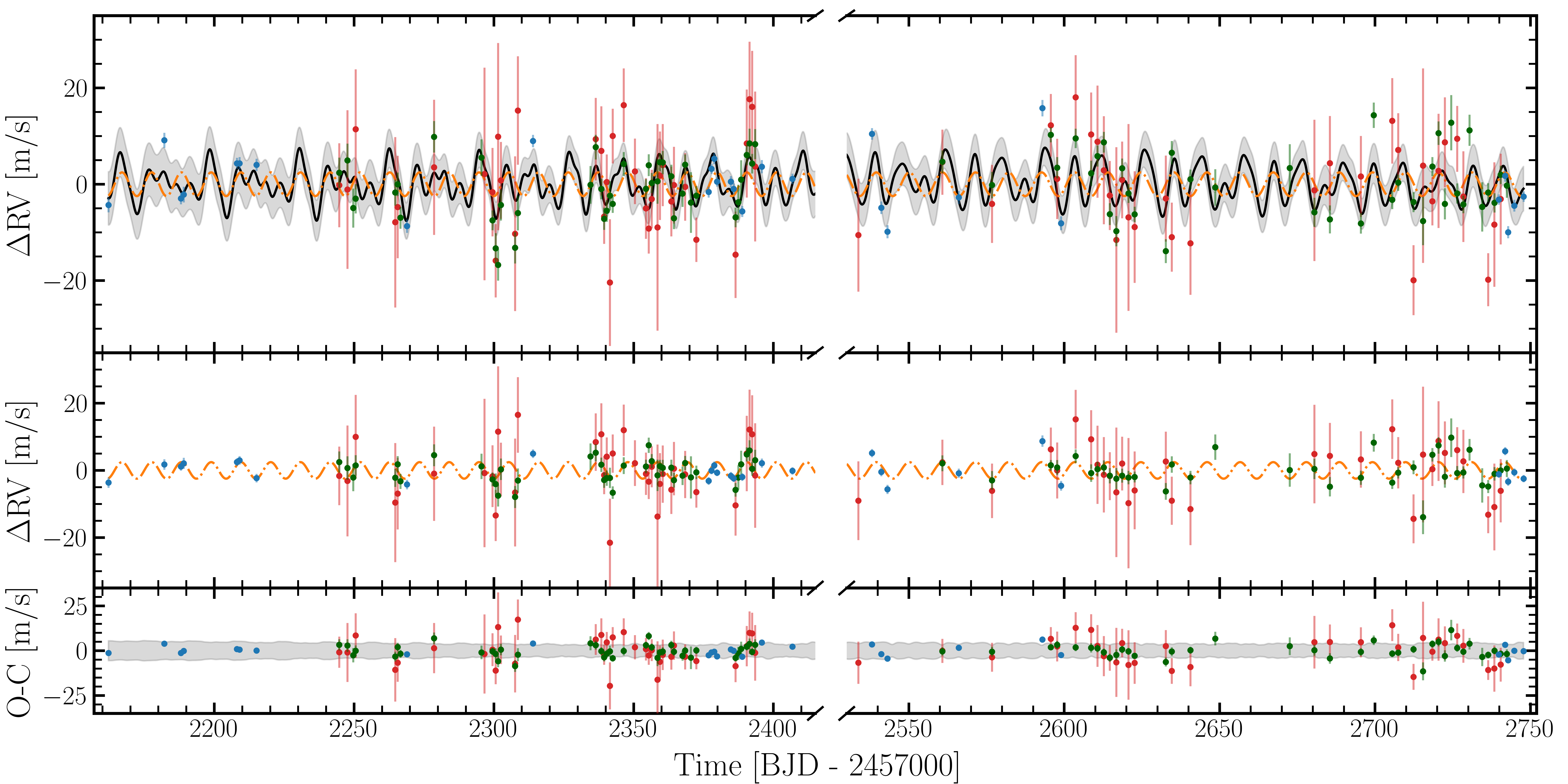}
\caption{CARMENES VIS, CARMENES NIR, and HIRES RV data for TOI-1801 (green, red, and blue dots, respectively). \textit{Top panel}: Combined model (black line) with its 1$\sigma$ level of confidence (grey shadow), and Keplerian model for the planet (dashed orange line). \textit{Middle panel}: Keplerian model alone (dashed orange line) after subtracting the best activity model. \textit{Bottom panel}: Residuals for the best-fit.  
\label{fig:RV_curve}}
\end{figure*}

The RV curve is shown in Fig.\,\ref{fig:RV_curve} with its best-fit model (Sect.\,\ref{sec:jfit} for details). As in the photometric TESS data, we searched for RV measurements affected by flares by measuring the relative intensity of a set of emission lines (H$\alpha$, \ion{Ca}{II}\,IRT, \ion{Na}{I}, \ion{K}{I}, and \ion{He}{I}) usually associated with chromospheric activity \citep{Fuhrmeister2019, Fuhrmeister2022}. We compared these lines with each other for all spectra to search for significant variations, but none of the lines seem to be affected by flares. In the appendices, Table\,\ref{tab:CARMV_RV} and Table\,\ref{tab:CARMN_RV} give the time stamps of the spectra in BJD$_{\mathrm{TDB}}$ and the relative RVs measured with {\tt serval} along with their $1\sigma$ error bars.

\subsection{HIRES spectroscopic observations}
\label{sec:hiresrvs}

Using the HIRES spectrometer on the Keck I telescope on Mauna Kea, we collected 29 RVs spanning 2.1 years. The spectrometer resolutions 60\,000 and the median S/N per pixel is 163 at 550 nm resulting in a median internal uncertainty of 1.21 m\,s$^{-1}$. The star is sufficiently faint, and therefore, we used the C2 decker (0.87\arcsec{}\,$\times$\,14.0\arcsec{}) in order to remove sky-background flux uniquely for every column in each order in the 2D raw CCD image. RVs were calculated using a forward model that uses a Fourier transform spectrum of the iodine cell spectrum that is imprinted on each RV observation, an iodine-free observation of the target star, and a model of the point spread function. Further details of the raw reduction and RV pipeline were presented by \citet{Howard2010}. The RVs were originally collected as part of the TESS-Keck Survey \citep{Chontos2022}. Table \ref{tab:HIRES_RV} lists the time stamps and the relative RVs with their 1$\sigma$ error bars.

\subsection{iSHELL spectroscopic observations}
\label{sec:ishell}

We obtained 287 spectra of TOI-1801 with on-source integration times of 300\,s, repeated 8--16 times on 25 nights in variable seeing conditions, spanning over 385 days with the iSHELL spectrometer on the NASA Infrared Telescope Facility (IRTF, \citealp{Rayner2016}). Each night, a cumulative S/N of 102--175 (median: 146) per spectral pixel at approximately 2.2 $\mu$m at centre blaze was obtained. A per night RV precision of 5--30 m s$^{-1}$ (median 10 m s$^{-1}$) is obtained. Spectra were reduced and RVs extracted using the methods outlined by \citet{Cale2019}. Three nights were discarded due to non-convergent modelling, resulting in 23 useful RV measurements. The iSHELL time stamps and the relative RVs together with their 1$\sigma$ error bars are listed in Table \ref{tab:iSHELL_RV}. These RVs were ultimately discarded in subsequent analyses due to their poor temporal cadence.

\section{Stellar properties}
\label{sec:properties}

From the high S/N stellar template generated by {\tt serval}, which combined the 80 CARMENES spectra, the stellar atmospheric parameters ($T_{\rm eff}$, $\log{g}$, and [Fe/H])) were derived with the {\sc SteParSyn}\footnote{\url{https://github.com/hmtabernero/SteParSyn/}} code \citep{tab22} using the line list and model grid described in \citet{marfil21}. The spectroscopically derived values of TOI-1801's $T_{\rm eff}$ (3863 $\pm$ 77 K), $\log{g}$ (4.78 $\pm$ 0.05 dex), and [Fe/H] ($-$0.19 $\pm$ 0.04 dex) are given in Table\,\ref{tab:stellar_parameters}. The high surface gravity suggests that the star belongs to the main sequence. The {\sc SteParSyn} method has been compared to different methods and procedures available in the literature by \citet{marfil21} and yields reliable values, in particular, for early-M dwarfs.

\begin{figure}[ht!]
\includegraphics[width=1\linewidth]{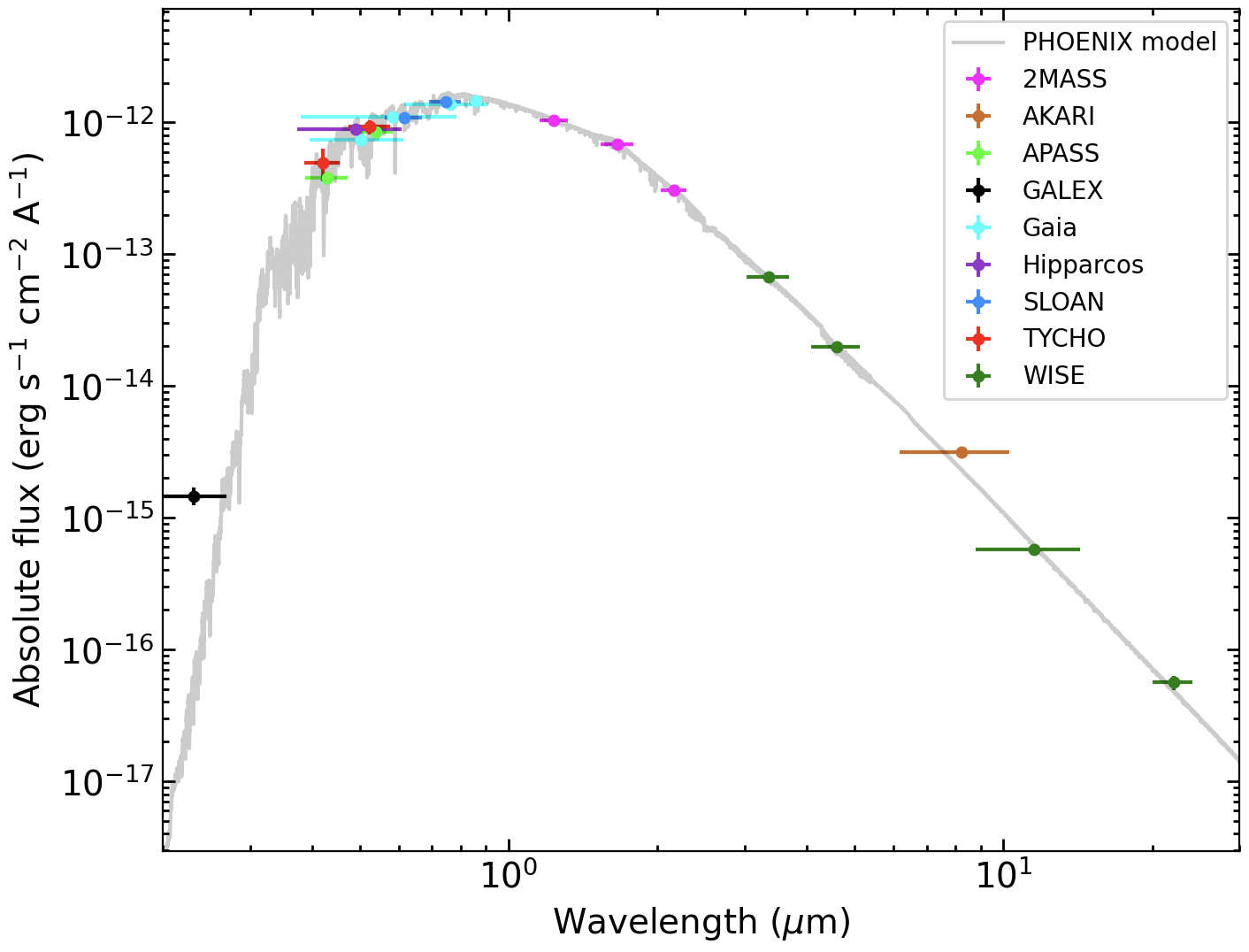}
\caption{Photometric SED of TOI-1801. The grey line corresponds to the PHOENIX solar metallicity model \citep{husser13} with $T_{\rm eff}$\,$=$\,3900 K and $\log{g}$\,$=$\,5.0 cm\,s$^{-2}$.  
\label{fig:TOI-1801_SED}}
\end{figure}

We obtained the stellar bolometric luminosity from photometric spectral energy distribution (SED) of TOI-1801 following a similar procedure to that of \citet{cifuentes20}, where all apparent photometry was converted into absolute fluxes using the {\sl Gaia} DR3 parallax. The SED (Fig.\,\ref{fig:TOI-1801_SED}) was built using data from various publicly available catalogues: the Galaxy Evolution Explorer (GALEX; \citealp{bianchi2017}), the American Association of Variable Star Observers Photometric All-Sky Survey (APASS; \citealp{henden_2014}), the Hipparcos and Tycho-2 catalogues \citep{perryman_1997, hog_2000}, {\sl Gaia} DR3 \citep{gaia2016, gaiadr3}, the Sloan Digital Sky Survey (SDSS; \citealp{york_2000}), the Two Micron All-Sky Survey (2MASS; \citealp{skrutskie2006}), the AKARI mid-infrared all-sky survey \citep{ishihara_2010}, and the Wide-field Infrared Survey Explorer (WISE; \citealp{wright_2010}). These data are also available through the Virtual Observatory SED analyser (VOSA; \citealp{bayo_2008}). The SED of TOI-1801 covers wavelengths from $\sim$\,0.25 through $\sim$\,25 $\mu$m and is well reproduced by the PHOENIX solar metallicity model \citep{husser13} with an effective temperature of 3900 K and a surface gravity of log\,$g$\,$=$\,5.0 cm\,s$^{-2}$, which is compatible with the values derived from the spectral analysis of the CARMENES data. At short wavelengths, TOI-1801 shows higher fluxes than expected from purely photometric emission, as expected for an M dwarf \citep{cifuentes20}, which is a clear indication of stellar activity. At long wavelengths, there is no evidence of IR flux excesses up to $\sim$\,25 $\mu$m. Therefore, TOI-1801 does not host a warm debris disk. The photometric SED (excluding the GALEX data) was integrated using the trapezoidal rule to derive the photospheric bolometric luminosity $L = (588.8 \pm 13.7) \times 10^{-4}$ L$_\odot$, where the error bar comes from the uncertainty in the trigonometric parallax and the photometric error bars. We did not complete the SED below 0.25 $\mu$m and above 25 $\mu$m using theoretical models because the contribution of these fluxes to the global luminosity is estimated to be less than a few percent at temperatures of about 3900 K, that is, below the quoted luminosity uncertainty. 

The stellar radius ($R_{\star}$\,=\,0.542$^{+0.029}_{-0.027}$\,R$_\odot$) follows from the Stefan-Boltzmann law, and the stellar mass ($M_\star$\,=\,0.548$^{+0.048}_{-0.045}$\,M$_\odot$) was derived from the linear mass-radius relation of \citet{Schweitzer2019}. We point out that the mass-radius relation of \cite{Schweitzer2019} is valid for main-sequence objects older than a few hundred million years, and therefore, it is applicable to TOI-1801 (Sect.\,\ref{sec:age}). As a consistency check, we also determined the TOI-1801 surface gravity to be log\,$g$ = 4.71 $\pm$ 0.08 cm\,s$^{-2}$ by using the bolometric luminosity-based stellar mass and radius and Newton's law. The spectroscopic measurement of log\,$g$ agrees at better than 1\,$\sigma$, which supports the determinations of the stellar mass and radius, which are critical for the characterization of the planetary system.

\begin{figure}[ht!]
\includegraphics[width=1\linewidth]{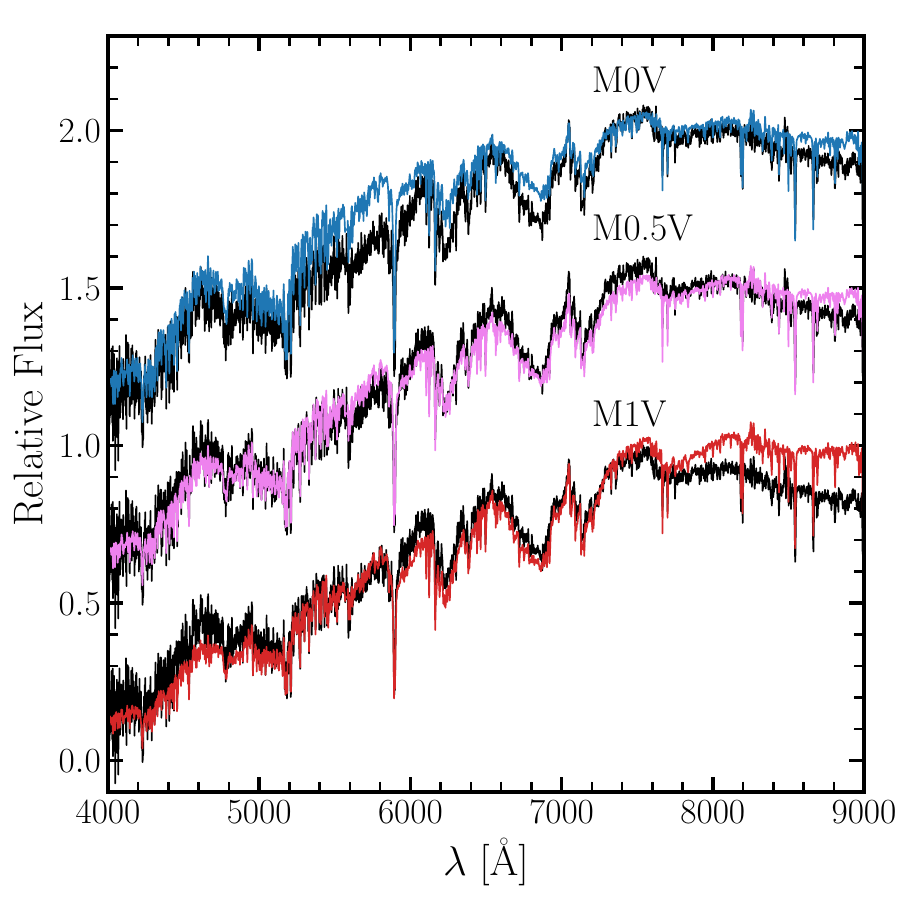}
\caption{Spectral type determination of TOI-1801. The black and coloured lines show the LAMOST spectra of TOI-1801 and the templates from SDSS, respectively.  
\label{fig:TOI-1801_SpT}}
\end{figure}

We searched for spectra in large-scale surveys and telescope archives. We found a good-quality low-resolution ($\mathcal{R}$\,$\sim$\,1800 at 5500 \AA{}) optical (3690--9100 \AA{}) spectrum in the 6th data release of the LAMOST survey\footnote{\url{http://dr6.lamost.org/}}, where TOI-1801 was automatically classified as field dM0. Using the LAMOST spectrum, in comparison with the SDSS template spectra \citep{Bochansky2007}, we determine a spectral type of M0.5V for TOI-1801 with an uncertainty of half a subclass (see Fig.\,\ref{fig:TOI-1801_SpT}), which agrees with the previous classification of \citet{Lepine2013} and the derived $T_{\rm eff}$ from the CARMENES high-resolution spectra. The retrieved stellar parameters are given in Table\,\ref{tab:stellar_parameters}.

\begin{table}
\caption{Stellar parameters of TOI-1801.}\label{tab:stellar_parameters}
\centering
\begin{tabular}{lcr}
\hline\hline
Parameter & Value & Reference\\
\hline
Name & LP~375-23 & Luy79\\
     & TIC~119584412  & TIC\\
\noalign{\smallskip}
$\alpha$ (J2016) & 11:42:18.1 & \textit{Gaia} DR3\\
$\delta$ (J2016) & +23:01:36.3  & \textit{Gaia} DR3\\
Sp. type & M0.5\,$\pm$\,0.5\,V & This work \\
$\varpi$ [mas] & 32.370\,$\pm$\,0.024 & \textit{Gaia} DR3\\
$d$ [pc] & 30.893\,$\pm$\,0.023 & \textit{Gaia} DR3\\
RUWE & $1.041$ & \textit{Gaia} DR3\\
$T_{\text{eff}}$ [K] & 3863\,$\pm$\,77 & This work \\ 
$\log{g}$ [cgs] & 4.78\,$\pm$\,0.05 & This work \\
{[Fe/H]} [dex] & --0.19\,$\pm$\,0.04 & This work \\
$M_{\star}$ [$M_{\odot}$] & 0.548\,$\pm$\,0.048 & This work \\
$R_{\star}$ [$R_{\odot}$] & 0.542\,$\pm$\,0.029 & This work \\
$L_{\star}$ [$10^{-4}\, L_{\odot}$] &  588.8\,$\pm$\,13.7 & This work\\ 
$v\sin i$ [km\,s$^{-1}$] & $< 3$ & This work\\
$P_{\text{rot}}$ [d] & 15.98\,$\pm$\,0.66 & This work\\
$U$ [km s$^{-1}$] & --28.94\,$\pm$\,0.04 & This work\\
$V$ [km s$^{-1}$] & --6.67\,$\pm$\,0.04 & This work\\
$W$ [km s$^{-1}$] & --7.48\,$\pm$\,0.18 & This work\\
Gal. population & Young disk & This work\\
Age [Myr] & 600--800 & This work\\
$NUV$ [mag] & 20.317\,$\pm$\,0.165 & \textit{GALEX}\\
$B_{p}$ [mag] & 11.801\,$\pm$\,0.003 & \textit{Gaia} DR3\\
$G$ [mag] & 10.830\,$\pm$\,0.003 & \textit{Gaia} DR3\\
$R_{p}$ [mag] & 9.857\,$\pm$\,0.004 & \textit{Gaia} DR3\\
$J$ [mag] & 8.649\,$\pm$\,0.026 & 2MASS\\
$L_{\rm X}$ [$10^{28}$erg\,s$^{-1}$] & $\sim$1.2 & This work\\
\noalign{\smallskip}
\hline
\end{tabular}
\tablebib{
Luy79: \citet{1979nlcs.book.....L}; 
TIC: \citet{Stassun2019};
\textit{Gaia} DR3: \cite{gaia2016, gaiadr3}; 
\textit{GALEX}: \citet{bianchi2017};
2MASS: \citet{skrutskie2006}.
}
\end{table}

\subsection{Rotation period}
\label{sec:prot}

Active regions (i.e.\ spots and faculae) on the stellar surface appear and disappear as the star rotates around its axis, inducing a quasi-periodic photometric variability. The rotation period of FGKM stars can thus be determined by performing a frequency analysis of their light curves.

\begin{figure}
\includegraphics[clip, trim=0.5cm 1.9cm 0.0cm 2.5cm, width=0.5\textwidth]{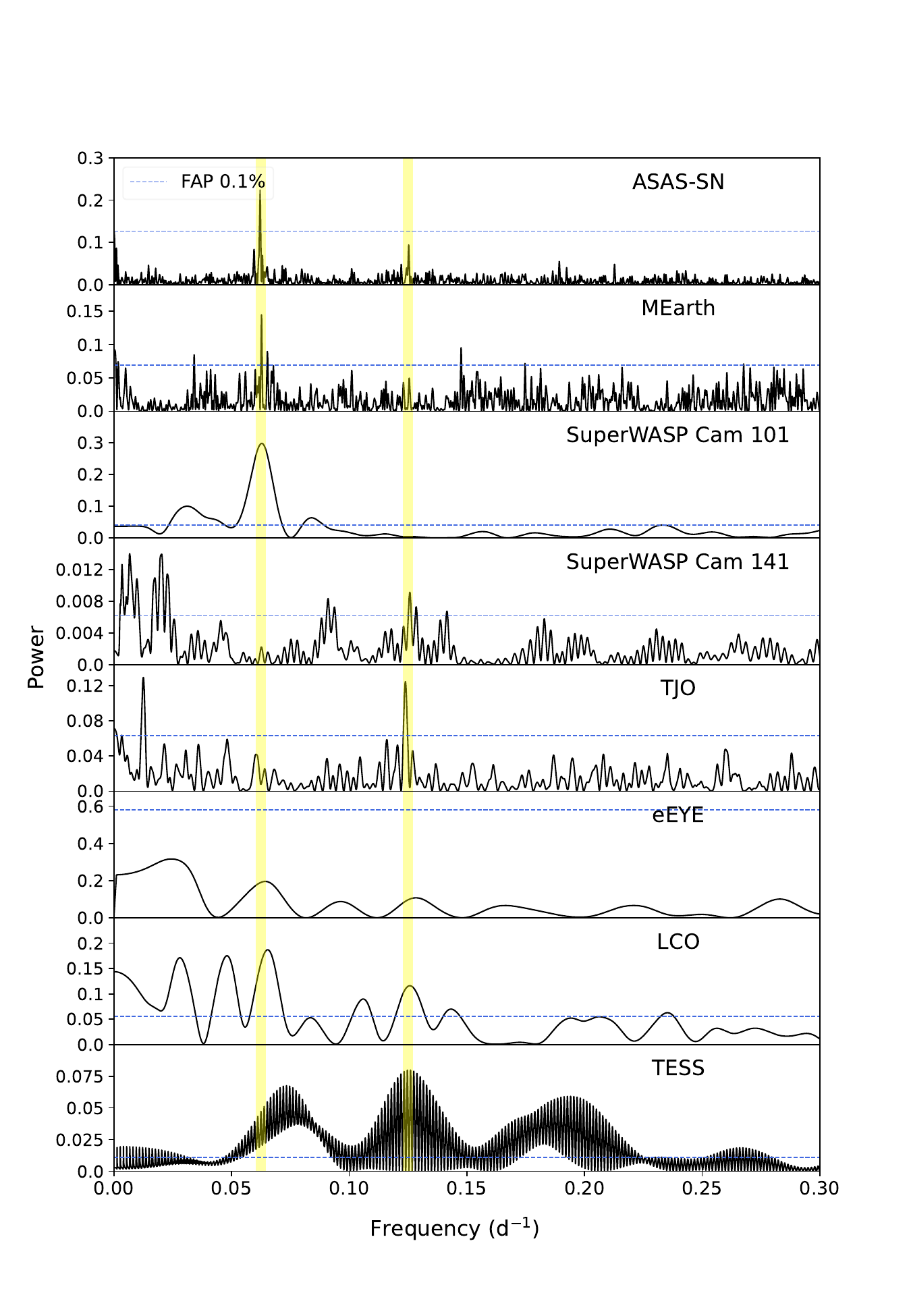}
\caption{GLS periodogram analysis of the photometric data for TOI-1801. In all panels, the two shaded vertical yellow bands indicate the stellar rotation period, centred at 0.0625~d$^{-1}$ (16 days), and its first harmonic, centred at 0.125~d$^{-1}$ (8 days).  The  dashed  horizontal  blue  lines corresponds to the FAP level of 0.1\%.
\label{fig:GLS_photometry_RotationPeriod}}
\end{figure}

Figure \ref{fig:GLS_photometry_RotationPeriod} shows the generalised Lomb-Scargle (GLS) periodograms \citep{GLS} of the ASAS-SN, MEarth, SuperWASP-North, TJO, e-EYE, LCOGT, and TESS time-series photometry of TOI-1801 (Sect.\,\ref{sec:longphot}). The dashed blue lines mark the 0.1\% false-alarm probability (FAP), which was estimated following the bootstrap method described by \cite{murdoch_1993}, that is, by computing the GLS periodogram of 10$^{5}$ simulated time series obtained by randomly shuffling the photometric measurements and their uncertainties while keeping the time stamps fixed.

The ASAS-SN periodogram shows the highest peak at a frequency of 0.0621\,$\pm$\,0.0006\,d$^{-1}$, corresponding to a period of 16.09\,$\pm$\,0.16\,d, while the highest peak in the MEarth data sets is at 0.0627\,$\pm$\,0.0007\,d$^{-1}$ (15.94\,$\pm$\,0.18\,d). Since SuperWASP-North observed TOI-1801 with two different cameras, camera 101 in 2004 and camera 141 between April 2006 and May 2007, we analysed the two data sets separately. The highest peak of the SuperWASP-North camera 101 periodogram occurs at 16.6\,$\pm$\,4.7\,d (0.060\,$\pm$\,0.018\,d$^{-1}$). The periodogram of Super WASP camera 141 data has the strongest peak at about 99 d and a second strong peak at 7.981\,$\pm$\,0.025\,d (0.1253\,$\pm$\,0.0004\,d$^{-1}$), which is likely the first harmonic of the rotation period. The TJO periodogram has its highest peak at a frequency of 0.124\,$\pm$\,0.002\,d$^{-1}$ (8.08\,$\pm$\,0.13\,d). The periodogram of e-EYE shows no significant peak, unlike that of the LCOGT, which shows a significant peak at 15.3\,$\pm$\,2.4\,d (0.065\,$\pm$\,0.01\,d$^{-1}$). In the TESS periodogram, the highest and most significant peak occurs at 7.93\,$\pm$\,0.26\,d (0.126\,$\pm$\,0.004\,d$^{-1}$).
From these data, we conclude that TOI-1801 has a rotation period of 15.98\,$\pm$\,0.66\,d, where the value and its error have been derived as the average of the most significant peaks in the GLS and the average of their FWHMs, respectively. The two vertical yellow bands in Fig.\,\ref{fig:GLS_photometry_RotationPeriod} indicate the regions in which most of the photometric time series have their highest GLS peaks, at about 16 d, and the first harmonic of the rotation period, that is, 8 d. We note that this period is consistent with the rotation period we recovered in our analysis of the stellar activity indicators (Sect.\,\ref{sec:activity}).

We furthermore performed a seasonal frequency analysis for which we split the data of different observational seasons to investigate the possibility of differential rotation. Figures \ref{fig:GLS_ASASSN_seasons} and \ref{fig:GLS_MEarth_seasons} show the GLS periodogram of the ASAS-SN and MEarth photometric data, divided for each observing season. The dashed blue line marks the 0.1\% FAP. The two vertical yellow bands indicate the regions around the stellar rotation period $\sim$16\,d and its first harmonic $\sim$8\,d. Our analysis of the stellar rotation period for each season shows that the observed variation in the period of the signals around the rotation period are consistent between them within 1$\sigma$. Therefore, we conclude that there is no appreciable differentiable rotation in these data.

\subsection{Wide companion}
\label{sec:widecomp}
We searched for common proper motion companions to TOI-1801 in the $Gaia$ DR3 catalogue up to a radius of 2 degrees. We imposed a restriction on the parallax with a range of 30--35 mas bracketing the parallax of TOI-1801 (Table \ref{tab:stellar_parameters}). The query returned only two stars with similar parallaxes and proper motions in right ascension and declination: the known system composed of TOI-1801 and LP\,375-24 \citep{Weis1991} at about 30 pc from the Sun, separated by 101.14\arcsec{}. No other wide common proper motion star is identified down to the depth of $Gaia$, corresponding to a spectral type of approximately L2--3, which is close to the substellar boundary at the distance of TOI-1801.

We followed the same procedure to derive the spectral type of LP\,375-24 as in Sect.\,\ref{sec:properties}. The LAMOST survey automatically classified the star as field dM4. Using the SDSS template spectra, we determined a spectral type of M4V for LP\,375-24, with an uncertainty of one subtype (Fig.\,\ref{fig:LP_375-24_SpT}). According to the spectral type-mass relation of Mamajek\footnote{\url{http://www.pas.rochester.edu/~emamajek/EEM_dwarf_UBVIJHK_colors_Teff.txt}} based on Table 5 of \citet{Pecaut2013}, an M4 spectral type corresponds to effective temperatures of 3210\,K, a radius of 0.274\,R${_\odot}$, and a mass of 0.23\,M${_\odot}$. The main stellar parameters for LP\,375-24 can be found in Table\,\ref{tab:stellar_parameters_comp}.

\subsection{Age}
\label{sec:age}

Young stars can have a higher rotation rate because they still preserve angular momentum from their formation. As a consequence, they show high chromospheric activity and prominent coverage by spots. As they become older, stellar rotation slows down through magnetic mechanisms \citep[][and references therein]{Gallet15}. We estimated the age of the system TOI-1801 and its wide companion LP\,375-24 using different age indicators such as kinematics, rotation period, and the near-UV ($NUV$) excess.

\subsubsection{$UVW$}

\begin{figure}[ht!]
\includegraphics[width=1\linewidth]{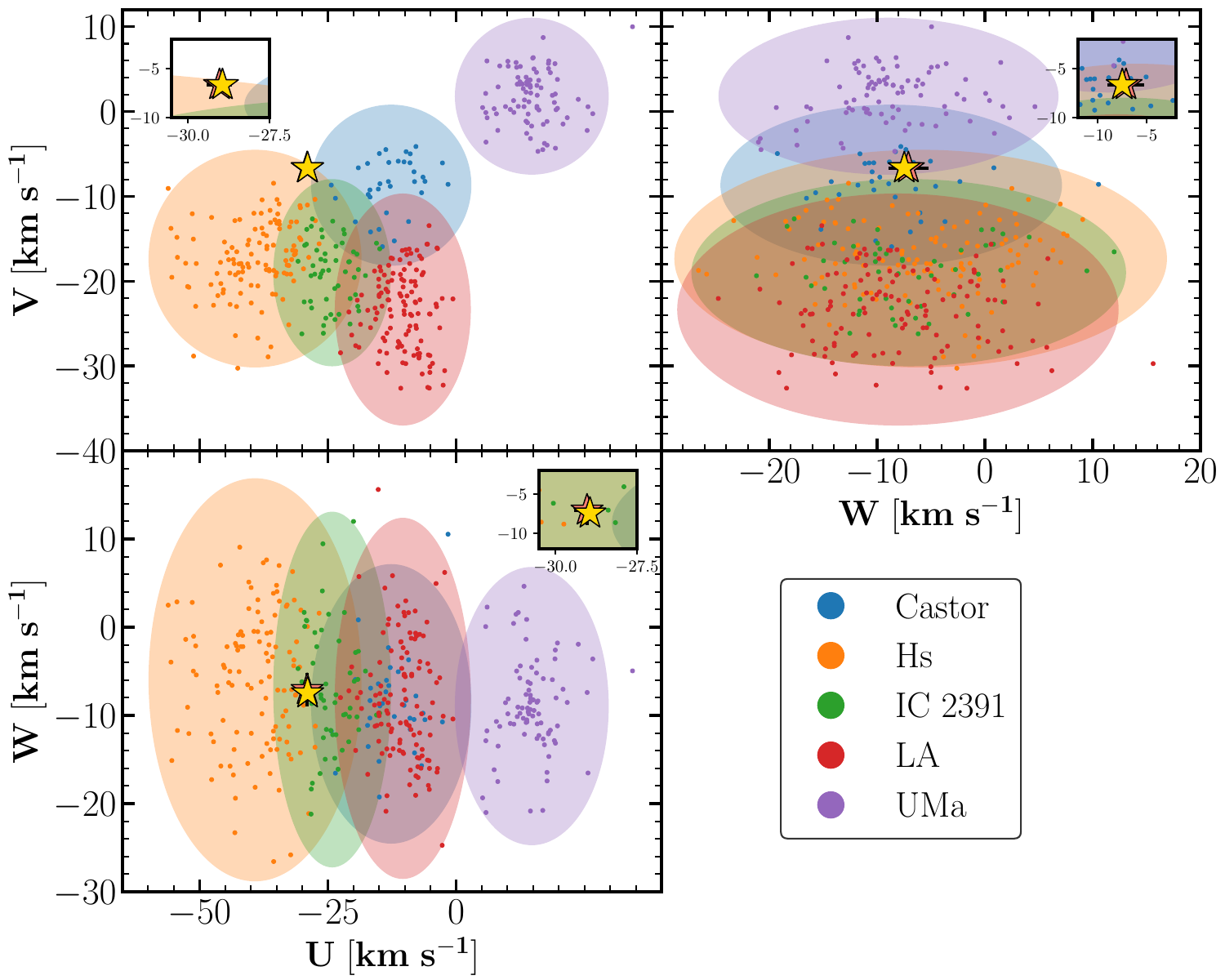}
\caption{$UVW$ velocity diagram for TOI-1801 (gold star) and LP\,375-24 (orange star). The members of the Castor moving group (200--400\,Myr), the Hyades supercluster (Hs; 600--800 Myr), the IC 2391 supercluster (35--55 Myr), the Local Association (LA; 10--300 Myr), and the Ursa Major group (UMa; $\sim$400 Myr) from \cite{montes2001b} are included. The ellipses represent the 3$\sigma$ values of $UVW$ for each young moving group. The inset in the upper left panel shows a zoom-in on the region with the stars. The locations of both stars agree with that of members of the Hyades supercluster.  
\label{fig:UVW}}
\end{figure}

The kinematics, in particular the $UVW$ galactocentric space velocities, allow us to know whether an object is associated with a moving group, a star formation region, or in a more general way, if the object belongs to the young disk. We found no previous association in the literature of TOI-1801 or LP\,375-24 with any of these. Hence, we derived the $UVW$ galactocentric space velocities using the \textit{Gaia} astrometry as in \cite{johnson1987}, that is, $U$ is positive toward the Galactic centre, $V$ is positive in the direction of Galactic rotation, and $W$ positive toward the North Galactic Pole. The velocities can be found in Tables \ref{tab:stellar_parameters} and \ref{tab:stellar_parameters_comp} for TOI-1801 and LP\,375-24, respectively, where both velocities are compatible with each other, which confirms a common origin of the two. The values for both stars indicate that the system displays young kinematics ($<$\,1 Gyr), and they could be consistent with the Hyades supercluster within 3$\sigma$ (Fig.\,\ref{fig:UVW}). Some authors indicated that the stars of this kinematic group could be associated with the Hyades cluster \citep{montes2001b}, which has an estimated age of between 600--800\,Myr \citep{brandt2015, lodieu18}.

\subsubsection{Gyrochronology}

\begin{figure}[ht!]
\includegraphics[width=1\linewidth]{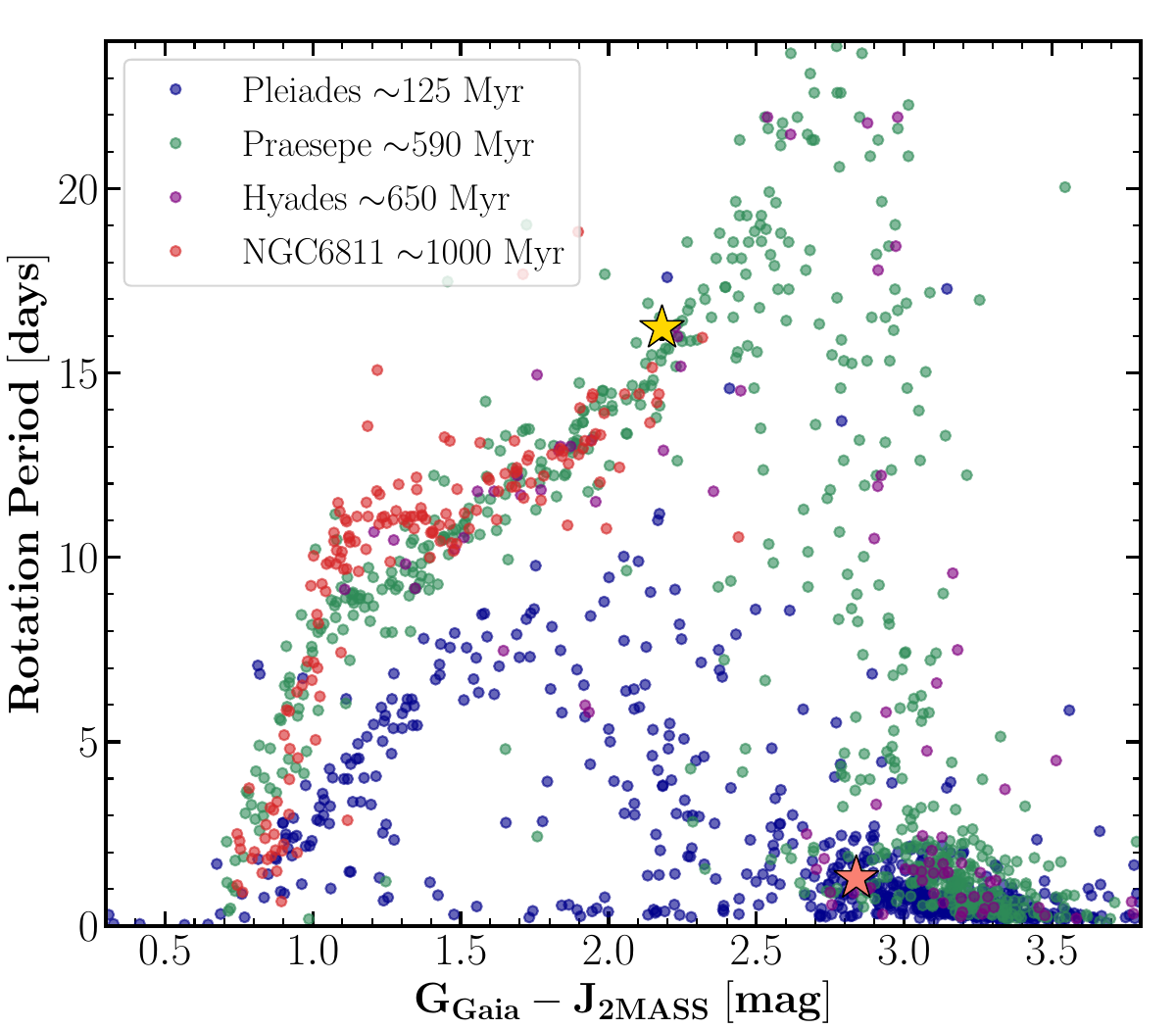}
\caption{Rotation period distribution as a function of colour $G$--$J$ for clusters in the Pleiades ($\sim$125 Myr; \citealp{rebull2016}), Praesepe ($\sim$590 Myr; \citealp{douglas2017}), Hyades ($\sim$650 Myr; \citealp{douglas2019}), and NGC~6811 ($\sim$1000 Myr; \citealp{curtis2019}). The gold star represents TOI-1801, and the orange star represents LP 275-24. 
\label{fig:prot_age}}
\end{figure}

We derived the rotation period of TOI-1801 ($P_{\text{rot}}$\,$=$15.98\,$\pm$\,0.66\,d) from the multiple light curves described in Sect.\,\ref{sec:prot}. For LP\,375-24, we used the light curves extracted from the quick-look pipeline (QLP; \citealt{Huang2020, Kunimoto2021}) that are available as high-level science products (HLSPs), and then we applied a GLS periodogram as for TOI-1801. From this, we obtained a rotation period of $P_{\text{rot}}$\,=\,1.30\,$\pm$\,0.06 d (the phase-folded curve is depicted in Fig.\,\ref{fig:LP_375-24_Prot}), which agrees with the value from the literature \citep{newton16}. Fig.\,\ref{fig:prot_age} shows the distribution of rotation periods as a function of colour $G$\,--\,$J$ for the clusters of the Pleiades ($\sim$125\,Myr; \citealp{rebull2016}), Praesepe ($\sim$590\,Myr; \citealp{douglas2017}), Hyades ($\sim$650\,Myr; \citealp{douglas2019}), and NGC~6811 ($\sim$1000\,Myr; \citealp{curtis2019}). TOI-1801 agrees with the members of Praesepe, Hyades, and NGC~6811 (i.e. with an age in the range 590--1000\,Myr), whereas LP\,375-24 is consistent with members of the Pleiades, Praesepe, and Hyades (i.e. with an age in the range 125--650\,Myr). Nevertheless, in the case of LP\,375-24, the dependence of the rotation period on age is not a clear age criterion because the rotation is faster at low masses. Moreover, the scarcity of stars of NGC~6811 with spectral types similar to that of LP\,375-24 did not allow us to set an older upper limit on the age of this star.

Various groups have widely used the distance-independent gyrochronology method (which is based solely on rotation periods and stellar colours) and the chromospheric method (which is based on the intensity of the Ca\,{\sc ii} H, and K emission) to date stars. \cite{Barnes2007} calibrated the age dependence of the gyrochronology technique using the Sun, stars from known open clusters, and field stars with well-determined ages. These authors claimed that age errors associated with this technique are about 15\,\%~for solar-type stars and early-M dwarfs.  By applying the calibration of \cite{Barnes2007} to TOI-1801, we derived an age of 565$\pm$75 Myr, which is fully consistent with what is derived from our colour-period diagram. Unfortunately, there are no $\log R'_{HK}$ (Ca\,{\sc ii} H, and K emission) measurements for TOI-1801 and LP\,375-24. Therefore, the chromospheric method (e.g., \citealt{Suarez2015}) cannot be applied to any of these stars.

\subsubsection{NUV excess} 

\begin{figure}[ht!]
\includegraphics[width=1\linewidth]{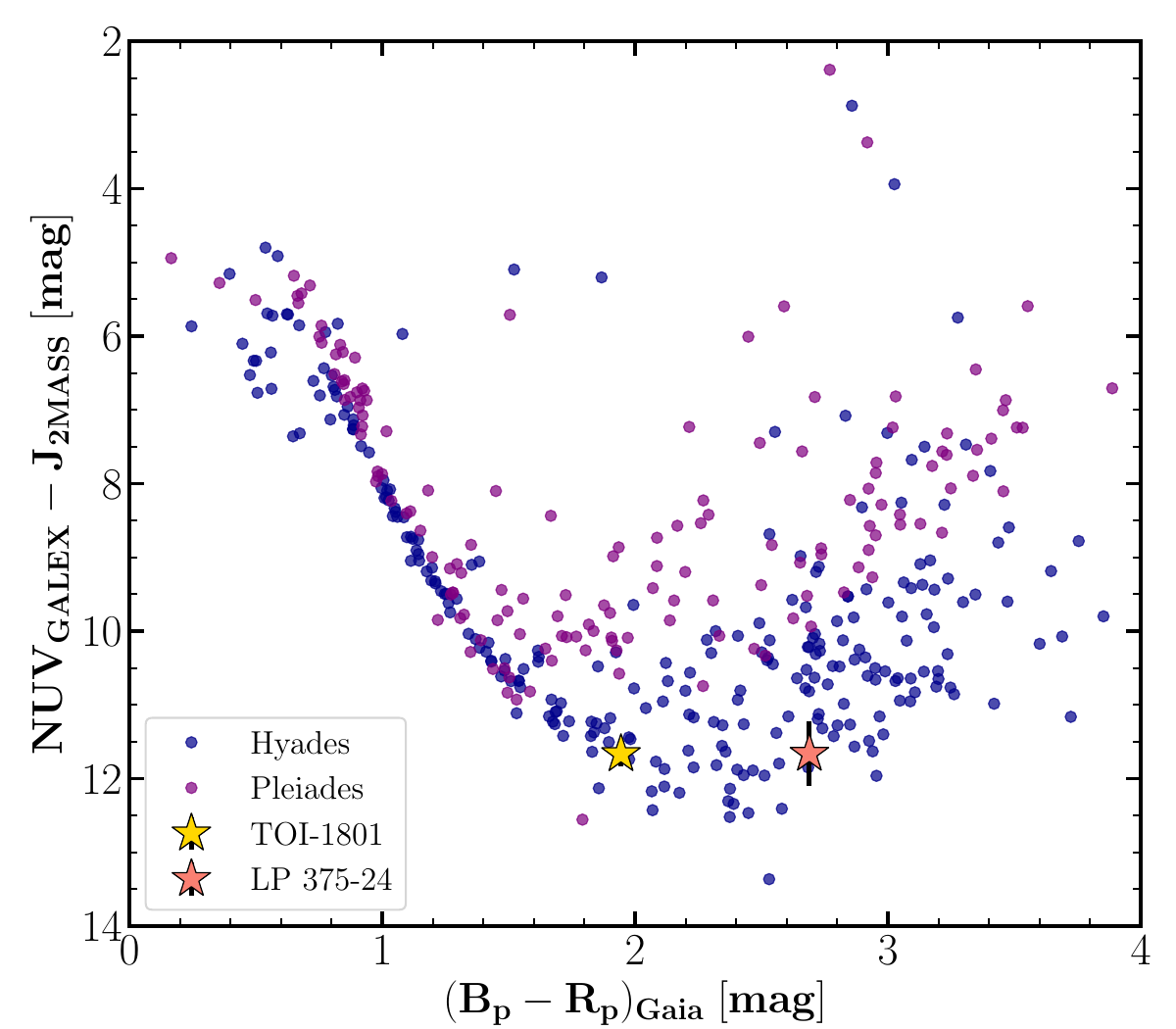}
\caption{Distribution of the $NUV$\,--\,$J$ colour as a function of colour $B_p$\,--\,$R_p$ for clusters in the Pleiades ($\sim$125 Myr; \citealp{olivares2018}) and Hyades ($\sim$650 Myr; \citealp{roser2019}). The gold and orange stars represent TOI-1801 and LP\,375-24, respectively. 
\label{fig:nuv_age}}
\end{figure}

The stellar UV (ultraviolet) emission can be an indicator of youth because it seems to decrease with age \citep{findeisen2011}. We employed the $NUV$ band of the GALEX all-sky catalogue \citep{bianchi2017} and Eq.\,10 in \cite{findeisen2011} and estimate the age of TOI-1801 to be of 500$^{+720}_{-290}$\,Myr. Unfortunately, the large dispersion in the $NUV$\,--\,$J$ colour versus $Gaia$ colours at low masses does not allow us to place a stringent constraint on the age of the wide companion. Similarly, \cite{shkolnik2011} and \cite{rodriguez2011} proposed a method for identifying young low-mass stars based on the flux ratio $F_{NUV}/F_{J}$ or $m_{NUV}$\,--\,$m_{J}$ colour, respectively. Fig.\,\ref{fig:nuv_age} shows the difference between the UV magnitude \citep{bianchi2017} and NIR magnitude \citep{cutri2003} as a function of $B_p$\,--\,$R_p$ colour for members of the clusters in the Pleiades \citep{olivares2018} and Hyades \citep{roser2019}. Our targets lie below the sequence of the Pleiades ($\sim$125\,Myr) and above the sequence of Hyades ($\sim$650\,Myr) members.

All these indicators point towards an age older than 125 Myr and younger than or equal to 1 Gyr and most probably similar to the ages of the Praesepe and Hyades clusters. Therefore, we adopt an age of 600--800 Myr for the system TOI-1801 and LP\,375-24 in the analysis below.

\section{Analysis}
\label{sec:Analysis}

\subsection{Photometric analysis}

\subsubsection{Transit search \label{sect:transit_search}}

As mentioned before, young stars show high levels of stellar activity, which makes the detection of possible transits difficult. First of all, we modelled the photometric stellar activity using Gaussian process regression \citep[GP;][]{Rasmussen2006}, using the simple harmonic oscillator (SHO) kernel implemented in the \texttt{celerite} package \citep{celerite}, which has the form

\begin{equation}
\begin{aligned}
k_{\mathrm{SHO}}(\tau; \eta_{\sigma}, \eta_{L}, \eta_{P}) = &\\
               = & \ \eta_{\sigma}^{2}  e^{-\frac{\tau}{\eta_{L}}}  \left[ \cos \left(\eta \frac{2\pi\tau}{\eta_{P}} \right) + \eta \frac{\eta_{P}}{2\pi \eta_{L}} \sin \left(\eta \frac{2\pi\tau}{\eta_{P}} \right) \right],
\end{aligned}
\label{eq:sho}
\end{equation}

\noindent where $\tau$\,$\equiv$\,$|t_i - t_j|$ is the time difference between two data points, $\eta$\,$\equiv$\,$|1 - (2\pi \eta_{L}/\eta_{P})^{-2} |)^{1/2}$ and $\eta_{\sigma}$, $\eta_{L}$, and $\eta_{P}$ are the hyperparameters that represent the standard deviation of the process, the decay timescale, and the period of the process (assumed to be the rotation period of the star), respectively. This kernel is defined in Eq.\,\ref{eq:sho} as long as $\eta_{P}$\,$<$\,$2\pi \eta_{L}$, a reasonable assumption in young stars where a clear quasi-periodic behaviour is observed in the time series. Moreover, our model also includes an instrumental offset ($\gamma_{\text{TESS}}$) for the TESS data set, as well as a jitter term ($\sigma_{\text{jit,TESS}}$) that was added in quadrature to the error bars. We set uniform wide priors for $\eta_{P}$ and $\eta_{L}$ based on the stellar rotation period. However, GPs can be too flexible and can model transits as activity. To constrain our model, we therefore set the $\eta_{\sigma}$ hyperparameter to $\sim$4.9 ppt (the dispersion in TESS data) with a restrictive normal prior and fixed the jitter term to $\sim$2.0 ppt (the maximum depth for the searched transit). With this set-up, the GP only models the stellar variability at longer timescales. To explore the parameter space, we applied the \texttt{emcee}\footnote{\url{https://github.com/dfm/emcee}} code \citep{emcee} that uses a Markov chain Monte Carlo procedure (MCMC). Subsequently, we flattened the SAP light curve by subtracting the best activity model. Then, we searched for transits using the box least-squares periodogram (BLS; \citealp{kovacs2002, hartman2016}). The main transit signal identified by the BLS lies at an orbital period of 10.6438 days, with three transits in sector 22 and two in sector 49. This confirms that the true orbital period is half the period on which the initial alert by TSO was based. Next, we masked out the signal, and we again applied the BLS algorithm to search for additional signals, but we did not find any significant signal. Finally, we visually inspected the light curve to find isolated transits, but we did not find variations compatible with them.

\subsubsection{Transit-only fit}
\label{sec:tronly}

After the transit was identified, we proceeded to create our photometric model as a combination of stellar activity and one planetary transit signal. The stellar activity was modelled with a GP for the TESS data set (as in the previous section) and with a jitter term for the LCO and MuSCAT2 data. To model the planetary transit, we used \texttt{PyTransit}\footnote{\url{https://github.com/hpparvi/PyTransit}} \citep{Parviainen2015}, which requires the following parameters: the planet--star radius ratio ($R_{p}/R_{\star}$), the quadratic limb-darkening coefficients ($u_1$ and $u_2$, which were parametrised after \cite{kipping02013}: $q_1$, $q_2$), the time-of-transit centre ($T_c$), the orbital period ($P$), the semi-major axis divided by the stellar radius ($a/R_{\star}$; which depends on $P$, the stellar mass $M_\star$, and the stellar radius $R_\star$, according to the third Kepler law), the orbital inclination ($i$, which we sampled from impact parameter, $b$, and $a/R_{\star}$), the eccentricity ($e$), and the argument of periastron ($\omega$). Therefore, our final transit-only model samples the planetary parameters of $T_c$, $P$, $R_p$, and, $b$. We also evaluated the possibility of non-circular orbits, including $e$ and $\omega$ with the parametrisation proposed by \cite{anderson2011}($\sqrt{e}\sin{\omega}$, $\sqrt{e}\cos{\omega}$). The stellar parameters are included as normal priors, and they are the $M_\star$, $R_\star$, and $q_1$, $q_2$ (where the limb-darkening coefficients are different for each instrument, and the initial values were previously calculated using the \texttt{Python limb darkening toolkit}\footnote{\url{https://github.com/hpparvi/ldtk}} (PyLDTk; \citealp{Parviainen2015b}). An instrumental offset and a jitter term, included as free parameters, were added for each individual photometric band. The prior and posterior results are presented in Table\,\ref{tab:fit-only}. For TOI-1801\,b we found $P$=$10.64387$\,$\pm$\,$0.00006$ d, $T_c$=$2458903.5435$\,$\pm$\,$0.0033$\,d, $R_p$=$2.08$\,$\pm$\,$0.12$ R$_{\oplus}$, and $b$=$0.27$\,$\pm$\,$0.18$. The errors correspond to 1$\sigma$ uncertainties. 

To assess the impact of using TESS SAP fluxes instead of the PDCSAP fluxes, which are corrected for instrumental systematics and stellar crowding, we repeated the same analysis by employing the PDCSAP data that were previously flattened, as explained in Section~\ref{sect:transit_search}. We note that the PDCSAP fluxes only have four planetary transits (see Section~\ref{sec:tesslc}). The new planetary radius is $R_p$=$2.13$\,$\pm$\,$0.11$ R$_{\oplus}$, that is, there is a difference of 0.05 R$_{\oplus}$ between the SAP and PDCSAP fluxes. This is twice smaller than the quoted error bar. Therefore, we conclude that there is no significant stellar contamination in the TESS SAP fluxes of TOI-1801.

\begingroup

\renewcommand{\arraystretch}{1.22} 
\begin{table*}
\caption{Prior and posterior parameters for the transit-only fit and RV-only fit of TOI-1801\,b.}\label{tab:fit-only}
\centering
\resizebox{0.89\linewidth}{!}{%
\begin{tabular}{lccc}
\hline\hline
Parameter & Prior & Posterior ($e$\,$=$\,0, $\omega$\,$=$\,$\pi/2$) & Posterior ($e$, $\omega$ free)\\
\hline
\multicolumn{4}{c}{Transit-only fit parameters}\\
\hline
$T_{c}^{b}$[BJD] & $\mathcal{N}$(2458903.543, 0.05) & 2458903.54346$^{+0.00319}_{-0.00329}$ & 2458903.54333$^{+0.00337}_{-0.00340}$\\
$P^{b}$[d] & $\mathcal{N}$(10.644, 0.05) & 10.64387$^{+0.00005}_{-0.00006}$ & 10.64386$^{+0.00005}_{-0.00006}$\\
$R_{p}^{b}$[R$_{\mathrm{Jup}}$] & $\mathcal{U}$(0, 1) & 0.185$^{+0.011}_{-0.010}$ & 0.187$^{+0.013}_{-0.011}$\\
$b^{b}$ & $\mathcal{U}$(0, 1) & 0.265$^{+0.181}_{-0.177}$ & 0.349$^{+0.220}_{-0.234}$\\
$(\sqrt{e}\sin\omega)^{b}$ & $\mathcal{U}$(--1, 1) & ... & --0.151$^{+0.272}_{-0.294}$\\
$(\sqrt{e}\cos\omega)^{b}$ & $\mathcal{U}$(--1, 1) & ... & --0.018$^{+0.542}_{-0.534}$\\
$\gamma_{\mathrm{TESS}}$[ppt] & $\mathcal{U}$(--3$\sigma_{\mathrm{TESS}}$, 3$\sigma_{\mathrm{TESS}}$) & 0.585$^{+1.628}_{-1.554}$ & 0.602$^{+1.586}_{-1.534}$\\
$\sigma_{\mathrm{jit,TESS}}$[ppt] & $\mathcal{U}$(0, 3$\sigma_{\mathrm{TESS}}$) & 0.146$^{+0.015}_{-0.015}$ & 0.146$^{+0.015}_{-0.015}$\\
$q_{1,\mathrm{TESS}}$ & $\mathcal{N}$(0.2, 0.1) & 0.288$^{+0.097}_{-0.096}$ & 0.292$^{+0.098}_{-0.098}$\\
$q_{2,\mathrm{TESS}}$ & $\mathcal{N}$(0.3, 0.1) & 0.357$^{+0.100}_{-0.099}$ & 0.359$^{+0.100}_{-0.099}$\\
$\gamma_{\mathrm{LCOGT}\ z_{s}}$[ppt] & $\mathcal{U}$(--3$\sigma_{\mathrm{LCOGT}\ z_{s}}$, 3$\sigma_{\mathrm{LCOGT}\ z_{s}}$) & --0.400$^{+0.220}_{-0.218}$ & --0.410$^{+0.219}_{-0.215}$\\
$\sigma_{\mathrm{jit,LCOGT}\ z_{s}}$[ppt] & $\mathcal{U}$(0, 3$\sigma_{\mathrm{LCOGT}\ z_{s}}$) & 0.274$^{+0.253}_{-0.188}$ & 0.269$^{+0.255}_{-0.186}$\\
$q_{1,\mathrm{LCOGT}\ z_{s}}$ & $\mathcal{N}$(0.2, 0.1) & 0.379$^{+0.190}_{-0.181}$ & 0.377$^{+0.194}_{-0.175}$\\
$q_{2,\mathrm{LCOGT}\ z_{s}}$ & $\mathcal{N}$(0.3, 0.1) & 0.367$^{+0.197}_{-0.185}$ & 0.356$^{+0.194}_{-0.182}$\\
$\gamma_{\mathrm{MuSCAT2}\ g'}$[ppt] & $\mathcal{U}$(--3$\sigma_{\mathrm{MuSCAT2}\ g'}$, 3$\sigma_{\mathrm{MuSCAT2}\ g'}$) & --0.101$^{+0.135}_{-0.136}$ & --0.099$^{+0.138}_{-0.138}$\\
$\sigma_{\mathrm{jit,MuSCAT2}\ g'}$[ppt] & $\mathcal{U}$(0, 3$\sigma_{\mathrm{MuSCAT2}\ g'}$) & 0.140$^{+0.142}_{-0.098}$ & 0.138$^{+0.140}_{-0.097}$\\
$q_{1,\mathrm{MuSCAT2}\ g'}$ & $\mathcal{N}$(0.5, 0.1) & 0.343$^{+0.187}_{-0.173}$ & 0.345$^{+0.188}_{-0.175}$\\
$q_{2,\mathrm{MuSCAT2}\ g'}$ & $\mathcal{N}$(0.3, 0.1) & 0.327$^{+0.192}_{-0.177}$ & 0.324$^{+0.193}_{-0.173}$\\
$\gamma_{\mathrm{MuSCAT2}\ r'}$[ppt] & $\mathcal{U}$(--3$\sigma_{\mathrm{MuSCAT2}\ r'}$, 3$\sigma_{\mathrm{MuSCAT2}\ r'}$) & --0.090$^{+0.090}_{-0.092}$ & --0.089$^{+0.090}_{-0.092}$\\
$\sigma_{\mathrm{jit,MuSCAT2}\ r'}$[ppt] & $\mathcal{U}$(0, 3$\sigma_{\mathrm{MuSCAT2}\ r'}$) & 0.158$^{+0.115}_{-0.104}$ & 0.157$^{+0.116}_{-0.104}$\\
$q_{1,\mathrm{MuSCAT2}\ r'}$ & $\mathcal{N}$(0.4, 0.1) & 0.288$^{+0.182}_{-0.163}$ & 0.280$^{+0.183}_{-0.161}$\\
$q_{2,\mathrm{MuSCAT2}\ r'}$ & $\mathcal{N}$(0.3, 0.1) & 0.302$^{+0.187}_{-0.169}$ & 0.304$^{+0.191}_{-0.172}$\\
$\gamma_{\mathrm{MuSCAT2}\ z_{s}}$[ppt] & $\mathcal{U}$(--3$\sigma_{\mathrm{MuSCAT2}\ z_{s}}$, 3$\sigma_{\mathrm{MuSCAT2}\ z_{s}}$) & --0.042$^{+0.102}_{-0.102}$ & --0.038$^{+0.100}_{-0.103}$\\
$\sigma_{\mathrm{jit,MuSCAT2}\ z_{s}}$[ppt] & $\mathcal{U}$(0, 3$\sigma_{\mathrm{MuSCAT2}\ z_{s}}$) & 0.129$^{+0.130}_{-0.090}$ & 0.129$^{+0.130}_{-0.090}$\\
$q_{1,\mathrm{MuSCAT2}\ z_{s}}$ & $\mathcal{N}$(0.2, 0.1) & 0.297$^{+0.183}_{-0.163}$ & 0.298$^{+0.181}_{-0.164}$\\
$q_{2,\mathrm{MuSCAT2}\ z_{s}}$ & $\mathcal{N}$(0.3, 0.1) & 0.312$^{+0.190}_{-0.170}$ & 0.309$^{+0.189}_{-0.173}$\\
$\eta_{\sigma, \mathrm{TESS}}$ & $\mathcal{N}$(5, 1) & 5.417$^{+0.742}_{-0.682}$ & 5.385$^{+0.728}_{-0.669}$\\
$\eta_{L, \mathrm{TESS}}$ & $\mathcal{U}$(2.5, 750) & 4.430$^{+1.885}_{-1.152}$ & 4.426$^{+1.890}_{-1.154}$\\
$\eta_{P_{\mathrm{rot}}}$ & $\mathcal{U}$(12, 20) & 15.410$^{+2.268}_{-1.884}$ & 15.324$^{+2.298}_{-1.856}$\\
$M_{\star}$[M$_{\odot}]$ & $\mathcal{N}$(0.548, 0.048) & 0.556$^{+0.043}_{-0.035}$ & 0.557$^{+0.044}_{-0.036}$\\
$R_{\star}$[R$_{\odot}]$ & $\mathcal{N}$(0.542, 0.029) & 0.547$^{+0.024}_{-0.022}$ & 0.547$^{+0.027}_{-0.025}$\\
\hline
\multicolumn{4}{c}{RV-only fit parameters}\\
\hline
$T_{c}^{b}$[BJD] & $\mathcal{N}$(2458903.543, 0.05) & 2458903.54104$^{+0.05049}_{-0.04945}$ & 2458903.53590$^{+0.04954}_{-0.04970}$\\
$P^{b}$[d] & $\mathcal{N}$(10.644, 0.05) & 10.64503$^{+0.00574}_{-0.00595}$ & 10.63473$^{+0.00860}_{-0.01002}$\\
$K^{b}$[m s$^{-1}$] & $\mathcal{U}$(0, 50) & 2.426$^{+0.624}_{-0.619}$ & 2.685$^{+0.794}_{-0.702}$\\
$(\sqrt{e}\sin\omega)^{b}$ & $\mathcal{U}$(--1, 1) & ... & 0.007$^{+0.379}_{-0.368}$\\
$(\sqrt{e}\cos\omega)^{b}$ & $\mathcal{U}$(--1, 1) & ... & 0.003$^{+0.378}_{-0.375}$\\
$\gamma_{\mathrm{CARMENES\ VIS}}$[m s$^{-1}$] & $\mathcal{U}$(--3$\sigma_{\mathrm{CARMENES\ VIS}}$, 3$\sigma_{\mathrm{CARMENES\ VIS}}$) & 0.482$^{+1.956}_{-2.003}$ & 0.436$^{+1.972}_{-2.102}$\\
$\sigma_{\mathrm{jit,CARMENES\ VIS}}$[m s$^{-1}$] & $\mathcal{U}$(0, 3$\sigma_{\mathrm{CARMENES\ VIS}}$) & 2.735$^{+0.780}_{-0.757}$ & 2.810$^{+0.829}_{-0.790}$\\
$\gamma_{\mathrm{CARMENES\ NIR}}$[m s$^{-1}$] & $\mathcal{U}$(--3$\sigma_{\mathrm{CARMENES\ NIR}}$, 3$\sigma_{\mathrm{CARMENES\ NIR}}$) & --0.130$^{+2.557}_{-2.655}$ & --0.147$^{+2.587}_{-2.644}$\\
$\sigma_{\mathrm{jit,CARMENES\ NIR}}$[m s$^{-1}$] & $\mathcal{U}$(0, 3$\sigma_{\mathrm{CARMENES\ NIR}}$) & 2.506$^{+1.997}_{-1.690}$ & 2.531$^{+2.015}_{-1.712}$\\
$\gamma_{\mathrm{HIRES}}$[m s$^{-1}$] & $\mathcal{U}$(--3$\sigma_{\mathrm{HIRES}}$, 3$\sigma_{\mathrm{HIRES}}$) & --2.189$^{+2.025}_{-2.056}$ & --2.257$^{+2.029}_{-2.067}$\\
$\sigma_{\mathrm{jit,HIRES}}$[m s$^{-1}$] & $\mathcal{U}$(0, 3$\sigma_{\mathrm{HIRES}}$) & 4.325$^{+1.323}_{-1.378}$ & 3.845$^{+1.274}_{-1.210}$\\
$\eta_{\sigma, \mathrm{CARMENES\ VIS}}$ & $\mathcal{N}$(6.3, 1.2) & 5.900$^{+0.974}_{-0.881}$ & 5.847$^{+0.978}_{-0.899}$\\
$\eta_{\sigma, \mathrm{CARMENES\ NIR}}$ & $\mathcal{N}$(9, 2) & 6.890$^{+1.705}_{-1.614}$ & 6.855$^{+1.752}_{-1.625}$\\
$\eta_{\sigma, \mathrm{HIRES}}$ & $\mathcal{N}$(6.3, 1.2) & 5.823$^{+1.034}_{-1.025}$ & 5.857$^{+1.039}_{-1.014}$\\
$\eta_{L, \mathrm{RV}}$ & $\mathcal{U}$(2.5, 500) & 109.615$^{+72.006}_{-41.314}$ & 111.358$^{+92.562}_{-46.393}$\\
$\eta_{\omega, \mathrm{RV}}$ & $\mathcal{U}$(0.1, 1.0) & 0.294$^{+0.064}_{-0.053}$ & 0.300$^{+0.068}_{-0.055}$\\
$\eta_{P_{\mathrm{rot}}}$ & $\mathcal{U}$(12, 20) & 16.043$^{+0.073}_{-0.079}$ & 16.045$^{+0.081}_{-0.085}$\\
$\Delta \ln \mathcal{Z}$ & ... & --602.7 & --605.3\\

\hline
\end{tabular}}
\tablefoot{The prior label of $\mathcal{N}$ and $\mathcal{U}$ represents the normal and uniform distribution, respectively.}\\
\end{table*}
\endgroup

Finally, we determined the chromaticity of the transit using a different planetary radius parameter for each available filter from the different instruments used in our analysis. The results are shown along with their 1$\sigma$ uncertainties in Table\,\ref{tabla:Rp}. All of them are consistent with an achromatic transit within their 1$\sigma$ error bars.

\begingroup
\renewcommand{\arraystretch}{1.22} 
\begin{table}[htbp]
\caption{Posteriors and their 1$\sigma$ uncertainties for planetary radii of TOI-1801\,b in different filters.}\label{tabla:Rp}
\begin{center}
\begin{tabular}{l c c }
\hline
\hline
Instrument & Filter & $R_p$ [R$_{\oplus}$]\\
\hline
TESS & $T$ & 2.06$^{+0.13}_{-0.11}$ \\
LCOGT & $z_{s}$ & 2.20$^{+0.34}_{-0.37}$ \\
MuSCAT2 & $g'$ & 2.01$^{+0.22}_{-0.24}$ \\
MuSCAT2 & $r'$ & 2.08$^{+0.18}_{-0.18}$ \\
MuSCAT2 & $z_{s}$ & 2.09$^{+0.17}_{-0.19}$ \\
All & -- & 2.08$^{+0.12}_{-0.11}$ \\
\hline
\end{tabular}
\end{center}
\end{table}
\endgroup

\subsection{Spectral stellar activity indicators}
\label{sec:activity}

\begin{figure}[ht!]
\includegraphics[width=1\linewidth]{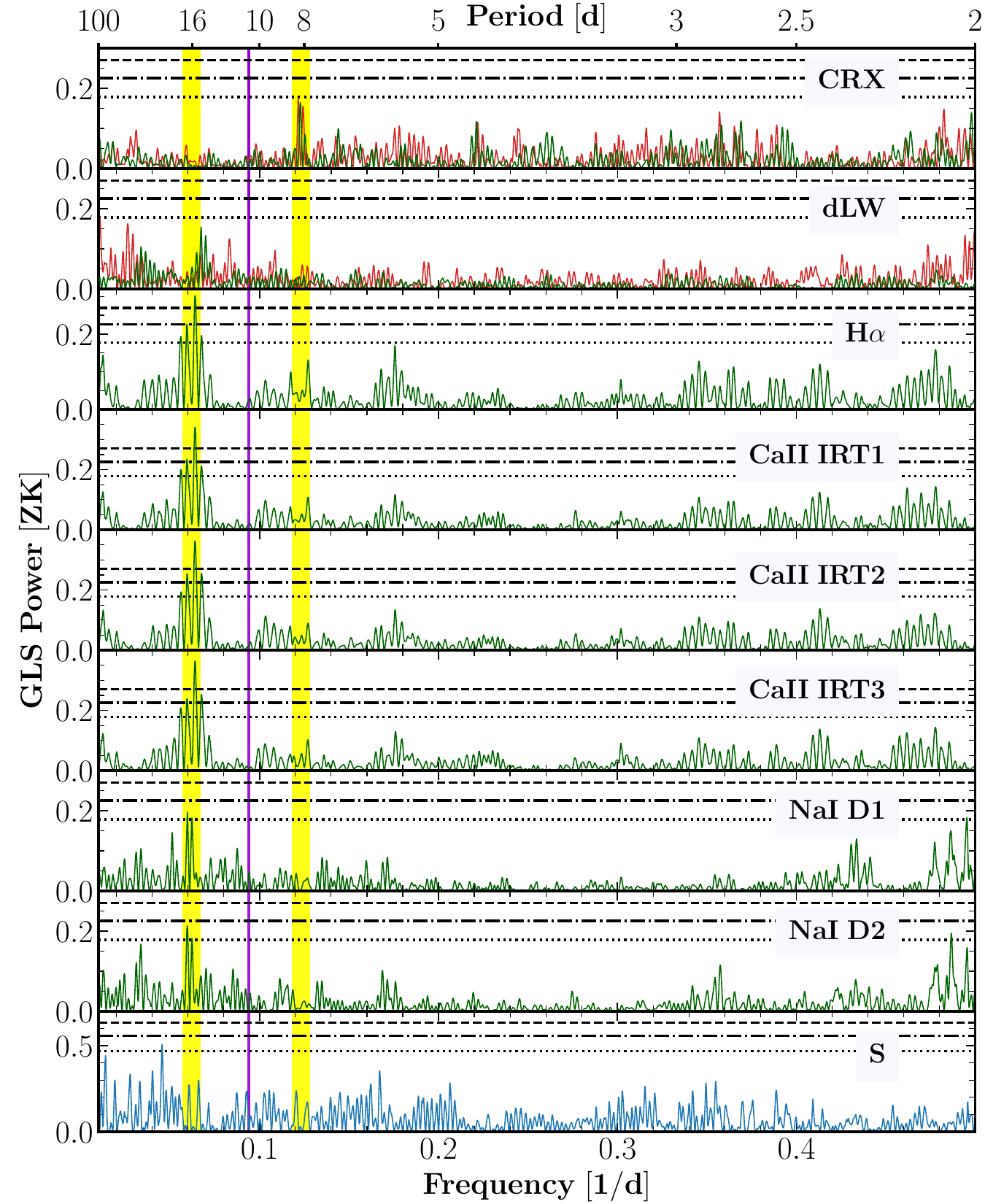}
\caption{GLS periodograms of the spectral activity indicators from CARMENES VIS (green) and NIR data (red), and the $S$ index from HIRES data (blue). In all panels, the solid vertical purple line indicates the orbital period (10.64 d) of the planet, and the vertical yellow shaded bands indicate the rotation period derived from the combined photometry data sets and half the rotation period. The dashed horizontal black lines correspond to FAP levels of 10\%, 1\%, and 0.1\% (from bottom to top).  
\label{fig:rv_act_ind}}
\end{figure}

We inspected periodic signals in the stellar activity indicators of the CARMENES data generated by the \texttt{serval} pipeline and in the $S$ index in the HIRES data set. The GLS periodograms were computed and are shown in Fig.\,\ref{fig:rv_act_ind}. In the CRX periodograms, we see a non-significant signal (close to 10\% of FAP) at $\sim$8 days. For the H$\alpha$, \ion{Ca}{II}, and \ion{Na}{I} activity indices, all periodograms show a significant signal (lower in the case of \ion{Na}{I} indices) centred between 15 and 17 days (marked as a vertical yellow band in Fig.\,\ref{fig:rv_act_ind}), consistent with the stellar rotation period (Sect.\,\ref{sec:prot}). Lastly, the periodograms of dLW and the $S$ index do not show any significant signal. The purple line in all panels indicates the transiting planetary signal, but no signal is seen in the activity indicators at this period. Furthermore, we computed the Pearson $r$ coefficient to study the correlation between the CARMENES RV data and the activity indicators. We found no significant correlation. In conclusion, the activity indices show a significant signal between 15--17 days that is compatible with the stellar rotation.

\subsection{Radial velocity analysis}
\label{sec:rvonly}

First of all, we determined whether our RV observations fell on transits, which could alter the measurements because of the Rossiter-McLaughlin (RM) effect \citep{rossiter, mclaughlin}. Indeed, four CARMENES RVs were obtained during a transit event (BJD\,=\,2459616.7499, 2459648.5406, 2459680.4698, and 2459712.3640). However, the expected upper limit of the semi-amplitude of the RM effect \citep{gaudi07} is $\sim$1.5 m\,s$^{-1}$, which is less than the uncertainties in our measurements. For this reason, we decided to use all the available data. Overall, the final combined data set comprises 80 RV measurements from CARMENES VIS plus 68 from CARMENES NIR and 29 from HIRES.

\begin{figure}[ht!]
\includegraphics[width=1\linewidth]{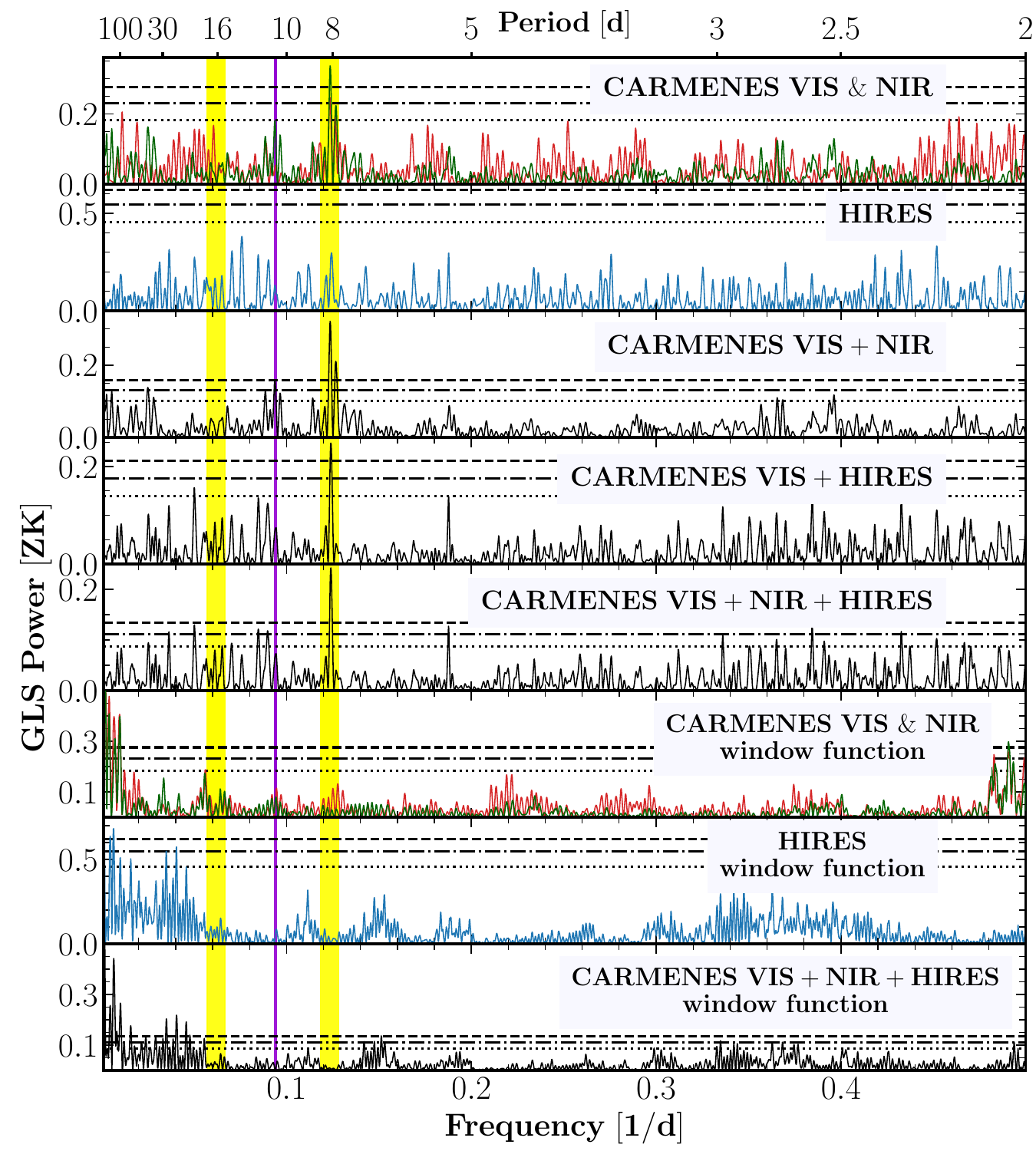}
\caption{GLS periodograms for CARMENES VIS (green line) and CARMENES NIR (red line) in the upper panel and for HIRES (blue line) in the second panel. The third, fourth, and fifth panels (from top to bottom) show different combinations for these data sets (black line). The last three panels show the window functions. In all panels, the solid vertical purple line indicates the transiting planetary signal (10.6 d), and the shaded vertical yellow bands indicate the stellar rotation period (15--17\,d) and half the rotation period. The dashed horizontal black lines correspond to FAP levels of 10\%, 1\%, and 0.1\% (from bottom to top).
\label{fig:RV_periodograms}}
\end{figure}

We explored periodic signals in the CARMENES and HIRES data by computing GLS periodograms in the same way as for the activity indicators (Fig.\,\ref{fig:RV_periodograms}). The first five panels (from top to bottom) show the RV periodograms of the CARMENES VIS and NIR data, HIRES data, and their combinations. In the first panel, the most significant signal is at $\sim$8 d ($<$\,0.1\% FAP in the VIS and $\sim$1\% FAP in the NIR data sets; vertical yellow band), where the signal of the planet (10.64 d; purple line) is close to 10\% FAP for the VIS data. In the second panel from the top, no significant signals are visible in the HIRES data alone. In the panels that combine the data sets (third, fourth, and fifth panels), the signal at half the rotation period is always present with a high significance. In addition, the signal of the planet increased ($<$\,0.1\% FAP) for the combination of CARMENES VIS and NIR data sets and decreased when we included the HIRES data. The next three panels show the window functions for these three data sets. To conclude, the combination of the CARMENES data shows the signal of the transiting planet ($\sim$10.64 d) and a signal that is related to half the stellar rotation period ($\sim$8.1 d) with high significance.

After the identification of the signals, we performed an RV-only fit by modelling the stellar activity in three ways. The first model only used a jitter term. The second model included a jitter term plus two sinusoidal functions centred on the rotation period and half of the rotation period. The last model used a jitter term plus a quasi-periodic (QP) kernel from \citet{aigrain12},

\begin{equation}
k_{\mathrm{QP}}(\tau)  =  \eta_{\sigma}^2 \exp \left[ -\frac{\tau^2}{2\eta_{L}^2} -\frac{\sin^2{ \left( \frac{\pi \tau}{\eta_{P}} \right)}}{2\eta_{\omega}^2} \right], \\
\label{eq:qp}
\end{equation}

\noindent where $\tau$, $\eta_{\sigma}$, $\eta_{L}$, and $\eta_{P}$ are defined as in Eq.\,\ref{eq:sho}, and $\eta_{\omega}$ works as the balance between the periodic and non-periodic part of the kernel. The QP kernel has been widely used in the literature to model the stellar activity of young stars \citep{barr19, klein21, cale21, zicher22, nardiello22, barr22, Mallorquin23}. As in the photometric fit, we set normal priors on the covariance amplitudes to create a smooth model of the stellar activity. The planet signal was modelled as a circular Keplerian orbit with the \texttt{RadVel}\footnote{\url{https://github.com/California-Planet-Search/radvel}} package \citep{fultonradvel}. The Keplerian model parameters were $T_c$, $P$, and the RV amplitude of the planet ($K$), where the initial parameters of $T_c$ and $P$ were obtained from the transit-only fit (Sect.\,\ref{sec:tronly}). The RV-only models also included an instrumental offset ($\gamma_{\mathrm{RV}}$) and a jitter term added in quadrature to the error bars ($\sigma_{\mathrm{jit,RV}}$). In addition, we also explored the possibility of not including a planetary signal in our models.

\begin{table*}[htbp]
\caption{Model comparison for RV-only analysis of TOI-1801\,b using the difference between the Bayesian log-evidence ($\Delta \ln \mathcal{Z}$).}
\label{tabla:logZ}
\begin{center}
\begin{tabular}{c l c c}
\hline
\hline
data set & Activity model & K$^b$[m\,s$^{-1}$] & $\Delta \ln \mathcal{Z}$ \\
\hline
CARMENES VIS & jitter & ... & --12.9\\
CARMENES VIS & jitter & 3.25\,$\pm$\,0.97 & --10.7\\
CARMENES VIS & 2 sin (P$_{1}\sim16$\,d, P$_{2}\sim8$\,d) + jitter & 2.17\,$\pm$\,0.89 & --8.1\\
CARMENES VIS & GP (QP kernel) + jitter & ... & --4.0\\
CARMENES VIS & GP (QP kernel) + jitter & 2.60\,$\pm$\,0.75 & 0\\
\hline
CARMENES VIS + CARMENES NIR & jitter & ... & --16.3\\
CARMENES VIS + CARMENES NIR & jitter & 3.25\,$\pm$\,0.87 & --12.7\\
CARMENES VIS + CARMENES NIR & 2 sin (P$_{1}\sim16$\,d, P$_{2}\sim8$\,d) + jitter & 2.06\,$\pm$\,0.76 & --9.6\\
CARMENES VIS + CARMENES NIR & GP (QP kernel) + jitter & ... & --7.4\\
CARMENES VIS + CARMENES NIR & GP (QP kernel) + jitter & 2.58\,$\pm$\,0.69 & 0\\
\hline
CARMENES VIS + CARMENES NIR + HIRES & jitter & ... & --22\\
CARMENES VIS + CARMENES NIR + HIRES & jitter & 2.85\,$\pm$\,0.75 & --18.2\\
CARMENES VIS + CARMENES NIR + HIRES & 2 sin (P$_{1}\sim16$\,d, P$_{2}\sim8$\,d) + jitter & 2.08\,$\pm$\,0.69 & --13.7\\
CARMENES VIS + CARMENES NIR + HIRES & GP (QP kernel) + jitter & ... & --7.4\\
\textbf{CARMENES VIS + CARMENES NIR + HIRES} & \textbf{GP (QP kernel) + jitter} & \textbf{2.43\,$\pm$\,0.62} & \textbf{0}\\
\hline
\end{tabular}
\tablefoot{In the model name, "2 Sin" refers to two sinusoidal functions and their periods. All models assume circular orbits. The result in bold indicates the RV model we adopted.}\\
\end{center}
\end{table*}

To evaluate different models, we used the rules defined by \cite{trotta2008} based on the Bayesian log-evidence ($\ln \mathcal{Z}$, calculated as by \citet{diaz16}). The model with a larger log-evidence is strongly favoured if $|\Delta \ln \mathcal{Z}|$\,$>$\,5. However, if 2.5\,$<$\,$|\Delta \ln \mathcal{Z}|$\,$<$\,5, the evidence in favour of one of the models is moderate, it is weak if 1\,$<$\,$|\Delta \ln \mathcal{Z}|$\,$<$\,2.5, and it is indistinguishable if $|\Delta \ln \mathcal{Z}|$\,$<$\,1. Although no significant signals appear in the GLS periodogram with the HIRES data alone, we explored the possibility of combining the HIRES data with CARMENES RVs. The results of a model comparison are provided in Table \ref{tabla:logZ}. The amplitude we obtained for the planet varies between 2.1 and 3.3 m\,s$^{-1}$, and all values are consistent within the error bars. The Bayesian log-evidence indicates that the models that include the planet are always favoured over those without a Keplerian model. Moreover, the best model in all the cases is obtained when the activity is modelled with GP. Finally, we considered as the best RV model the one with a Keplerian fit to the transiting planet and a GP model for the stellar activity obtained for the combined CARMENES VIS plus the CARMENES NIR and HIRES data sets (K$^b$=2.43$\pm$0.62\,m\,s$^{-1}$). The planet is more significant when we included the CARMENES NIR data (Fig.\,\ref{fig:RV_periodograms}). Furthermore, and although the HIRES data do not show any significant signal (probably due to the low number of measurements), the dispersion of the data and the error bars is comparable to that in the CARMENES data and adds more cadence to the time series. The planetary parameters for the RV-only fit including non-eccentric orbits are listed in Table \ref{tab:fit-only}, where the Bayesian log-evidence shows that the circular model is moderately better than with non-zero eccentricity.

\subsection{Joint fit}
\label{sec:jfit}

\begin{figure}[ht!]
\includegraphics[width=1\linewidth]{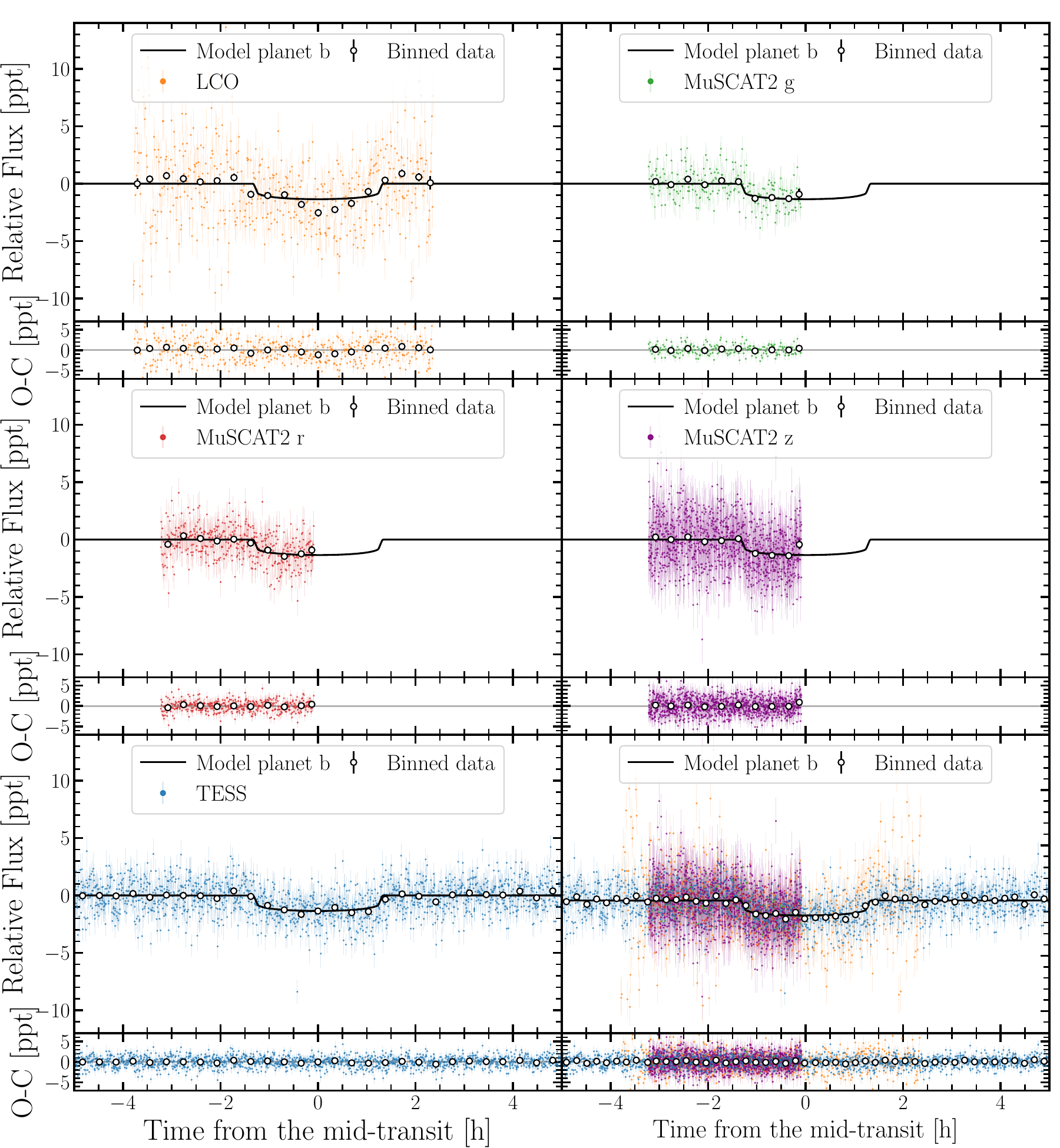}
\caption{Phase-folded light curves of TOI-1801\,b for data of LCO $z_{s}$, MuSCAT2 $g'$, MuSCAT2 $r'$, MuSCAT2 $z_{s}$, TESS, and the combination of all them. In each sub-panel, the photometric data (coloured dots) are shown, along with the binned data (white dots), the best transit-fit model (black line) in the top, and the residuals for the best fit in the bottom.
\label{fig:PH_folded}}
\end{figure}

\begin{figure}[ht!]
\includegraphics[width=1\linewidth]{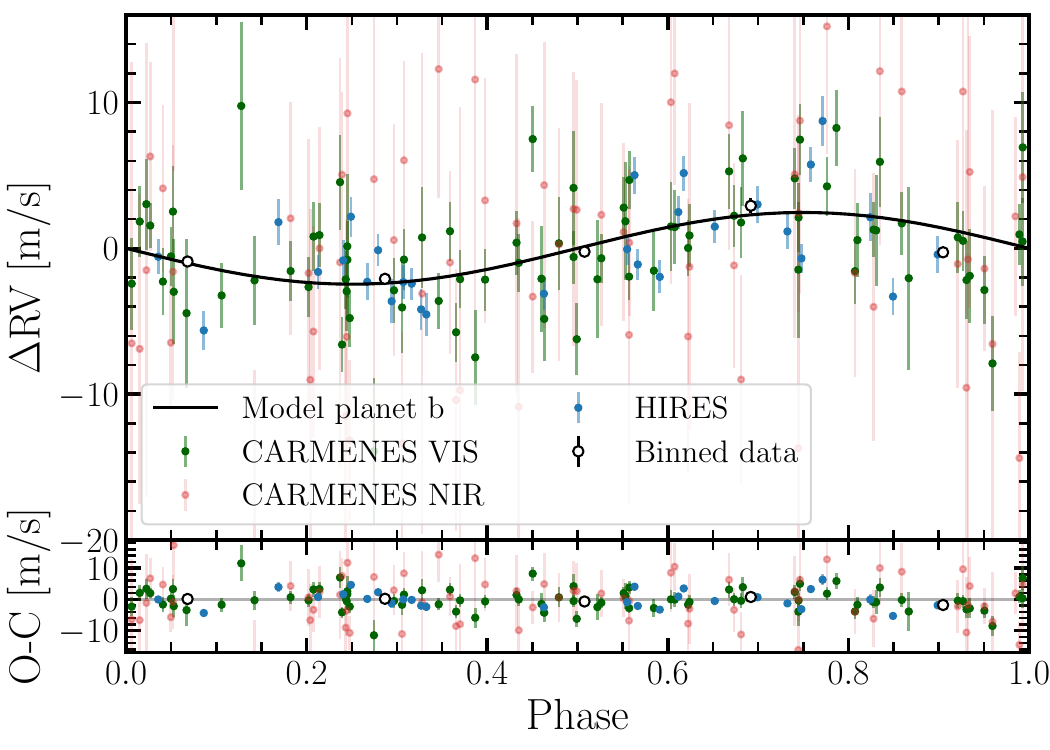}
\caption{Phase-folded RVs for TOI-1801\,b. \textit{Top panel:} CARMENES VIS, CARMENES NIR, and HIRES data (green, red, and blue dots, respectively), binned data (white dots), and the Keplerian model of the joint fit (black line). \textit{Bottom panel:} Residuals for the best fit. Due to the size of the error bars of CARMENES NIR, and in order to properly appreciate the Keplerian model of the planet, a zoom-in has been performed on y-axis of the figure, even when some CARMENES NIR points were out of the plot.
\label{fig:RV_folded}}
\end{figure}

Finally, we combined all the data from TESS, the ground-based transit follow-ups, CARMENES VIS, CARMENES NIR, and HIRES to obtain more precise parameters of the TOI-1801 system. Our global fit includes the photometric (Fig.\,\ref{fig:PH_TESS}) and RV stellar activity models (Fig.\,\ref{fig:RV_curve}; Sects.\,\ref{sec:tronly} and \ref{sec:rvonly}, respectively), and transit (see the phase-folded transits in Fig.\,\ref{fig:PH_folded}) and Keplerian models (the phase-folded RVs in Fig.\,\ref{fig:RV_folded}) to obtain the planetary parameters. These planetary parameters are as follows: $T_c$ and $P$ are in common to all the data sets (with normal priors), $b$ and $R_p$ are in common to the photometry data sets (uniform priors), and $K_p$ is in common in the RV data sets (uniform prior). In the hyperparameters that model the stellar activity, $\eta_P$ is in common to TESS and the RV data sets, while $\eta_L$ and $\eta_\omega$ are in common to the RV data sets. Therefore, the global model consists of a combination of a 10.6 d planet and activity, the former being computed for a circular orbit and an eccentric orbit (where $e$ and $\omega$ also are in common to all the data sets). The priors and posterior results can be found in Table \ref{tab:joint-fit}, and the derived parameters can be found in Table \ref{tab:joint-fit_der}.

\begingroup
\renewcommand{\arraystretch}{1.22} 
\begin{table*}
\caption{Prior and posterior parameters of the joint fit for TOI-1801\,b.}\label{tab:joint-fit}
\centering
\begin{tabular}{lccc}
\hline\hline
Parameter & Prior & Posterior ($e$\,$=$\,0, $\omega$\,$=$\,$\pi/2$) & Posterior ($e$, $\omega$ free)\\
\hline
$T_{c}^{b}$[BJD] & $\mathcal{N}$(2458903.543, 0.05) & 2458903.54351$^{+0.00325}_{-0.00331}$ & 2458903.54335$^{+0.00334}_{-0.00337}$\\
$P^{b}$[d] & $\mathcal{N}$(10.644, 0.05) & 10.64387$^{+0.00005}_{-0.00006}$ & 10.64387$^{+0.00005}_{-0.00006}$\\
$R_{p}^{b}$[R$_{\mathrm{Jup}}$] & $\mathcal{U}$(0, 1) & 0.186$^{+0.011}_{-0.010}$ & 0.186$^{+0.011}_{-0.010}$\\
$b^{b}$ & $\mathcal{U}$(0, 1) & 0.265$^{+0.182}_{-0.179}$ & 0.282$^{+0.180}_{-0.189}$\\
$K^{b}$[m\,s$^{-1}$] & $\mathcal{U}$(0, 50) & 2.459$^{+0.624}_{-0.612}$ & 2.496$^{+0.643}_{-0.638}$\\
$(\sqrt{e}\sin\omega)^{b}$ & $\mathcal{U}$(--1, 1) & ... & --0.012$^{+0.201}_{-0.200}$\\
$(\sqrt{e}\cos\omega)^{b}$ & $\mathcal{U}$(--1, 1) & ... & --0.005$^{+0.249}_{-0.245}$\\
\hline
$\gamma_{\mathrm{TESS}}$[ppt] & $\mathcal{U}$(--3$\sigma_{\mathrm{TESS}}$, 3$\sigma_{\mathrm{TESS}}$) & 0.672$^{+1.659}_{-1.682}$ & 0.620$^{+1.704}_{-1.673}$\\
$\sigma_{\mathrm{TESS}}$[ppt] & $\mathcal{U}$(0, 3$\sigma_{\mathrm{TESS}}$) & 0.146$^{+0.015}_{-0.016}$ & 0.146$^{+0.015}_{-0.015}$\\
$q_{1,\mathrm{TESS}}$ & $\mathcal{N}$(0.2, 0.1) & 0.288$^{+0.097}_{-0.097}$ & 0.287$^{+0.099}_{-0.097}$\\
$q_{2,\mathrm{TESS}}$ & $\mathcal{N}$(0.3, 0.1) & 0.359$^{+0.098}_{-0.097}$ & 0.359$^{+0.097}_{-0.098}$\\
$\gamma_{\mathrm{LCOGT}\ z_{s}}$[ppt] & $\mathcal{U}$(--3$\sigma_{\mathrm{LCOGT}\ z_{s}}$, 3$\sigma_{\mathrm{LCOGT}\ z_{s}}$) & --0.391$^{+0.218}_{-0.222}$ & --0.401$^{+0.220}_{-0.219}$\\
$\sigma_{\mathrm{LCOGT}\ z_{s}}$[ppt] & $\mathcal{U}$(0, 3$\sigma_{\mathrm{LCOGT}\ z_{s}}$) & 0.268$^{+0.256}_{-0.184}$ & 0.263$^{+0.255}_{-0.181}$\\
$q_{1,\mathrm{LCOGT}\ z_{s}}$ & $\mathcal{N}$(0.2, 0.1) & 0.380$^{+0.193}_{-0.180}$ & 0.375$^{+0.192}_{-0.180}$\\
$q_{2,\mathrm{LCOGT}\ z_{s}}$ & $\mathcal{N}$(0.3, 0.1) & 0.367$^{+0.196}_{-0.187}$ & 0.362$^{+0.198}_{-0.185}$\\
$\gamma_{\mathrm{MuSCAT2}\ g'}$[ppt] & $\mathcal{U}$(--3$\sigma_{\mathrm{MuSCAT2}\ g'}$, 3$\sigma_{\mathrm{MuSCAT2}\ g'}$) & --0.102$^{+0.134}_{-0.136}$ & --0.099$^{+0.137}_{-0.138}$\\
$\sigma_{\mathrm{MuSCAT2}\ g'}$[ppt] & $\mathcal{U}$(0, 3$\sigma_{\mathrm{MuSCAT2}\ g'}$) & 0.132$^{+0.140}_{-0.092}$ & 0.137$^{+0.143}_{-0.096}$\\
$q_{1,\mathrm{MuSCAT2}\ g'}$ & $\mathcal{N}$(0.5, 0.1) & 0.342$^{+0.184}_{-0.172}$ & 0.350$^{+0.185}_{-0.177}$\\
$q_{2,\mathrm{MuSCAT2}\ g'}$ & $\mathcal{N}$(0.3, 0.1) & 0.329$^{+0.191}_{-0.178}$ & 0.321$^{+0.192}_{-0.177}$\\
$\gamma_{\mathrm{MuSCAT2}\ r'}$[ppt] & $\mathcal{U}$(--3$\sigma_{\mathrm{MuSCAT2}\ r'}$, 3$\sigma_{\mathrm{MuSCAT2}\ r'}$) & --0.090$^{+0.090}_{-0.092}$ & --0.092$^{+0.091}_{-0.092}$\\
$\sigma_{\mathrm{MuSCAT2}\ r'}$[ppt] & $\mathcal{U}$(0, 3$\sigma_{\mathrm{MuSCAT2}\ r'}$) & 0.157$^{+0.116}_{-0.103}$ & 0.159$^{+0.113}_{-0.104}$\\
$q_{1,\mathrm{MuSCAT2}\ r'}$ & $\mathcal{N}$(0.4, 0.1) & 0.283$^{+0.185}_{-0.161}$ & 0.287$^{+0.182}_{-0.161}$\\
$q_{2,\mathrm{MuSCAT2}\ r'}$ & $\mathcal{N}$(0.3, 0.1) & 0.303$^{+0.191}_{-0.167}$ & 0.308$^{+0.185}_{-0.172}$\\
$\gamma_{\mathrm{MuSCAT2}\ z_{s}}$[ppt] & $\mathcal{U}$(--3$\sigma_{\mathrm{MuSCAT2}\ z_{s}}$, 3$\sigma_{\mathrm{MuSCAT2}\ z_{s}}$) & --0.041$^{+0.104}_{-0.100}$ & --0.042$^{+0.103}_{-0.102}$\\
$\sigma_{\mathrm{MuSCAT2}\ z_{s}}$[ppt] & $\mathcal{U}$(0, 3$\sigma_{\mathrm{MuSCAT2}\ z_{s}}$) & 0.125$^{+0.126}_{-0.087}$ & 0.130$^{+0.130}_{-0.091}$\\
$q_{1,\mathrm{MuSCAT2}\ z_{s}}$ & $\mathcal{N}$(0.2, 0.1) & 0.300$^{+0.186}_{-0.167}$ & 0.298$^{+0.184}_{-0.167}$\\
$q_{2,\mathrm{MuSCAT2}\ z_{s}}$ & $\mathcal{N}$(0.3, 0.1) & 0.317$^{+0.183}_{-0.169}$ & 0.314$^{+0.185}_{-0.169}$\\
$\gamma_{\mathrm{CARMENES\ VIS}}$[m $s^{-1}$] & $\mathcal{U}$(--3$\sigma_{\mathrm{CARMENES\ VIS}}$, 3$\sigma_{\mathrm{CARMENES\ VIS}}$) & 0.462$^{+1.916}_{-2.062}$ & -0.492$^{+1.967}_{-2.045}$\\
$\sigma_{\mathrm{CARMENES\ VIS}}$[m $s^{-1}$] & $\mathcal{U}$(0, 3$\sigma_{\mathrm{CARMENES\ VIS}}$) & 2.692$^{+0.773}_{-0.748}$ & 2.750$^{+0.811}_{-0.781}$\\
$\gamma_{\mathrm{CARMENES\ NIR}}$[m $s^{-1}$] & $\mathcal{U}$(--3$\sigma_{\mathrm{CARMENES\ NIR}}$, 3$\sigma_{\mathrm{CARMENES\ NIR}}$) & --0.146$^{+2.528}_{-2.623}$ & --0.123$^{+2.601}_{-2.671}$\\
$\sigma_{\mathrm{CARMENES\ NIR}}$[m $s^{-1}$] & $\mathcal{U}$(0, 3$\sigma_{\mathrm{CARMENES\ NIR}}$) & 2.500$^{+2.038}_{-1.712}$ & 2.561$^{+2.024}_{-1.738}$\\
$\gamma_{\mathrm{HIRES}}$[m $s^{-1}$] & $\mathcal{U}$(--3$\sigma_{\mathrm{HIRES}}$, 3$\sigma_{\mathrm{HIRES}}$) & --2.222$^{+1.995}_{-2.036}$ & --2.250$^{+2.013}_{-2.064}$\\
$\sigma_{\mathrm{HIRES}}$[m $s^{-1}$] & $\mathcal{U}$(0, 3$\sigma_{\mathrm{HIRES}}$) & 3.945$^{+1.253}_{-1.132}$ & 3.951$^{+1.256}_{-1.150}$\\
$\eta_{\sigma, \mathrm{TESS}}$ & $\mathcal{N}$(5, 1) & 5.460$^{+0.639}_{-0.541}$ & 5.465$^{+0.633}_{-0.533}$\\
$\eta_{\sigma, \mathrm{CARMENES\ VIS}}$ & $\mathcal{N}$(6.3, 1.2) & 5.906$^{+0.974}_{-0.893}$ & 5.882$^{+1.003}_{-0.884}$\\
$\eta_{\sigma, \mathrm{CARMENES\ NIR}}$ & $\mathcal{N}$(9, 2) & 6.844$^{+1.746}_{-1.627}$ & 6.880$^{+1.728}_{-1.602}$\\
$\eta_{\sigma, \mathrm{HIRES}}$ & $\mathcal{N}$(6.3, 1.2) & 5.783$^{+1.023}_{-1.012}$ & 5.840$^{+1.020}_{-1.009}$\\
$\eta_{L, \mathrm{TESS}}$ & $\mathcal{U}$(2.5, 750) & 4.263$^{+1.536}_{-1.025}$ & 4.259$^{+1.544}_{-1.033}$\\
$\eta_{L, \mathrm{RV}}$ & $\mathcal{U}$(2.5, 500) & $107.117^{+67.794}_{-40.286}$ & 109.381$^{+79.426}_{-43.783}$\\
$\eta_{P_{\mathrm{rot}}}$  & $\mathcal{U}$(12, 20) & 16.039$^{+0.073}_{-0.082}$ & 16.042$^{+0.078}_{-0.084}$\\
$\eta_{\omega, \mathrm{RV}}$ & $\mathcal{U}$(0.1, 1.0) & 0.294$^{+0.062}_{-0.054}$ & 0.298$^{+0.066}_{-0.054}$\\
$M_{\star}$[M$_{\odot}]$ & $\mathcal{N}$(0.548, 0.048) & 0.556$^{+0.042}_{-0.035}$ & 0.556$^{+0.042}_{-0.035}$\\
$R_{\star}$[R$_{\odot}]$ & $\mathcal{N}$(0.542, 0.029) & 0.547$^{+0.025}_{-0.023}$ & 0.548$^{+0.025}_{-0.024}$\\
\hline
\end{tabular}
\tablefoot{The prior label of $\mathcal{N}$ and $\mathcal{U}$ represents the normal and uniform distribution, respectively.}\\
\end{table*}
\endgroup

\begingroup
\renewcommand{\arraystretch}{1.22} 
\begin{table}
\caption{Derived parameters of the joint fit for TOI-1801\,b.}\label{tab:joint-fit_der}
\centering
\begin{tabular}{l@{\hskip 0.01in}c@{\hskip 0.04in}c}
\hline\hline
Parameter & Posterior ($e$\,$=$\,0, $\omega$\,$=$\,$\pi/2$) & Posterior ($e$, $\omega$ free)\\
\hline
$R_{p}^{b}/R_{\star}$ & 0.034$^{+0.001}_{-0.001}$ & 0.034$^{+0.001}_{-0.001}$\\
$R_{p}^{b}$ [R$_{\oplus}$] & 2.079$^{+0.124}_{-0.110}$ & 2.084$^{+0.129}_{-0.115}$\\
$a^{b}/R_{\star}$ & 30.632$^{+1.337}_{-1.378}$ & 30.589$^{+1.501}_{-1.436}$\\
$i^{b}$ [$^\circ$] & 89.507$^{+0.333}_{-0.362}$ & 89.474$^{+0.354}_{-0.352}$\\
$e^{b}$ & 0 & 0.074$^{+0.071}_{-0.050}$\\
$\omega^{b}$ [rad] & $\pi/2$ & --0.108$^{+2.327}_{-2.198}$\\
$M_p^{b}$ [M$_{\oplus}$] & 5.738$^{+1.455}_{-1.428}$ & 5.789$^{+1.535}_{-1.500}$\\
$\rho^{b}$ [g\,cm$^{-3}$] & 3.701$^{+1.217}_{-1.051}$ & 3.712$^{+1.278}_{-1.081}$\\
$T_{\mathrm{eq}}^{b}(A$$=$$0)$ [K] & 493.7$^{+15.1}_{-14.3}$ & 493.9$^{+15.6}_{-15.0}$\\
$T_{\mathrm{eq}}^{b}(A$$=$$0.6)$ [K] & 392.6$^{+12.0}_{-11.4}$ & 392.7$^{+12.4}_{-12.0}$\\
\hline
$u_{1,\mathrm{TESS}}$ & 0.374$^{+0.127}_{-0.117}$ & 0.372$^{+0.129}_{-0.117}$\\
$u_{2,\mathrm{TESS}}$ & 0.144$^{+0.112}_{-0.101}$ & 0.144$^{+0.114}_{-0.101}$\\
$u_{1,\mathrm{LCOGT}\ z_{s}}$ & 0.423$^{+0.296}_{-0.236}$ & 0.414$^{+0.298}_{-0.232}$\\
$u_{2,\mathrm{LCOGT}\ z_{s}}$ & 0.147$^{+0.232}_{-0.219}$ & 0.150$^{+0.231}_{-0.219}$\\
$u_{1,\mathrm{MuSCAT2}\ g'}$ & 0.355$^{+0.260}_{-0.205}$ & 0.351$^{+0.264}_{-0.205}$\\
$u_{2,\mathrm{MuSCAT2}\ g'}$ & 0.180$^{+0.226}_{-0.199}$ & 0.189$^{+0.230}_{-0.203}$\\
$u_{1,\mathrm{MuSCAT2}\ r'}$ & 0.294$^{+0.233}_{-0.176}$ & 0.299$^{+0.234}_{-0.179}$\\
$u_{2,\mathrm{MuSCAT2}\ r'}$ & 0.186$^{+0.212}_{-0.180}$ & 0.180$^{+0.221}_{-0.175}$\\
$u_{1,\mathrm{MuSCAT2}\ z_{s}}$ & 0.320$^{+0.242}_{-0.190}$ & 0.308$^{+0.240}_{-0.182}$\\
$u_{2,\mathrm{MuSCAT2}\ z_{s}}$ & 0.176$^{+0.210}_{-0.175}$ & 0.182$^{+0.216}_{-0.179}$\\
\hline
\end{tabular}
\end{table}
\endgroup

\section{Discussion}
\label{sec:disc}

\subsection{Planet characterisation}

TOI-1801\,b is fully characterised with a planetary radius of $R_p^b$=\,2.08\,$\pm$\,0.12 R$_\oplus$, a semi-amplitude of $K_p^b$=\,2.46\,$\pm$\,0.62 m\,s$^{-1}$, and a derived mass of $M_p^b$=\,5.74\,$\pm$\,1.46 M$_\oplus$ with a 3.9$\sigma$ detection. We derive a bulk density of $\rho^{b}$\,=\,3.70\,$\pm$\,1.22 g\,cm$^{-3}$. Assuming planetary albedos ($A_{\rm Bond}$) in the 0.6--0.0 range, we calculate an equilibrium temperature ($T_\mathrm{eq}$) of 390--490\,K. Therefore, we classify TOI-1801\,b as a sub-Neptune planet slightly warmer than the Earth, but less dense.

To estimate the prospects for atmospheric characterisation of TOI-1801\,b, we followed the method established by \cite{Kempton2018PASP..130k4401K}. TOI-1801\,b has a transmission spectroscopy metric (TSM) of 63$_{-16}^{+25}$, which is below the cut-off value of 92 defining the first quartile of targets for the James Webb Space Telescope (JWST) with the strongest predicted atmospheric detection. Nonetheless, simulated spectra obtained with \texttt{TauREx3} \citep{Al-Refaie2021} and \texttt{ExoTETHyS} \citep{Morello2021} show absorption features larger than 100 parts per million (ppm) in case of an H$_2$-dominated atmosphere, which should be detectable with a single JWST transit observation, depending on the cloud coverage. A much flatter spectrum would be observed in case of a steam H$_2$O atmosphere, for which the predicted absorption features are 10--20\,ppm.

\subsection{Mass-radius diagram}

\begin{figure*}[ht!]
\includegraphics[width=1\linewidth]{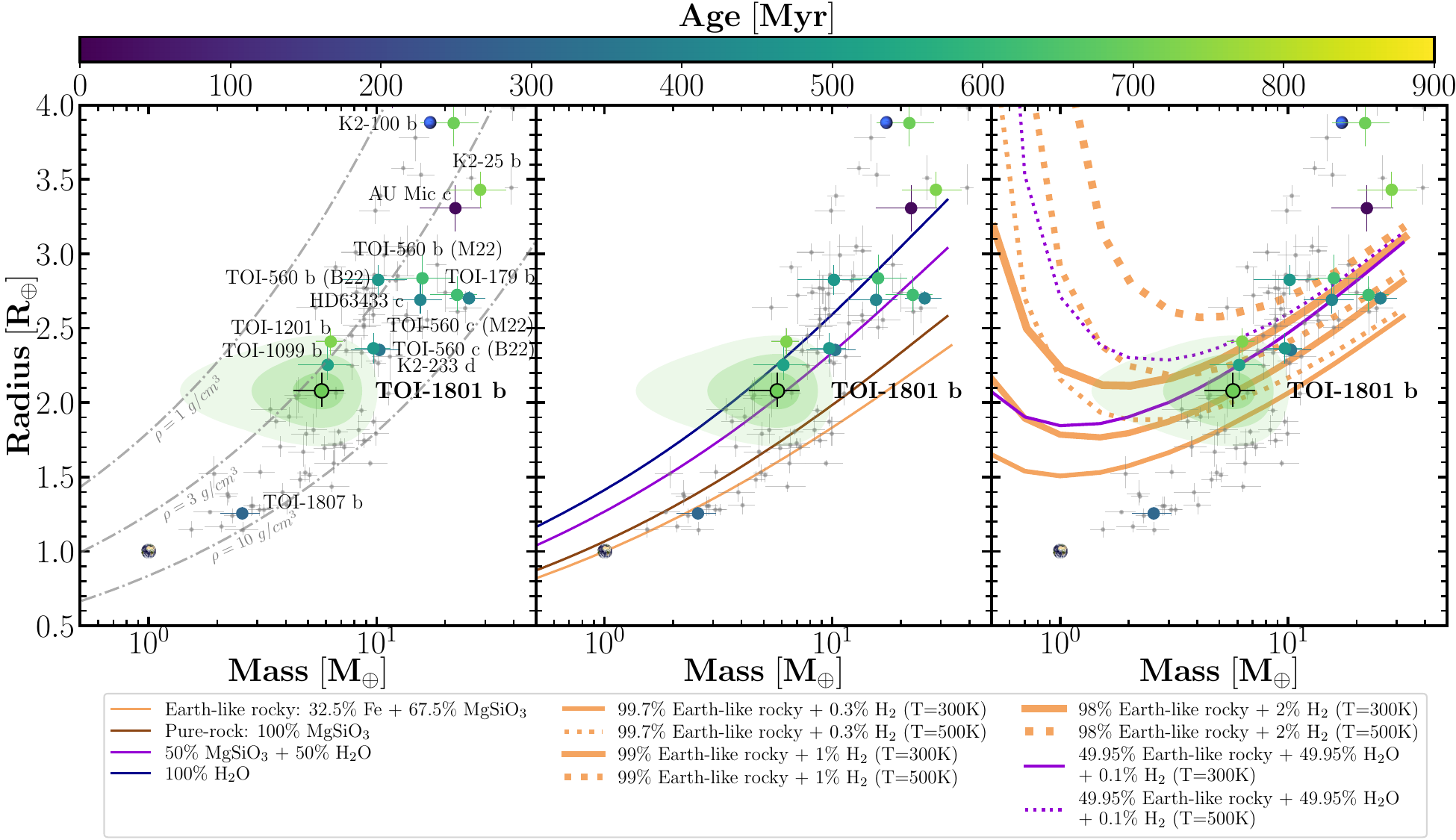}
\caption{Mass-radius diagram for TOI-1801\,b, together with all known exoplanets (grey dots) with a precision better than 8\% in radius (through transit) and 20\% in mass (from RV). The population of young transiting exoplanets ($<$ 900 Myr) with measured masses is plotted as coloured dots, according to their ages. The uncertainties on TOI-1801\,b are shown as coloured shaded regions with 1, 2, and 3$\sigma$ levels of confidence. In the left panel, the iso-density lines are displayed as dashed grey lines. In the remaining panels where the population of exoplanets is plotted along with TOI-1801\,b, coloured lines indicate different composition models without gas (middle panel) and with a gas envelope (right panel) from \cite{Zeng2019PNAS..116.9723Z}. In the right panel, the solid and dotted lines show the models with temperatures of 300\,K and 500\,K, respectively, of the corresponding specific entropy at 100 bar level in the gas envelope. The Earth and Neptune are also depicted as reference. We include B22 and M22 as references because different results have been published for the same planet, and this refers to \cite{barr22} and \cite{elmufti2023}, respectively.
\label{fig:RM}}
\end{figure*}

We show a mass-radius diagram in Fig.\,\ref{fig:RM} focused on the population of small planets ($R_p$\,$\le$\,4 $R_\oplus$), where known transiting planets (taken from the Extrasolar Planets Encyclopedia\footnote{\url{http://exoplanet.eu/}}) with radius uncertainties better than 8\% from the transit method and masses uncertainties better than 20\% from the RV method are plotted as grey dots. We overplot as coloured dots all known planets orbiting stars younger than 900 Myr from our own collection, where the colour scale represents the age of the system. We represent the radius and mass of TOI-1801\,b together with its uncertainty as green shaded regions with 1, 2, and 3$\sigma$ significance. The left panel clearly shows that TOI-1801\,b is the second-youngest planet with the smallest radius and lowest mass after the ultra-short-period planet TOI-1807\,b \citep{nardiello22}.

In the two remaining panels, different internal composition models are shown from \cite{Zeng2019PNAS..116.9723Z}. They do not include an atmosphere in the central panel and include an atmosphere in the right panel. The $\sim$33\% error in the average bulk density is large, and this means that the planetary interior and atmospheric composition is compatible with several composition scenarios. In the centre panel, an Earth-like rocky composition of TOI-1801\,b is ruled out with a confidence of 3$\sigma$ if it does not have an atmosphere. The planet is compatible with a mix of silicates and water or even a pure rock/water composition. On the other hand, the models in the right panel are consistent with a maximum of 2\% of H$_{2}$ in mass, while internally, the planet could be made up of rock, iron, and water, depending on the atmosphere.

\subsection{TOI-1801\,b within the context of M-dwarf planets}
\label{sec:Mdwarf}

\begin{figure}[ht!]
\includegraphics[width=0.99\linewidth]{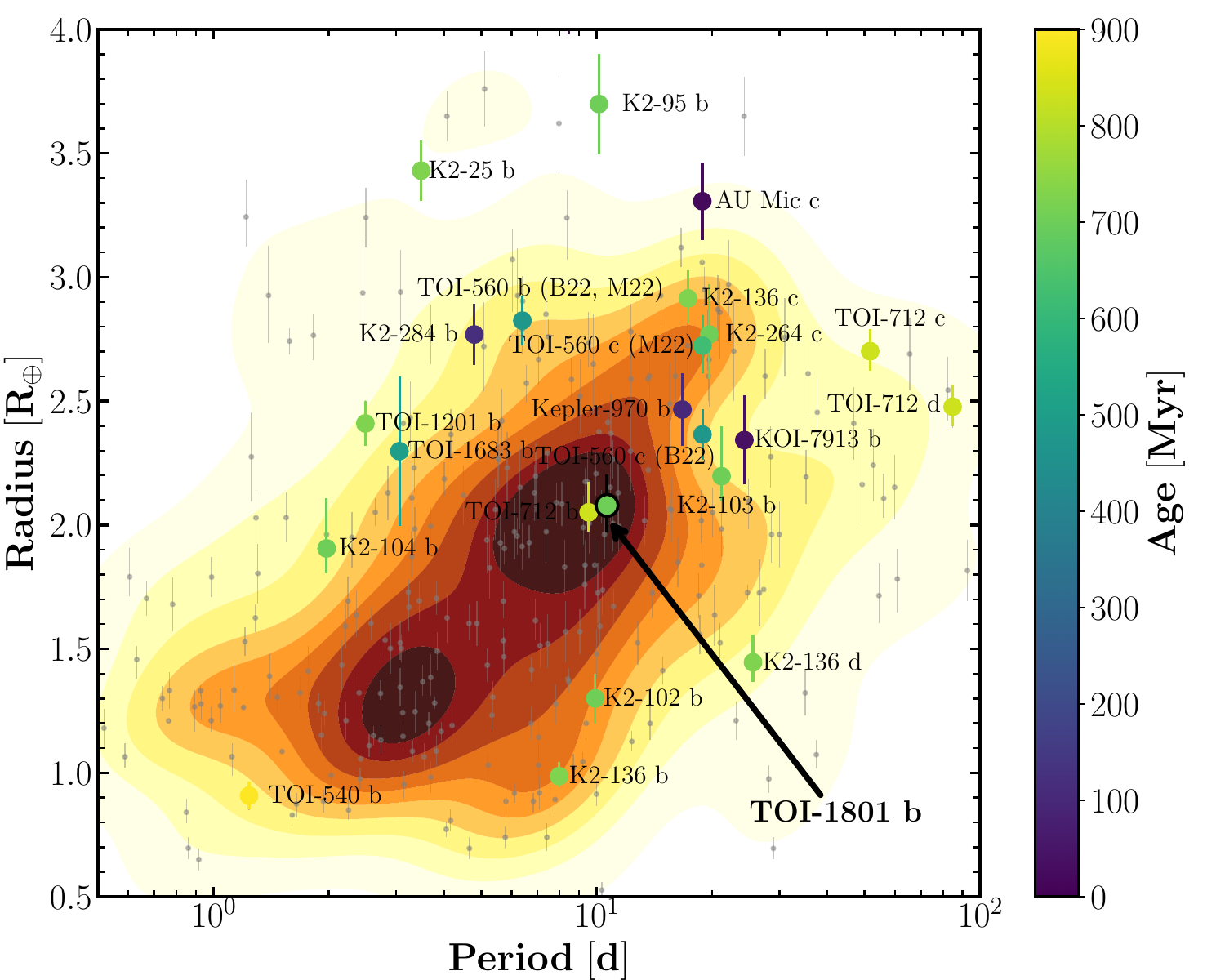}
\includegraphics[width=0.99\linewidth]{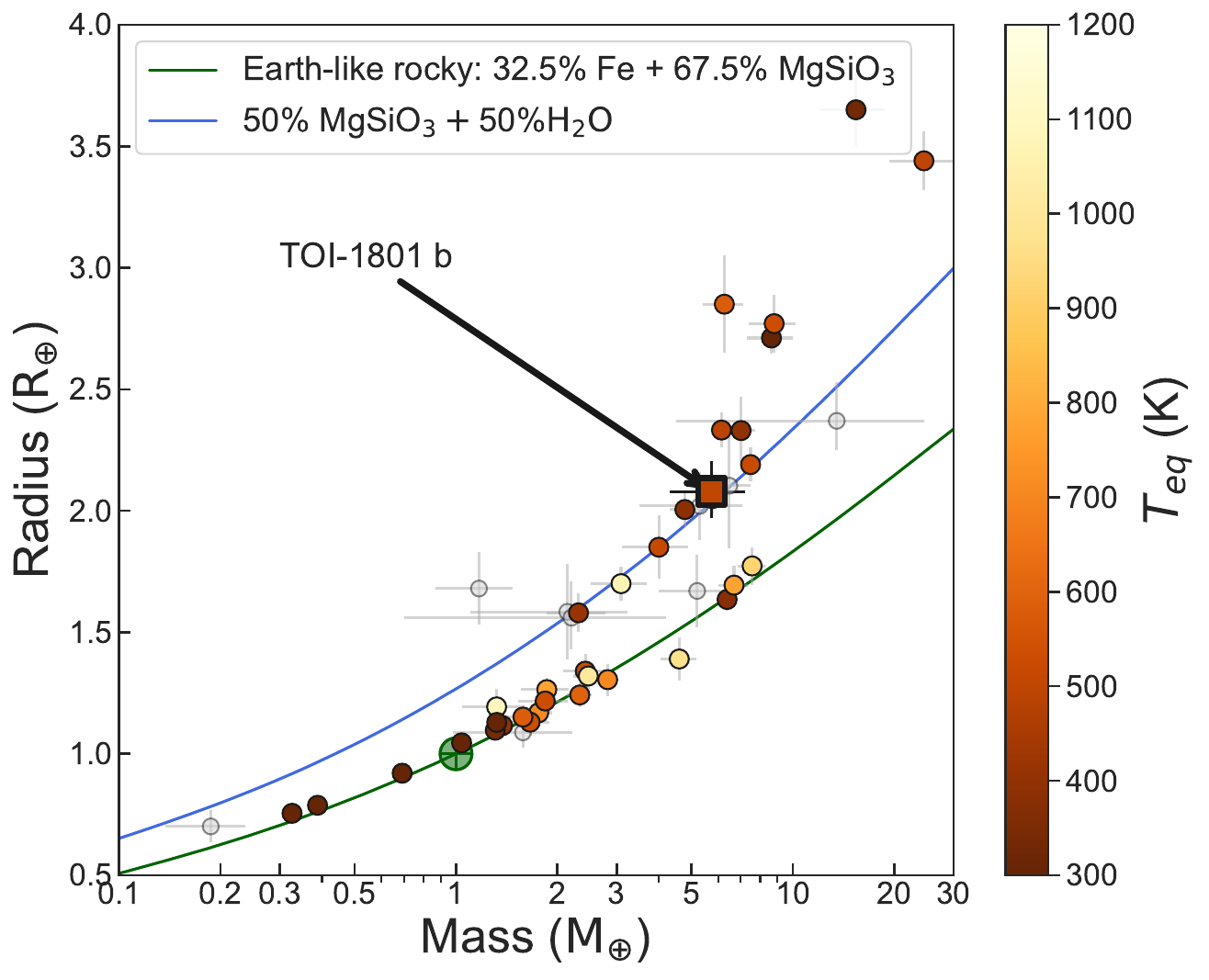}
\caption{Mass-period-radius diagram. \textit{Top:} Period-radius diagram with all known transiting planets around low-mass stars ($T_{\rm \star,eff}$\,$<$\,4700\,K, 0.5 $R_{\oplus}$\,$<$\,$R_p$\,$<$\,4 $R_{\oplus}$, 1 d\,$<$\,$P_{\rm orb}$\,$<$\,100 d, following \citealt{cloutier2020}). The contour lines represent the 2D distribution of the planet radius as a function of orbital period around low-mass stars, where the grey dots indicate each independent value. We overplot the population of young ($<$ 900 My) transiting planets with dots, and colour represents the planet age. The position of TOI-1801\,b is also marked.
\textit{Bottom:} Mass-radius diagram of all known small planets with precisely measured masses ($M_{p}$\,$<25$\%) and radii ($R_{p}$\,$<8$\%), around M-type stars ($T_{\rm eff} <4000$ K) from \citet{luque22b}. Two theoretical composition models \citep{Zeng2019PNAS..116.9723Z} are plotted: an Earth-like composition (mass fractions of 32.5\% iron and 67.5\% silicates, green curve) and a planet consisting of 50\% water-dominated ices and 50\% silicates (blue curve). Planets are colour-coded by their equilibrium temperature. TOI-1801\,b is marked with a square.
\label{fig:mp_diagram}}
\end{figure}

The physical properties of TOI-1801\,b seem to be consistent with the bulk population of transiting planets orbiting low-mass stars as measured by \citet{cloutier2020} using \textit{Kepler/K2} data (Fig.\,\ref{fig:mp_diagram}). These authors argued that the positive slope measured in the period-radius space, opposite in sign to the one measured for Sun-like stars \citep{Fulton18, Martinez19}, could be an indication that thermally driven atmospheric mass loss may not be the dominant mechanism shaping the demographic properties of the super-Earth and sub-Neptune populations around low-mass stars. Although there is no consensus about the accuracy of this slope measurement for low-mass stars \citep[see][for different values]{VanEylen21, Petigura22, luque22b}, a number of studies have challenged the interpretation that the bimodal size distribution of sub-Neptune planets is a consequence of atmospheric evolution (rocky planets with or without large hydrogen-dominated atmospheres), rather attributing it to a difference in planet composition (planets with rocky or water-rich cores; e.g. \citet{Zeng2019PNAS..116.9723Z, Venturini20, Burn21, Izidoro22}). In particular, for a refined sample of M-dwarf planets with precise and accurate bulk densities, \citet{luque22b} showed that the apparent scarcity of small planets with radii between 1.5--2.0\,$R_\oplus$ is due to a combination of the rocky population having a maximum mass of 10\,$M_\oplus$ and the water worlds a minimum mass of 2--3\,$M_\oplus$ (Fig.~\ref{fig:mp_diagram}). The bulk density of TOI-1801\,b makes it consistent with this emerging population of water worlds orbiting M~dwarfs having little or no primordial hydrogen-rich envelopes \citep{Diamond-Lowe22, Cadieux22, Piaulet23, Cherubim23}. 

\subsection{TOI-1801\,b as a young planet}

TOI-1801\,b joins the sample of small young exoplanets with a measured radius, mass, and density (Fig.\,\ref{fig:RM}): AU\,Mic\,c, TOI-1807\,b, TOI-179\,b, K2-233\,d, HD\,63433\,c, TOI-560\,b and c, TOI-1099\,b, K2-25\,b, K2-100\,b, and TOI-1201\,b. These represent less than 10\% of the total population of exoplanets shown in the diagram. So far, the densities of young planets lie in the 1--10 g\,cm$^{-3}$ range, although no young planet has been found with a mass lower than $\sim$5 M$_\oplus$ and the density of Neptune. The population of young planets characterised so far, including TOI-1801\,b, seems to follow the same distribution as the older planets. 

Young planetary systems are ideal for observationally constraining models of planet formation and evolution. The measurement of radius and mass, and therefore, of internal composition, obtained in this work for TOI-1801\,b, allows us to discuss it in this context. If TOI-1801\,b is indeed a water-world (with a water-mass fraction close to 50\%) without an envelope (or with an insignificant envelope), and it had a significant gaseous atmosphere when it formed. The lack of an atmosphere or the very thin atmosphere deduced from our measurements implies that evolutionary mechanisms have removed its atmosphere on timescales of hundreds of million years after it migrated inwards to its current location, and it lost its primordial hydrogen envelope by means of photo-evaporation \citep{owen17} or giant impacts \citep{Wyatt20} on a timescale shorter than 1\,Gyr. A similar conclusion was drawn for HD\,63433\,c, a transiting planet with similar characteristics orbiting a $\sim$400 Myr solar-type star \citep{Mallorquin23, damasso23}.

Assuming an energy-limited approach, we can evaluate the mass-loss rate of TOI-1801\,b due to XUV (X-ray+EUV, 1--920~\AA) stellar irradiation. However, no X-ray observations of TOI-1801 have been reported to date. To approximate the value of the XUV irradiation, we can use the value of the rotation period to calculate the X-ray luminosity, following \citet{Wright2011}, and then calculate the flux in two EUV ranges of interest by applying the relations of \citet{Sanz2022}. Following this procedure, we obtain $L_{\rm X}$(5--100~\AA)\,$=$\,1.2$\times 10^{28}$erg\,s$^{-1}$, $L_{\rm EUV}$(100--920~\AA)\,$=$\,6.2$\times 10^{28}$erg\,s$^{-1}$, and $L_{\rm EUV,He}$(100--504~\AA)\,$=$\,2.6$\times 10^{28}$erg\,s$^{-1}$. The calculated X-ray luminosity implies a value of $\log L_{\rm X}/L_{\rm bol}$\,$=$\,$-4.3$, indicating a moderate to high level of activity, consistent with the upper limit of the estimated age. Finally, the calculated XUV stellar luminosity implies an approximate mass-loss rate in the atmosphere of TOI-1801\,b of 0.08\,M$_\oplus$\,Gyr$^{-1}$ following \citet[][and references therein]{Sanz2011}.

\subsection{Star-planet interaction}

We present here predictions for the radio emission that is expected to arise from magnetic star-planet interaction between the planet TOI-1801\,b and its host star. The mechanism that causes this emission is the electron cyclotron maser (ECM) instability \citep{Melrose1982}, which can generate auroral radio emission in both the planet and its host star. The characteristic frequency of this emission is given by the electron gyrofrequency, $\nu_G$\,=\,2.8\,$B$ MHz, where $B$ is the local magnetic field in the source region, in Gauss. ECM emission is a coherent mechanism that can yield broadband ($\Delta$\,$\nu$$\sim$$\nu_G/2$) highly circularly polarized (sometimes reaching 100\%) amplified non-thermal radiation.

Because the expected magnetic field of TOI-1801\,b is unlikely to be more than just a few Gauss, the associated gyrofrequency falls below the ionosphere cut-off, and so it is not possible to detect radio emission from it. However, in the case of star-planet  interaction, the radio emission instead arises from the magnetosphere of the host star, induced by the crossing of the stellar magnetosphere by the exoplanet, and the  relevant magnetic field is that of the star,  $B_\star$, which may be much larger.   This interaction is expected to  yield detectable auroral radio emission via the cyclotron emission mechanism 
(e.g. \citealt{Zarka2007,Saur2013,Vedantham2020,PerezTorres2021}) because the magnetic field is large enough for the electron gyrofrequency to frequently reach several hundred MHz or even the GHz regime, to which current radio interferometry facilities are very sensitive. The only prerequisite is that this interaction proceeds in the sub-Alfv\'enic regime, so that the energy that feeds the ECM emission is transported from the planet to the star.

We estimated the stellar magnetic field of TOI-1801 using the relations of \citet{Reiners2022}. From this work, we have $P_{\text{rot}}$\,$=$15.98\,$\pm$\,0.66 d. The rotation period is thus shorter than the critical period, defined by \citet{Reiners2014}, $P_{\rm sat}$\,$=$1.6$\times (L_{\rm bol}/L_\odot)^{-1/2}$ d, which is $\sim$\,6.7 d. The resulting magnetic field is about 950 G, so that the ECM emission is expected to detectable at about 2.7 GHz.

We used the models from Appendix B of \citet{PerezTorres2021} to estimate the ECM emission arising from sub-Alfv\'enic star-planet interaction, using two different magnetic field geometries: a closed dipolar geometry, and an open Parker spiral geometry. In both cases, the interaction between the planet and its host star occurs in the sub-Alfv\'enic regime, essentially because the planet is close to its host star. We assumed an isothermal wind with $T$\,$=$\,2$\times 10^6$ K and a the solid angle covered by the ECM emission of 1.6 steradians. 

\begin{figure}[htb]
\includegraphics[width=0.93\linewidth]{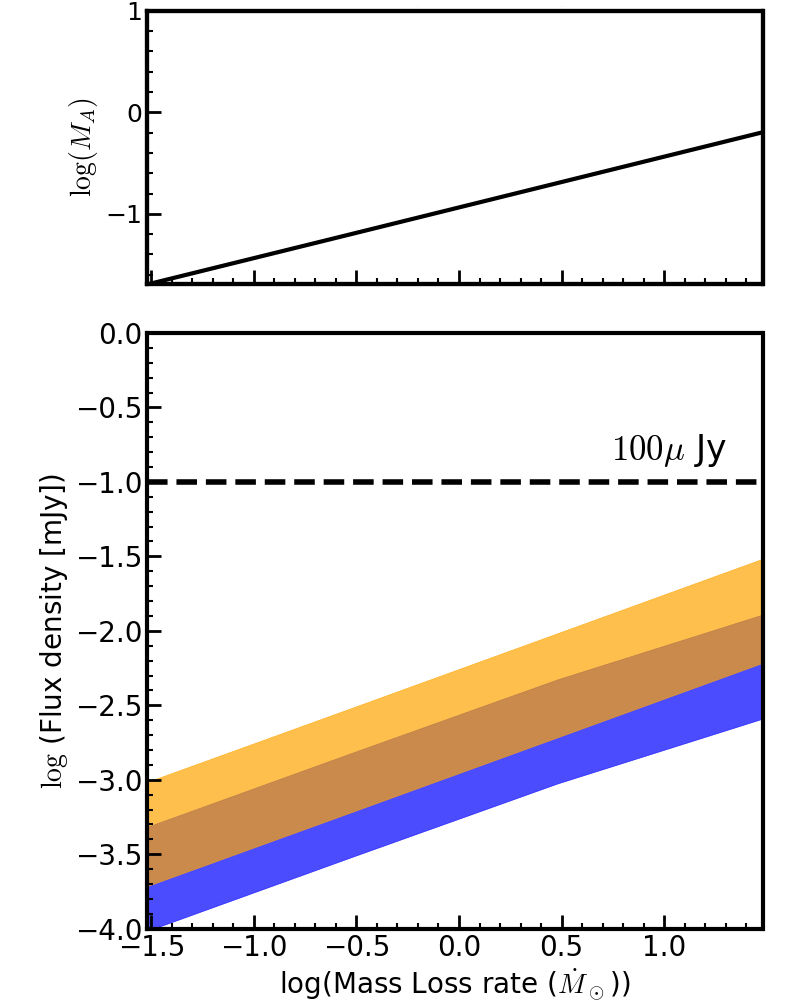}
\includegraphics[width=0.93\linewidth]{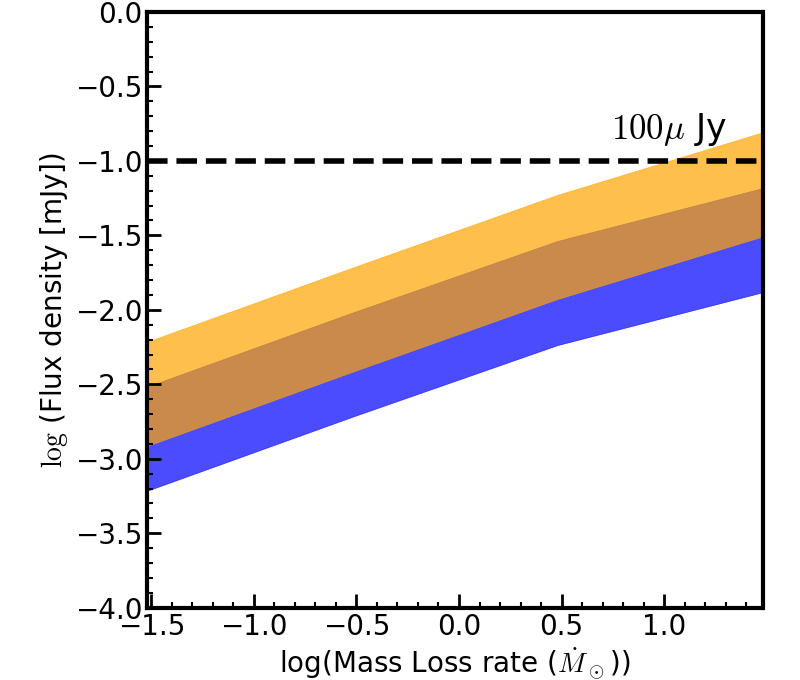}
\caption{Alfv\'en Mach number (upper panel) and flux density (for $B_{\rm pl}$\,$=$\,0 and 1 $G$ for middle and lower panels, respectively) arising from SPI as a function of the stellar mass-loss rate (in solar units). The emission expected from Saur-Turnpenney's model is shown in orange, and the emission expected from the Zarka-Lanz model is shown in blue. The overlap of the models is shown in brown. The dashed black line represents the assumed detection threshold of 100\,$\rm \mu Jy$.}
\label{fig:SPI}
\end{figure}

We determined the expected radio flux density as a function of the mass-loss rate of the star and found that for the open Parker spiral, the radio emission was far too low in all plausible cases to yield any significant radio emission. In Fig.\,\ref{fig:SPI} we show the results for a closed dipolar geometry, both for a non-magnetized planet and for $B_{\rm pl}$\,$=$\,1\,$G$. The  shaded areas encompass the range of values from 0.01 up to 0.05 for the efficiency factor, $\epsilon$, in converting Poynting flux into ECM radio emission, and we assumed a 5$\sigma$ detection of $\sim$\,100\,$\mu$Jy in both cases. If the planet is not magnetized (Fig.\,\ref{fig:SPI} middle panel), we do not expect to detect any significant radio emission due to star-planet interaction. Conversely, if the planet is magnetised (Fig.\,\ref{fig:SPI} lower panel), with a magnetic field several times that of the Earth, it is possible to detect emission arising from star-planet interaction if the mass-loss rate of TOI-1801 is also similar to that of the Sun, or even higher. The mass-loss rate of the star should be at least 2\,$\dot{M}_\odot$ for a clear radio-emitting signal. While the prospects for a radio detection of star-planet interaction in this system are not promising, we note that a detection of this radio emission could be used to constrain the mass-loss rate of the star and the magnetic field of the planet. 

\section{Conclusions}
\label{sec:Concl}

We presented the validation and characterisation of TOI-1801\,b, a mini-Neptune orbiting a moderately young M0.5\,V star with a period of 10.6 days. We simultaneously fit the TESS light curves with CARMENES and HIRES high-resolution spectroscopy data, and we determined a radius of 2.08\,$\pm$\,0.12 R$_\oplus$ and a mass of 5.74\,$\pm$\,1.46 M$_\oplus$, resulting in a bulk density of 3.70\,$\pm$\,1.22 g\,cm$^{-3}$. In addition, and using the $Gaia$ astrometry, we confirmed that TOI-1801 has a wide ($\sim$3000\,AU) M4V companion, LP\,375-24, forming a young system whose age has been constrained from several age indicators to the range of 600--800 Myr.

According to theoretical models and the observational results found by \citet{luque22b}, the planet TOI-1801\,b is mostly made of silicates and water, perhaps with a gaseous envelope with less than 2\% of H$_{2}$ by mass. Our results suggest that if TOI-1801\,b most probably formed beyond the ice line and had a larger atmosphere in the past, it has already lost most of it during its 600--800 Myr life time. This result favours rapid evolutionary mechanisms of mass loss, such as photo-evaporation.

\begin{acknowledgements}
This paper includes data collected by the \tess{} mission. Funding for the \tess{} mission is provided by the NASA Explorer Program. We acknowledge the use of public TOI Release data from pipelines at the \tess{} Science Office and at the \tess{} Science Processing Operations Center. Resources supporting this work were provided by the NASA High-End Computing (HEC) Program through the NASA Advanced Supercomputing (NAS) Division at Ames Research Center for the production of the SPOC data products. This research has made use of the Exoplanet Follow-up Observation Program website, which is operated by the California Institute of Technology, under contract with the National Aeronautics and Space Administration under the Exoplanet Exploration Program.

This work has made use of data from the European Space Agency (ESA) mission {\it Gaia} (\url{https://www.cosmos.esa.int/gaia}), processed by the {\it Gaia} Data Processing and Analysis Consortium (DPAC, \url{https://www.cosmos.esa.int/web/gaia/dpac/consortium}). Funding for the DPAC has been provided by national institutions, in particular the institutions participating in the {\it Gaia} Multilateral Agreement.

CARMENES is an instrument at the Centro Astron\'omico Hispano en Andaluc\'ia (CAHA) at Calar Alto (Almer\'{\i}a, Spain), operated jointly by the Junta de Andaluc\'ia and the Instituto de Astrof\'isica de Andaluc\'ia (CSIC). CARMENES was funded by the Max-Planck-Gesellschaft (MPG), the Consejo Superior de Investigaciones Cient\'{\i}ficas (CSIC), the Ministerio de Econom\'ia y Competitividad (MINECO) and the European Regional Development Fund (ERDF) through projects FICTS-2011-02, ICTS-2017-07-CAHA-4, and CAHA16-CE-3978, and the members of the CARMENES Consortium (Max-Planck-Institut f\"ur Astronomie, Instituto de Astrof\'{\i}sica de Andaluc\'{\i}a, Landessternwarte K\"onigstuhl, Institut de Ci\`encies de l'Espai, Institut f\"ur Astrophysik G\"ottingen, Universidad Complutense de Madrid, Th\"uringer Landessternwarte Tautenburg, Instituto de Astrof\'{\i}sica de Canarias, Hamburger Sternwarte, Centro de Astrobiolog\'{\i}a and Centro Astron\'omico Hispano-Alem\'an), with additional contributions by the MINECO, the Deutsche Forschungsgemeinschaft through the Major Research Instrumentation Programme and Research Unit FOR2544 ''Blue Planets around Red Stars'', the Klaus Tschira Stiftung, the states of Baden-W\"urttemberg and Niedersachsen, and by the Junta de Andaluc\'{\i}a.

This work is partly supported by JSPS KAKENHI Grant Numbers JP18H05439 and JST CREST Grant Number JPMJCR176. This article is partly based on observations made with the MuSCAT2 instrument, developed by ABC, at Telescopio Carlos S\'{a}nchez operated on the island of Tenerife by the IAC in the Spanish Observatorio del Teide. 

Some of the observations in this paper made use of the High-Resolution Imaging instrument ‘Alopeke and were obtained under Gemini LLP Proposal Number: GN/S-2021A-LP-105. ‘Alopeke was funded by the NASA Exoplanet Exploration Program and built at the NASA Ames Research Center by Steve B. Howell, Nic Scott, Elliott P. Horch, and Emmett Quigley. Alopeke was mounted on the Gemini North telescope of the international Gemini Observatory, a program of NSF's OIR Lab, which is managed by the Association of Universities for Research in Astronomy (AURA) under a cooperative agreement with the National Science Foundation. On behalf of the Gemini partnership: the National Science Foundation (United States), National Research Council (Canada), Agencia Nacional de Investigaci\'on y Desarrollo (Chile), Ministerio de Ciencia, Tecnolog\'ia e Innovaci\'on (Argentina), Minist\'erio da Ci\^encia, Tecnologia, Inovaç\~oes e Comunicaç\~oes (Brazil), and Korea Astronomy and Space Science Institute (Republic of Korea).

PPP, BC, DV and MRM would like to acknowledge the following iSHELL observers: Claire Geneser, Ahmad Sohani, John Berberian, Patrick Nercessian, Jennah Fayaz, Kevin I Collins and Ian Helm. PPP would like to acknowledge support from NASA (Exoplanet Research Program Award \#80NSSC20K0251, TESS Cycle 3 Guest Investigator Program Award \#80NSSC21K0349, JPL Research and Technology Development, and Keck Observatory Data Analysis) and the NSF (Astronomy and Astrophysics Grants \#1716202 and 2006517), and the Mt Cuba Astronomical Foundation.

This work makes use of observations from the Las Cumbres Observatory global telescope network. Part of the LCOGT telescope time was granted by NOIRLab through the Mid-Scale Innovations Program (MSIP). MSIP is funded by NSF.

The Joan Or\'o Telescope (TJO) of the Montsec Observatory (OdM) is owned by the Catalan Government and operated by the Institute for Space Studies of Catalonia (IEEC).

Ariel Postdoctoral Fellowship program of the Swedish National Space Agency (SNSA).

The results reported herein benefitted from collaborations and/or information exchange within NASA’s Nexus for Exoplanet System Science (NExSS) research coordination network sponsored by NASA’s Science Mission Directorate under Agreement No. 80NSSC21K0593 for the program “Alien Earths”. 

G.N. thanks for the research funding from the Ministry of Education and Science programme the "Excellence Initiative - Research University" conducted at the Centre of Excellence in Astrophysics and Astrochemistry of the Nicolaus Copernicus University in Toru\'n, Poland.

This work is partly financed by the Spanish Ministry of Economics and Competitiveness through projects PGC2018-098153-B-C31, PID2019-109522GB-C5[1:4]. 

E. G. acknowledges the generous support from the Deutsche Forschungsgemeinschaft (DFG) of the grant HA3279/14-1.

P. D. acknowledges support from a 51 Pegasi b Postdoctoral Fellowship from the Heising-Simons Foundation.

D. H. acknowledges support from the Alfred P. Sloan Foundation, the National Aeronautics and Space Administration (80NSSC21K0652) and the Australian Research Council (FT200100871).

\end{acknowledgements}

\bibliographystyle{aa.bst} 
\bibliography{biblio.bib}

\begin{appendix}

\section{Radial velocity data}

\longtab[1]{
\begin{longtable}{crr}
\caption{RV data from CARMENES VIS.} \label{tab:CARMV_RV}\\ 
\hline\hline
Time & \multicolumn{1}{c}{RV} & \multicolumn{1}{c}{$\sigma$} \\

[BJD] & [m\,s$^{-1}$] & [m\,s$^{-1}$] \\ 
\hline
\endfirsthead
\caption{continued} \\ 
\hline
Time & \multicolumn{1}{c}{RV} & \multicolumn{1}{c}{$\sigma$} \\

[BJD] & [m\,s$^{-1}$] & [m\,s$^{-1}$] \\ 
\hline
\endhead
\hline
\endfoot
\hline\hline
\endlastfoot
2459244.7009 & 2.86 & 3.13\\ 
2459247.6372 & 5.41 & 3.50\\ 
2459249.7038 & --4.47 & 4.05\\ 
2459250.5704 & --2.54 & 3.04\\ 
2459264.6986 & --1.26 & 2.70\\ 
2459265.5968 & 0.45 & 2.44\\ 
2459266.5624 & --6.49 & 2.25\\ 
2459278.6018 & 10.33 & 3.26\\ 
2459295.5650 & 6.02 & 3.80\\ 
2459299.5185 & --7.02 & 2.05\\ 
2459300.6209 & --12.85 & 3.14\\ 
2459301.4828 & --16.30 & 3.25\\ 
2459302.4694 & --3.84 & 3.20\\ 
2459307.5801 & --12.73 & 3.26\\ 
2459308.5734 & --5.50 & 3.61\\ 
2459334.5737 & 0.33 & 3.91\\ 
2459336.4071 & 8.17 & 2.56\\ 
2459338.4429 & --0.50 & 2.14\\ 
2459339.4163 & --6.70 & 2.31\\ 
2459340.3770 & --5.05 & 2.28\\ 
2459341.4571 & --1.92 & 3.05\\ 
2459342.4889 & --3.62 & 1.86\\ 
2459346.4100 & 4.67 & 2.51\\ 
2459354.4050 & --0.49 & 2.02\\ 
2459355.3798 & 4.43 & 2.27\\ 
2459356.4524 & 0.75 & 2.09\\ 
2459358.5112 & 1.08 & 4.66\\ 
2459359.3980 & 5.05 & 1.88\\ 
2459360.3908 & 4.98 & 2.44\\ 
2459363.4402 & 2.13 & 2.39\\ 
2459364.3888 & --6.61 & 2.22\\ 
2459367.4516 & --1.47 & 2.78\\ 
2459368.3975 & 4.54 & 2.12\\ 
2459370.4566 & --3.32 & 6.29\\ 
2459372.4031 & --1.91 & 2.00\\ 
2459386.4085 & --6.42 & 2.04\\ 
2459387.4150 & --3.26 & 3.63\\ 
2459388.4044 & 1.43 & 4.04\\ 
2459390.4002 & 6.53 & 2.11\\ 
2459391.4041 & 8.93 & 3.06\\ 
2459392.3836 & 4.75 & 2.05\\ 
2459393.4016 & 8.81 & 3.12\\ 
2459560.7477 & 5.17 & 2.37\\ 
2459576.7076 & 0.30 & 2.76\\ 
2459595.6807 & 10.73 & 1.57\\ 
2459597.6748 & 3.90 & 2.22\\ 
2459603.6574 & 10.01 & 2.02\\ 
2459608.6485 & 2.74 & 2.63\\ 
2459610.6421 & 6.33 & 2.18\\ 
2459612.6807 & 9.20 & 2.14\\ 
2459614.6311 & --5.73 & 2.28\\ 
2459616.7499 & --9.26 & 3.14\\ 
2459618.6230 & 3.78 & 2.05\\ 
2459620.6209 & --1.45 & 1.98\\ 
2459622.6130 & --5.79 & 1.62\\ 
2459632.6398 & --13.40 & 2.49\\ 
2459634.5765 & 7.02 & 2.44\\ 
2459640.5604 & 1.52 & 2.01\\ 
2459648.5406 & --0.20 & 3.78\\ 
2459672.5104 & 3.84 & 4.96\\ 
2459680.4698 & --5.36 & 3.02\\ 
2459685.4754 & --6.84 & 2.91\\ 
2459695.4228 & --7.67 & 2.12\\ 
2459699.5636 & 14.81 & 2.63\\ 
2459705.5196 & --2.77 & 1.91\\ 
2459707.4364 & 0.80 & 1.69\\ 
2459712.3640 & --3.23 & 2.07\\ 
2459715.4000 & --7.16 & 5.02\\ 
2459718.4110 & 4.13 & 2.01\\ 
2459720.4219 & 11.08 & 2.45\\ 
2459722.4232 & --3.59 & 3.22\\ 
2459724.4797 & 13.26 & 5.74\\ 
2459726.3978 & --1.45 & 2.10\\ 
2459728.3969 & --3.73 & 1.82\\ 
2459730.3920 & 11.67 & 3.24\\ 
2459734.4793 & --4.22 & 5.12\\ 
2459736.4023 & --1.25 & 1.95\\ 
2459738.3955 & --3.39 & 2.00\\ 
2459740.3907 & 2.42 & 1.90\\ 
2459742.3866 & 0.20 & 2.53\\ 
\hline
\end{longtable}
}

\longtab[2]{
\begin{longtable}{crr}
\caption{RV data from CARMENES NIR.} \label{tab:CARMN_RV}\\ 
\hline\hline
Time & \multicolumn{1}{c}{RV} & \multicolumn{1}{c}{$\sigma$} \\

[BJD] & [m\,s$^{-1}$] & [m\,s$^{-1}$] \\ 
\hline
\endfirsthead
\caption{continued} \\ 
\hline
Time & \multicolumn{1}{c}{RV} & \multicolumn{1}{c}{$\sigma$} \\

[BJD] & [m\,s$^{-1}$] & [m\,s$^{-1}$] \\ 
\hline
\endhead
\hline
\endfoot
\hline\hline
\endlastfoot
2459244.7014 & --0.35 & 8.71\\ 
2459247.6376 & --1.23 & 16.48\\ 
2459250.5705 & 11.28 & 12.46\\ 
2459264.6973 & --8.03 & 17.70\\ 
2459265.5961 & --4.86 & 10.65\\ 
2459278.6015 & 3.37 & 14.03\\ 
2459296.6471 & 2.00 & 22.07\\ 
2459299.5183 & --1.79 & 9.19\\ 
2459300.6205 & --16.01 & 7.60\\ 
2459301.4819 & 9.75 & 19.42\\ 
2459302.4690 & 0.67 & 7.99\\ 
2459307.5801 & --10.41 & 16.07\\ 
2459308.5727 & 15.17 & 11.26\\ 
2459336.4067 & 9.24 & 8.57\\ 
2459338.4424 & 6.79 & 9.27\\ 
2459339.4175 & --6.86 & 5.67\\ 
2459340.3768 & 0.32 & 5.74\\ 
2459341.4556 & --20.55 & 13.15\\ 
2459342.4881 & 9.87 & 5.69\\ 
2459346.4093 & 16.29 & 7.64\\ 
2459350.4255 & 2.54 & 6.80\\ 
2459354.4054 & --5.16 & 6.30\\ 
2459355.3801 & --9.33 & 5.21\\ 
2459356.4504 & --3.22 & 6.09\\ 
2459358.5106 & --9.09 & 21.45\\ 
2459359.3975 & 1.62 & 9.19\\ 
2459360.3900 & 3.85 & 8.51\\ 
2459363.4395 & --3.72 & 7.24\\ 
2459364.3872 & --0.32 & 7.96\\ 
2459368.3971 & --0.69 & 7.03\\ 
2459372.4024 & --11.64 & 4.61\\ 
2459386.4081 & --14.77 & 8.97\\ 
2459390.3998 & 8.32 & 11.23\\ 
2459391.4038 & 17.54 & 11.93\\ 
2459392.3833 & 15.92 & 11.63\\ 
2459393.4017 & 3.55 & 15.56\\ 
2459533.7039 & --10.70 & 11.73\\ 
2459560.7482 & 4.45 & 6.75\\ 
2459576.7081 & --4.21 & 8.10\\ 
2459595.6809 & 12.10 & 6.50\\ 
2459597.6763 & 0.96 & 8.33\\ 
2459603.6575 & 17.91 & 8.76\\ 
2459608.6495 & 10.22 & 8.70\\ 
2459610.6425 & 8.70 & 11.64\\ 
2459612.6805 & 2.77 & 11.21\\ 
2459614.6313 & --2.48 & 7.16\\ 
2459616.7494 & --11.70 & 19.27\\ 
2459618.6229 & 0.77 & 7.98\\ 
2459620.6211 & --7.02 & 19.39\\ 
2459622.6124 & --9.04 & 11.59\\ 
2459632.6387 & --3.10 & 8.93\\ 
2459634.5770 & --11.12 & 7.18\\ 
2459640.5596 & --12.38 & 10.73\\ 
2459680.4703 & --1.40 & 14.68\\ 
2459685.4748 & 4.23 & 9.81\\ 
2459695.4219 & 1.48 & 8.40\\ 
2459705.5195 & 13.04 & 8.86\\ 
2459707.4365 & 6.99 & 7.64\\ 
2459712.3644 & --20.05 & 7.26\\ 
2459715.4004 & 3.72 & 20.20\\ 
2459718.4107 & --3.67 & 5.00\\ 
2459720.4216 & 2.64 & 11.86\\ 
2459722.4227 & 8.57 & 9.34\\ 
2459726.3976 & 9.30 & 6.84\\ 
2459728.3966 & --2.71 & 9.43\\ 
2459736.4025 & --19.96 & 5.51\\ 
2459738.3948 & --8.50 & 12.92\\ 
2459740.3904 & --3.20 & 9.41\\ 
\hline
\end{longtable}
}

\longtab[3]{
\begin{longtable}{crr}
\caption{RV data from HIRES.} \label{tab:HIRES_RV} \\ 
\hline\hline
Time & \multicolumn{1}{c}{RV} & \multicolumn{1}{c}{$\sigma$} \\

[BJD] & [m\,s$^{-1}$] & [m\,s$^{-1}$] \\ 
\hline
\endfirsthead
\caption{continued} \\ 
\hline
Time & \multicolumn{1}{c}{RV} & \multicolumn{1}{c}{$\sigma$} \\

[BJD] & [m\,s$^{-1}$] & [m\,s$^{-1}$] \\ 
\hline
\endhead
\hline
\endfoot
\hline\hline
\endlastfoot
2459162.1275 & --6.51 & 1.51\\ 
2459182.0809 & 6.91 & 1.56\\ 
2459188.0805 & --5.17 & 1.21\\ 
2459189.0566 & --4.35 & 1.64\\ 
2459208.0827 & 2.13 & 1.14\\ 
2459209.0163 & 2.13 & 1.29\\ 
2459215.0592 & 1.82 & 1.28\\ 
2459268.9138 & --10.91 & 1.41\\ 
2459314.0047 & 6.79 & 1.28\\ 
2459376.7999 & --3.83 & 1.14\\ 
2459377.7809 & 0.99 & 1.06\\ 
2459378.8137 & 3.10 & 1.16\\ 
2459379.8349 & --1.69 & 1.00\\ 
2459384.7819 & --1.69 & 1.18\\ 
2459385.7866 & --3.17 & 1.15\\ 
2459388.8091 & --7.86 & 1.13\\ 
2459395.8121 & 1.44 & 1.36\\ 
2459406.7769 & --1.10 & 1.08\\ 
2459538.1061 & 8.26 & 1.20\\ 
2459541.0984 & --7.09 & 1.28\\ 
2459543.0925 & --12.06 & 1.36\\ 
2459566.0240 & --4.93 & 1.37\\ 
2459592.9620 & 13.59 & 1.70\\ 
2459598.9348 & --10.31 & 1.45\\ 
2459739.7976 & --5.39 & 1.07\\ 
2459741.8376 & --0.48 & 1.19\\ 
2459742.8058 & --12.17 & 1.27\\ 
2459744.7889 & --6.67 & 1.21\\ 
2459747.7768 & --4.78 & 1.10\\ 
\hline
\end{longtable}
}

\begin{table}
\caption{RV data of iSHELL.} \label{tab:iSHELL_RV}
\begin{tabular}{crr}
\hline\hline
Time & \multicolumn{1}{c}{RV} & \multicolumn{1}{c}{$\sigma$} \\

[BJD] & [m\,s$^{-1}$] & [m\,s$^{-1}$] \\ 
\hline
2458986.8506 & --36.33 & 9.56\\ 
2459002.7771 & --49.22 & 8.55\\ 
2459004.7726 & --55.76 & 10.14\\ 
2459014.7949 & --40.91 & 9.72\\ 
2459016.7501 & --42.07 & 7.23\\ 
2459021.7473 & --18.39 & 10.77\\ 
2459022.7458 & --56.44 & 7.84\\ 
2459217.1352 & --22.12 & 8.01\\ 
2459220.1138 & --27.07 & 5.24\\ 
2459221.1090 & --28.66 & 7.15\\ 
2459233.0900 & 4.01 & 9.60\\ 
2459255.0838 & --11.64 & 8.42\\ 
2459257.0615 & --18.93 & 7.67\\ 
2459261.1317 & --2.92 & 10.57\\ 
2459319.9487 & --57.88 & 8.37\\ 
2459320.9520 & --55.76 & 9.77\\ 
2459321.9525 & --41.00 & 10.31\\ 
2459322.9070 & --68.26 & 7.68\\ 
2459328.8462 & --37.31 & 7.52\\ 
2459362.8718 & --5.05 & 13.48\\ 
2459363.8252 & 18.53 & 12.47\\ 
2459370.8613 & 17.83 & 29.97\\ 
2459371.7731 & --10.27 & 21.07\\ 
\hline\hline
\end{tabular}
\end{table}

\section{Rotation period: Seasonal analysis}
\label{appendixB}

\begin{figure}[ht!]
\includegraphics[width=1.1\linewidth]{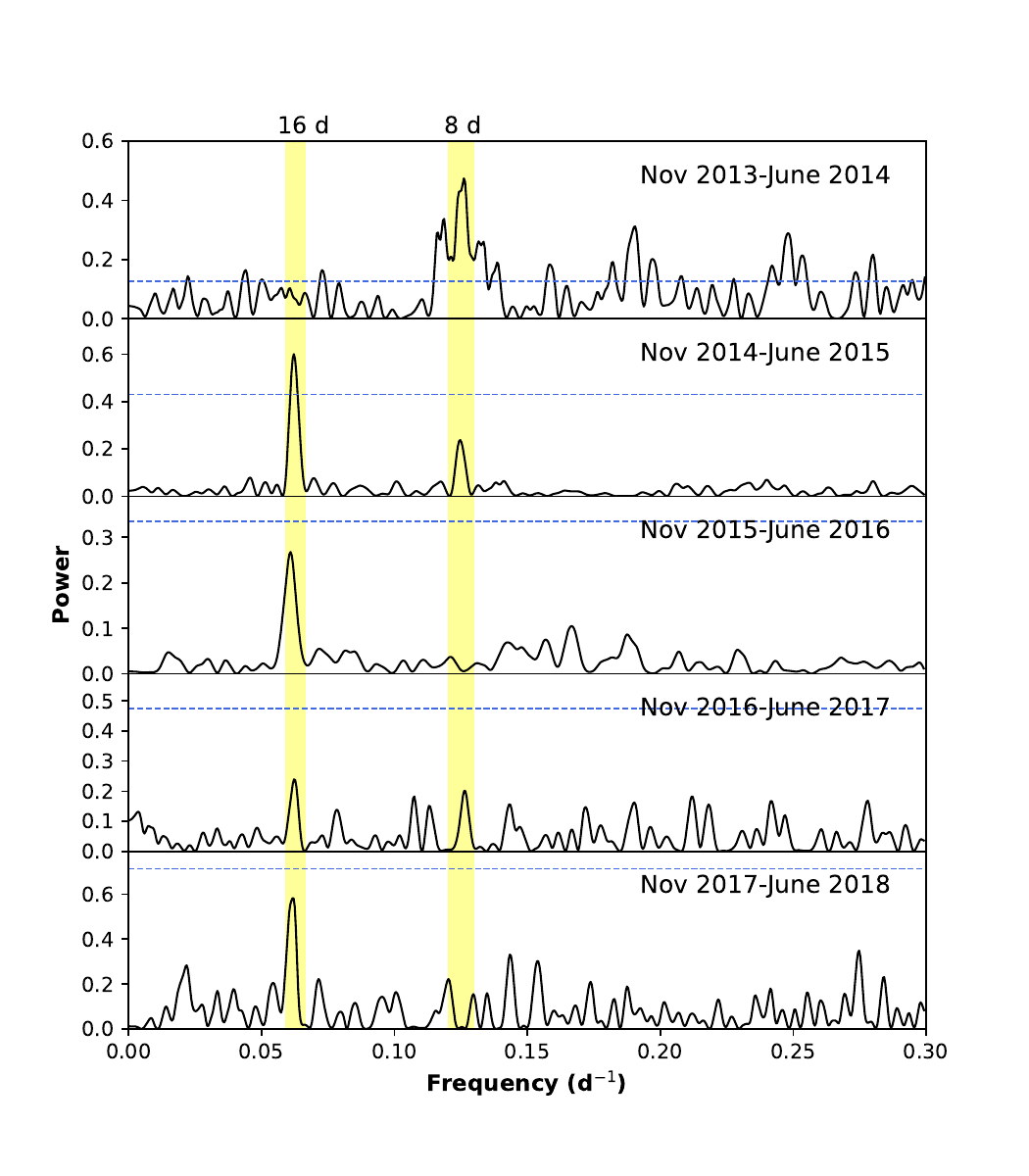}
\caption{GLS periodogram analysis of ASAS-SN photometry. In all the panels, the two shadow vertical yellow bands indicate the stellar rotation period (15--17 days) and its first harmonic. The  dashed  horizontal  blue  line  corresponds to FAP levels of 0.1\%.
\label{fig:GLS_ASASSN_seasons}}
\end{figure}

\begin{figure}[ht!]
\includegraphics[width=1.1\linewidth]{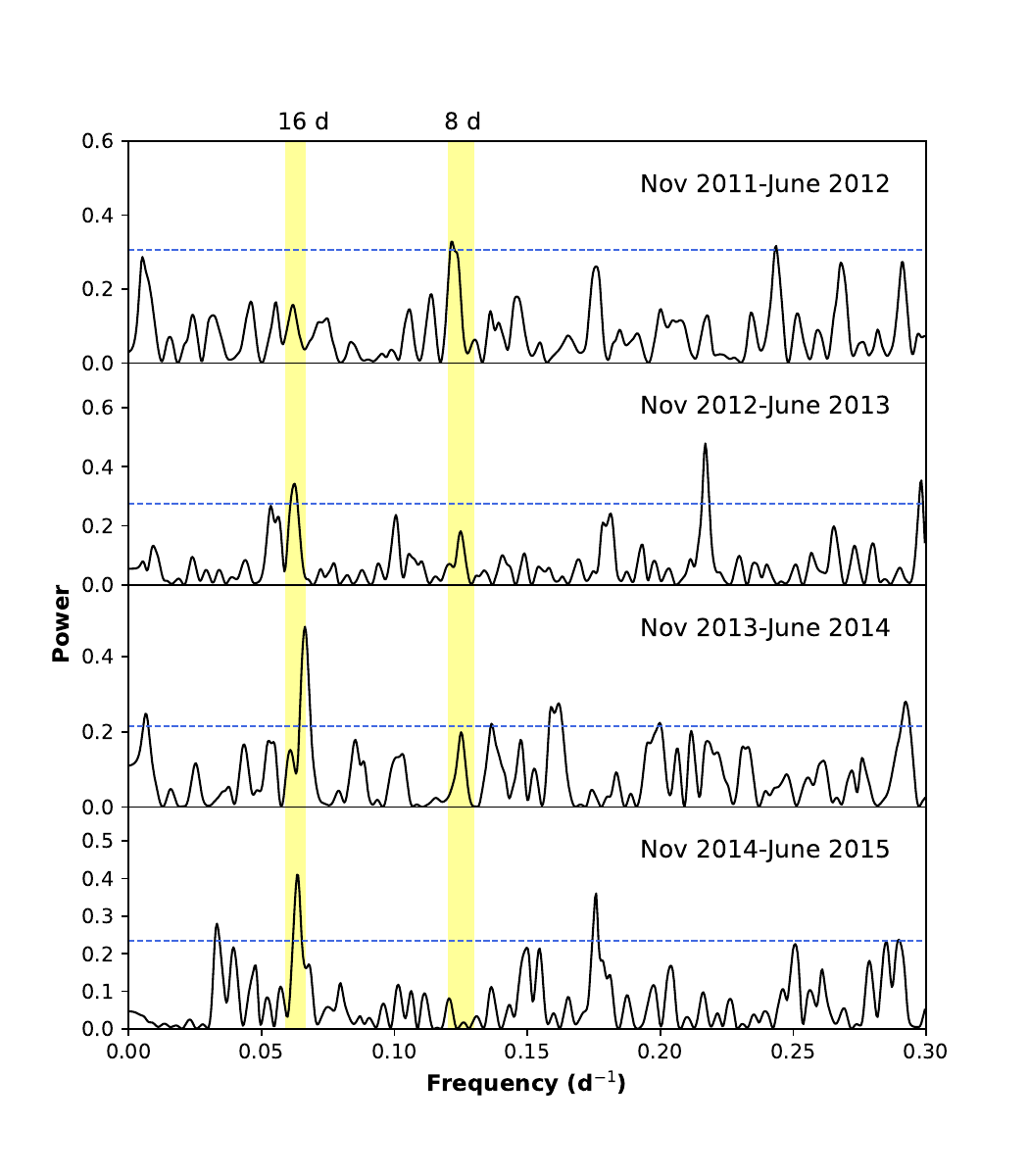}
\caption{GLS periodogram analysis of MEarth photometry. In all the panels, the two shadow vertical yellow bands indicate the stellar rotation period (15--17 days) and its first harmonic. The  dashed  horizontal  blue  line  corresponds to FAP levels of 0.1\%.
\label{fig:GLS_MEarth_seasons}}
\end{figure}

Figures\,\ref{fig:GLS_ASASSN_seasons} and \ref{fig:GLS_MEarth_seasons} show the seasonal analysis of the stellar rotation period in the ASAS-SN photometry data from 2014 to 2019 and in MEarth photometry from 2012 to 2015.

\section{Wide companion}
\label{appendixC}

\begin{figure}[ht!]
\includegraphics[width=1\linewidth]{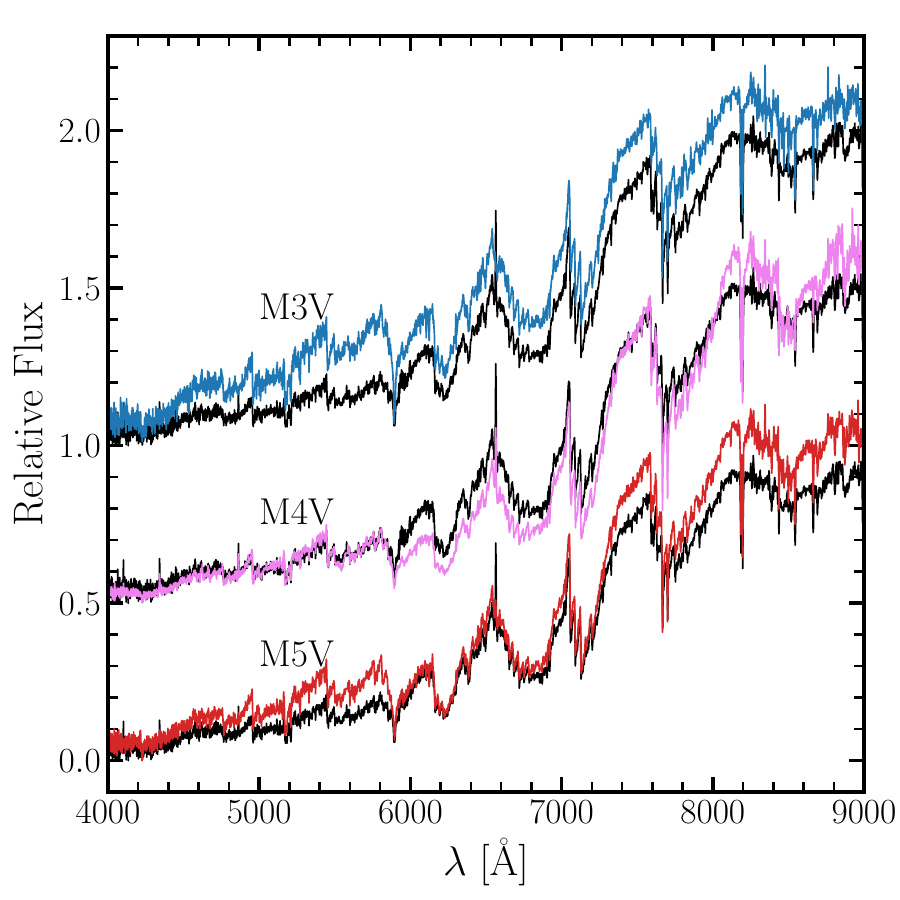}
\caption{Spectral type determination of LP\,375-24. The black and coloured lines show the LAMOST spectra of LP\,375-24 and the templates from SDSS, respectively.  
\label{fig:LP_375-24_SpT}}
\end{figure}

\begin{figure}[ht!]
\includegraphics[width=1\linewidth]{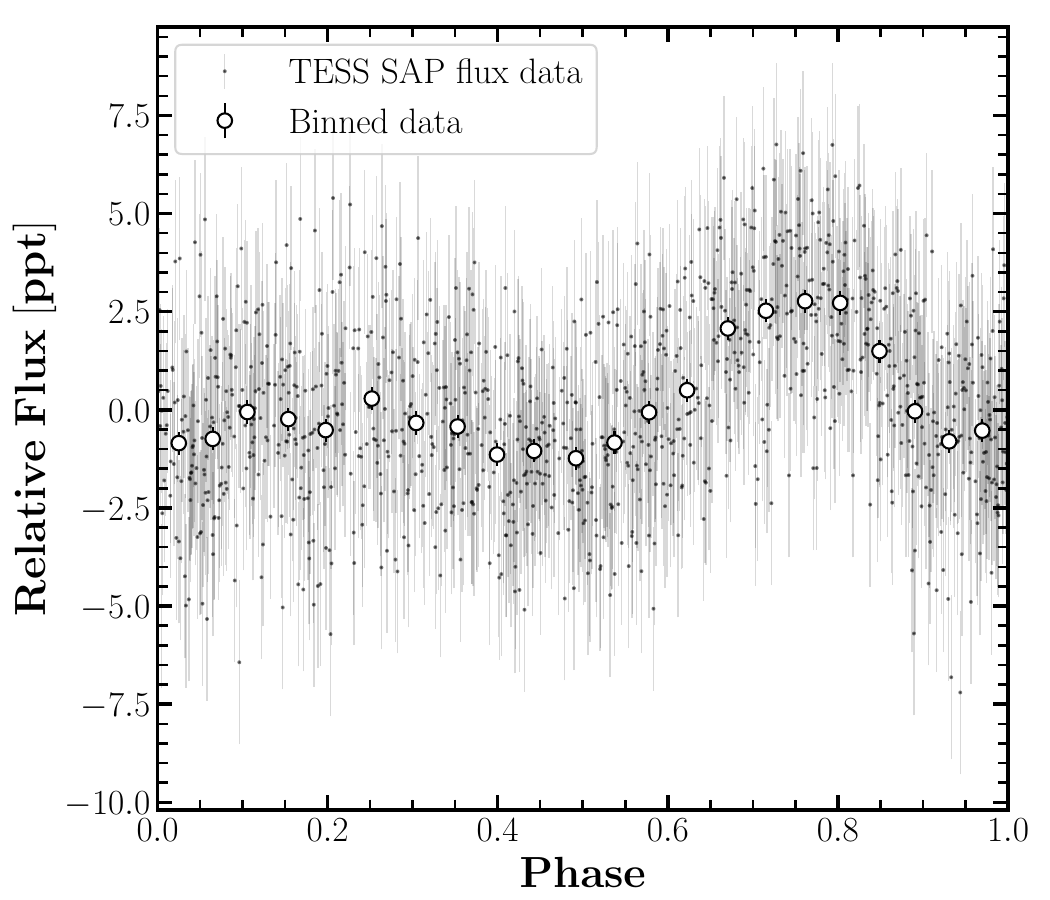}
\caption{TESS phased-folded light curve of LP\,375-24 in sector 22 and sector 49 at the period of 1.30 days.    
\label{fig:LP_375-24_Prot}}
\end{figure}

Figures\,\ref{fig:LP_375-24_SpT} and \ref{fig:LP_375-24_Prot} show the spectral determination of LP\,375-24 by comparison with the SDSS templates and the determination of its stellar rotation period from the TESS light curves, respectively. The retrieved stellar parameters are given in Table\,\ref{tab:stellar_parameters_comp}.

\begin{table}
\caption{Stellar parameters of LP\,375-24.} \label{tab:stellar_parameters_comp}
\centering
\begin{tabular}{lcr}
\hline\hline
Parameter & Value & Reference\\
\hline
Name & LP~375-24 & Luy79\\
     & TIC~119584394  & TIC\\
\noalign{\smallskip}
$\alpha$ (J2016) & 11:42:21.0 & \textit{Gaia} DR3\\
$\delta$ (J2016) & +23:03:14.8  & \textit{Gaia} DR3\\
Sp. type & M4V\,$\pm$\,1V & This work \\ 
\noalign{\smallskip}
$\varpi$ [mas] & 32.404\,$\pm$\,0.023 & \textit{Gaia} DR3\\
$d$ [pc] & 30.893\,$\pm$\,0.020 & \textit{Gaia} DR3\\
RUWE & $1.197$ & \textit{Gaia} DR3\\
$T_{\text{eff}}$ [K] & 3210\,$\pm$\,100 & This work \\
$M_{\star}$ [$M_{\odot}$] & 0.279 & Newton16 \\
$R_{\star}$ [$R_{\odot}$] & 0.285 & Newton16 \\
$v\sin i$ [km s$^{-1}$] & 11.0 & Newton16\\
$P_{\text{rot}}$ [d] & 1.30\,$\pm$\,0.06 & This work\\
$U$ [km s$^{-1}$] & --29.03\,$\pm$\,0.39 & This work\\
$V$ [km s$^{-1}$] & --6.66\,$\pm$\,0.37 & This work\\
$W$ [km s$^{-1}$] & --7.10\,$\pm$\,1.87 & This work\\
Gal. population & Young disk & This work\\
Age [Myr] & 600--800 & This work\\
$NUV$ [mag] & 22.017\,$\pm$\,0.442 & GALEX\\
$B_{p}$ [mag] & 14.691\,$\pm$\,0.004 & \textit{Gaia} DR3\\
$G$ [mag] & 13.193\,$\pm$\,0.003 & \textit{Gaia} DR3\\
$R_{p}$ [mag] & 12.002\,$\pm$\,0.004 & \textit{Gaia} DR3\\
$J$ [mag] & 10.355\,$\pm$\,0.020 & 2MASS\\
\noalign{\smallskip}
\hline
\end{tabular}
\tablebib{
Luy79: \citet{1979nlcs.book.....L}; 
TIC: \citet{Stassun2019};
\textit{Gaia} DR3: \cite{gaia2016, gaiadr3}; 
Newton16: \citet{newton16};
GALEX: \citet{bianchi2017};
2MASS: \citet{skrutskie2006}.
}

\end{table}

\end{appendix}

\end{document}